\documentclass[twocolumn]{aastex631}

\usepackage{rotating}
\usepackage{xcolor, soul}
\usepackage{booktabs}
\usepackage{longtable}

\newcommand\sn{SN~1987A}

\newcommand\spitzer{\textit{Spitzer}}
\newcommand\htwo{H$_{2}$}
\newcommand\kmps{km~s$^{-1}$}

\begin{document}

\title{The evolution of the mid-infrared spectrum of SN~1987A observed with the JWST/MIRI-MRS}

\correspondingauthor{P. J. Kavanagh}
\email{patrick.kavanagh@mu.ie}

\author[0000-0001-6872-2358]{P.\ J.\ Kavanagh}
\affil{Department of Physics, Maynooth University, Maynooth, Co. Kildare, Ireland}

\author[0000-0002-3875-1171]{M.\ J.\ Barlow}
\affiliation{Department of Physics and Astronomy, University College London (UCL), Gower Street, London WC1E 6BT, UK}

\author[0000-0001-8532-3594]{C.\ Fransson}
\affiliation{Department of Astronomy, Stockholm University, The Oskar Klein Centre, AlbaNova, SE-106 91 Stockholm, Sweden}

\author[0000-0003-0065-2933]{J.\ Larsson}
\affiliation{Department of Physics, KTH Royal Institute of Technology, The Oskar Klein Centre, AlbaNova, SE-106 91 Stockholm, Sweden}

\author[0000-0002-5529-5593]{M.\ Matsuura}
\affiliation{Cardiff Hub for Astrophysical Research and Technology (CHART), School of Physics and Astronomy, Cardiff University, The Parade, Cardiff CF24 3AA, UK}

\author[0000-0001-9855-8261]{B.\ Sargent}
\affiliation{Space Telescope Science Institute, 3700 San Martin Drive, Baltimore, MD 21218, USA}
\affiliation{Center for Astrophysical Sciences, The William H. Miller III Department of Physics and Astronomy, Johns Hopkins University, Baltimore, MD 21218, USA}

\author[0000-0003-4870-5547]{O.\ C.\ Jones}
\affiliation{UK Astronomy Technology Centre, Royal Observatory, Blackford Hill, Edinburgh, EH9 3HJ, UK}

\author[0000-0002-0522-3743]{M.\ Meixner}
\affil{Jet Propulsion Laboratory, California Institute of Technology, 4800 Oak Grove Dr., Pasadena, CA 91109, USA }

\author[0000-0002-4000-4394]{R.\ Wesson}
\affiliation{School of Physics and Astronomy, Cardiff University, Queen’s Buildings, The Parade, Cardiff, CF24 3AA, UK}
\affil{Department of Physics, Maynooth University, Maynooth, Co. Kildare, Ireland}

\author[0000-0002-5797-2439]{J.\ A.\ D.\ L.\ Blommaert}
\affiliation{Astronomy and Astrophysics Research Group, Department of Physics and Astrophysics, Vrije Universiteit Brussel, Pleinlaan 2, B-1050 Brussels, Belgium}

\author[0000-0002-6018-3393]{P.\ Bouchet}
\affiliation{Laboratoire AIM Paris-Saclay, CNRS, Universit\'e Paris Diderot, F-91191 Gif-sur-Yvette, France}
\affil{Université Paris-Saclay, Université Paris Cité, CEA, CNRS, AIM, F-91191 Gif-sur-Yvette, France }

\author[0000-0001-6492-7719]{A.\ Coulais}
\affiliation{LIRA, Observatoire de Paris, Universit\'e PSL, Sorbonne Universit\'e, CNRS, Paris, France}
\affiliation{Université Paris-Saclay, Université Paris Cité, CEA, CNRS, AIM, F-91191 Gif-sur-Yvette, France }

\author{R.\ Gastaud}
\affiliation{Université Paris-Saclay, CEA, DEDIP, 91191, Gif-sur-Yvette, France}

\author[0000-0003-1319-4089]{R.~D.\ Gehrz}
\affiliation{Minnesota Institute for Astrophysics, School of Physics and Astronomy, University of Minnesota, 116 Church Street, S.E., Minneapolis, MN 55455, USA}

\author[0000-0002-2667-1676]{N.\ Habel}
\affiliation{Jet Propulsion Laboratory, California Institute of Technology, 4800 Oak Grove Dr., Pasadena, CA 91109, USA}

\author[0000-0002-2954-8622]{A.\ S.\ Hirschauer}
\affiliation{Space Telescope Science Institute, 3700 San Martin Drive, Baltimore, MD 21218, USA}

\author[0000-0002-0577-1950]{J.\ Jaspers}
\affiliation{Department of Physics, Maynooth University, Maynooth, Co. Kildare, Ireland}
\affiliation{Dublin Institute for Advanced Studies, School of Cosmic Physics, Astronomy \& Astrophysics Section 31 Fitzwilliam Place, Dublin 2, Ireland}

\author[0000-0002-1966-3942]{R.~P.\ Kirschner}
\affiliation{Thirty Meter Telescope International Observatory, 100 West Walnut Street, Pasadena, CA 91124, USA}

\author[0000-0003-4023-8657]{L.\ Lenki\'{c}}
\affiliation{IPAC, California Institute of Technology, 1200 East California Boulevard, Pasadena, CA 91125, USA}

\author[0000-0001-6576-6339]{O.\ Nayak}
\affiliation{Space Telescope Science Institute, 3700 San Martin Drive, Baltimore, MD 21218, USA}

\author[0000-0002-2461-6913]{S.\ Rosu}
\affiliation{Department of Astronomy, University of Geneva, Chemin Pegasi 51, 1290 Versoix, Switzerland}

\author[0000-0001-7380-3144]{T.\ Temim}
\affiliation{Department of Astrophysical Sciences, Princeton University, Princeton, NJ 08544, USA}

\begin{abstract}

%max 250 words. -currently at this limit!
Supernova (SN) 1987A provides a unique laboratory for investigating many aspects of SN physics and evolution. An observation at Day 12927 (35.4 yr) since the explosion with the Mid-Infrared Instrument (MIRI) Medium Resolution Spectrometer (MRS) on {\it James Webb Space Telescope} (JWST) provided the first spatially resolved spectroscopic study of \sn\ in the mid-IR, yielding insights into the evolution of dust, the ejecta, the equatorial ring (ER), and shocks in the system. Here we present a second epoch with MIRI/MRS at Day 13311 (36.4 yr) allowing the mid-IR spatially resolved spectroscopic temporal evolution of \sn\ to be probed for the first time. Analysis of the ER-dominated dust continuum showed little evolution between Days 12927 and 13311. However, a spatial analysis reveals the inner ER to be fading while the outermost regions are brightening. Broad ejecta emission lines detected at Day 12927 are evolving rapidly, driven by the recent onset of the ejecta/equatorial ring interaction in the northeast and southwest of the ER. Most lines from the ER show no change during the 384 days between the epochs, though some such as [\ion{Ne}{2}] and [\ion{Ar}{2}] have faded. We identify mid-IR H$_2$ emission associated with the ejecta for the first time. Using the near- and mid-IR [\ion{Fe}{2}] lines as density and temperature diagnostics of the ejecta in the interaction region we find it likely that the dense inner Fe-rich ejecta has now reached the reverse shock. Continued monitoring of \sn\ is essential to observe the evolving ejecta/ER interaction and dust components.

\end{abstract}


%--------------- INTRODUCTION ----------------%
%--------------- INTRODUCTION ----------------%

\section{Introduction} 
\label{sec:intro}
Supernova (SN) 1987A in the Large Magellanic Cloud (LMC) has provided an unparalleled opportunity to probe many aspects of the physics and evolution of core-collapse supernovae (CCSNe). At a distance of 49.6~kpc \citep{Pietrzynski2019}, it is close enough that spatially resolved 
observations of its anisotropic ejecta and surrounding circumstellar material can be obtained across the electromagnetic spectrum, revealing changes on an almost yearly basis.

The \sn\ system is comprised of aysmmetrical ejecta expanding at velocities of up to $\sim$10,000~km~s$^{-1}$ in the outermost ejecta, with the inner metal-rich core reaching $\sim$4,500~km~s$^{-1}$ \citep{Fransson2013, Larsson2016} down to 1600-2000~km~s$^{-1}$ in the region containing dust \citep{Cigan2019}. The inner ejecta are surrounded by a three-ring system comprising a dense inner equatorial ring (ER) approximately 2$\arcsec$ in diameter and inclined at
43$^{\circ}$ from the line-of-sight, and two fainter outer rings (ORs) each about 5$\arcsec$ in diameter. 
The ring structures are believed to have been created during a mass-loss episode of the progenitor $\sim$20~000 years prior to the explosion, possibly resulting from a binary merger 
\citep{Morris2007, Morris2009}.

The ER is currently the dominant emission component at wavelengths from the X-ray to the mid-infrared (mid-IR). It was initially flash ionized by the ultraviolet (UV) radiation
pulse from the SN shock breakout on the progenitor star but the ER's emission is now dominated by the shock interaction with the outermost parts of the expanding ejecta, commencing from 1995 \citep{Sonneborn1998}. The interaction between the fast outer ejecta and the circumstellar medium (CSM) has created a complex system of shocks that produce bright emission across the electromagnetic spectrum, with X-rays in particular being a powerful source of photoionization \citep{McCray2016}.

As the forward shock interacted with the dense gas in the ER, $\sim$30 bright knots or hotspots formed and brightened at optical wavelengths. The brightness evolution of these hotspots, and the subsequent fading of some of them as the shock progressed through the ER, has been studied by 
\cite{Fransson2015, Larsson2019b, Kangas2023} and \cite{Tegkelidis2024},
while \cite{Larsson2013, Larsson2016, Larsson2019a, Kangas2022} and \cite{Rosu2024} have studied the ultraviolet, optical and near-IR morphological and spectroscopic evolution of the inner ejecta.

Dust formation in the expanding and cooling ejecta was detected 400-500 days after the explosion \citep{Lucy1989, Gehrz1990, Bouchet1991, Wooden1993}, with the dust mass at day 775 estimated to be $5\times10^{-4}$~M$_\odot$ by \cite{Wooden1993} and 2$\times10^{-3}$~M$_\odot$ by \cite{Wesson2015}. Ejecta dust emission has not been unambiguously detected at mid-IR wavelengths since day 1153 \citep{Dwek1992} but in 2010 emission from $\sim$0.4~M$_\odot$ of much cooler (17-23~K) dust was detected at far-IR and sub-mm wavelengths from \sn\ by the {\em Herschel Space Observatory} \citep{Matsuura2011}. This dust emission was subsequently confirmed to originate from the inner ejecta by high angular resolution sub-mm observations with the Atacama Large Millimeter/submillimeter Array (ALMA) \citep{Indebetouw2014,Cigan2019}.

Spatially resolved 10~$\mu$m dust emission was detected from the ER on day 6067 (year 16.6) by \cite{Bouchet2004} using the 8-m Gemini South telescope. Subsequent {\em Spitzer} IRS spectra revealed that the ER's mid-IR spectrum was dominated by strong 10 and 20~$\mu$m silicate emission features \citep{Bouchet2006}. \cite{Dwek2010} showed that the Day~7554 (20.7 years) dust emission could be modeled with 1.2$\times10^{-6}$~M$_\odot$ of 180~K silicate dust, collisionally heated by the hot X-ray emitting shocked gas, with gas-grain collisions dominating the cooling of the shocked gas.
The 8-24~$\mu$m emission from the ER was seen to brighten up to Day~7954 (21.9 years), after which only warm {\em Spitzer} observations at 3.6- and 4.5~$\mu$m continued, revealing a decline in brightness after Day~9024 (24.7 years) \citep{Fransson2015,Arendt2016, Arendt2020}, which was similar to the optical light curve of the ER and attributed by them to the passage of the forward shock causing the gradual destruction of the dust in the ER.

The launch of {\em the James Webb Space Telescope} ({\em JWST}) has provided a multi-generational leap forward in observing capabilities in the IR. For \sn, results from NIRCam filter imaging at Day~12975 have been reported by \cite{Arendt2023} and \cite{Matsuura2024}, who found that much of the near-IR emission was coming from a newly developing outer portion of the ER, while \cite{Bouchet2024} presented MIRI filter imaging at Day~12927 which showed that the mid-IR emission was also extended beyond the inner ER. NIRSpec/IFU 1-5~$\mu$m spectroscopic observations of \sn, also at Day~12927, were reported by \cite{Larsson2023}, who found evidence that the Fe-rich inner ejecta had started to interact with the ER. 

\cite{Jones2023} presented Day~12927 MIRI/MRS 5-28~$\mu$m spectroscopy of \sn's ER and surrounds. The broader emission lines (280-380~km~s$^{-1}$) seen from all singly ionized species were shown to originate from the ER, while the narrower emission lines (100-170~km~s$^{-1}$) associated with higher ionization species originated from a more extended lower density component that may have been flash-ionized by the original SN event. They also found, using both MIRI/MRS and NIRSpec/IFU spectra, that the 0.9-28~$\mu$m dust spectral energy distribution (SED) of the ER was best fit by two temperature components of \cite{Hensley2023} astrodust. 

\cite{Fransson2024} reported strong [Ar~{\sc ii}]~6.99~$\mu$m emission from the very center of the inner ejecta, along with associated emission from several other ionic species seen in the MRS and NIRSpec/IFU spectra of the inner ejecta. They attributed the excitation of these lines to photoionization by a neutron star or pulsar wind nebula located inside the inner ejecta.

Most of the above works were based on Cycle~1 {\em JWST} Guaranteed Time Observations (GTO) PID~\#1232 (PI: G. Wright) with the MIRI/MRS, MIRI imaging, and NIRSpec/IFU. More recently, a second epoch of MIRI/MRS observations of \sn\ was obtained as part of a Cycle~2 GTO programme PID~\#2763 (PI: M. Meixner) to monitor the evolution of the system. This new observation at Day~13311 post-explosion, is 384 days after the Cycle 1 observation. This paper presents this new epoch of MIRI/MRS observations, the mid-IR evolution of the ER, the interaction regions, and newly detected mid-IR H$_2$ lines. A separate paper by \citet{Larsson2025} reports on Cycle 1 and 2 observations of the inner ejecta with the MIRI/MRS and NIRSpec/IFU, along with their implications for neutron star/pulsar wind nebula photoionization models for the core of the inner ejecta.

The current paper is organized as follows: Sect.~2 describes the MIRI/MRS observations of \sn\ and the data processing steps used for the Cycle 1 and 2 data. Sect.~3 discusses the distribution and evolution of the dust emission from the ER, Sect.~4 describes the properties and evolution of the spectral lines emitted from the ER and ejecta regions, and Sect.~5 focuses on the properties of the ejecta-ER interaction regions. Sect.~6 presents our overall conclusions.

%--------------- OBS and DATA RED ----------------%
%--------------- OBS and DATA RED ----------------%
\section{Observations and data reduction}
\label{sec:obs}
\sn\ was observed with the MIRI/MRS on 2022-07-16 as part of the guaranteed time program PID~\#1232 (hereafter Cycle~1, PI: G.~Wright) and on 2023-08-04 as part of the GTO PID~\#2763 (hereafter Cycle~2, PI: M.~Meixner). An analysis of the former dataset was presented by \citet[][hereafter J23]{Jones2023}. Since that time, the JWST Calibration Pipeline and calibration reference files appropriate for MIRI/MRS have seen significant improvements, most notably in spectrophotometric flux and wavelength calibrations in channel 4. We therefore reprocessed the Cycle~1 data to include these updates. 

The Cycle~1 observations are described by J23, and were designed to provide complete spectral coverage of \sn\ from 4.9 to 27.9 $\mu$m at medium resolution (R $\sim$  4000--1500; \citealt{Jones2023res}). The same observing configuration was used for Cycle~2, namely taking 4 dithers (extended source-negative pattern), with 3 integrations of 94 groups per dither using the {\sc FASTR1} readout pattern across the 4 channels for the three MIRI/MRS bands (SHORT, MEDIUM, LONG, also referred to as bands A, B, and C), for a total exposure time of 3152.4~s per MIRI/MRS channel/band combination. 

%-------------------------------------------
\begin{figure}
\centering
\includegraphics[width=\hsize]{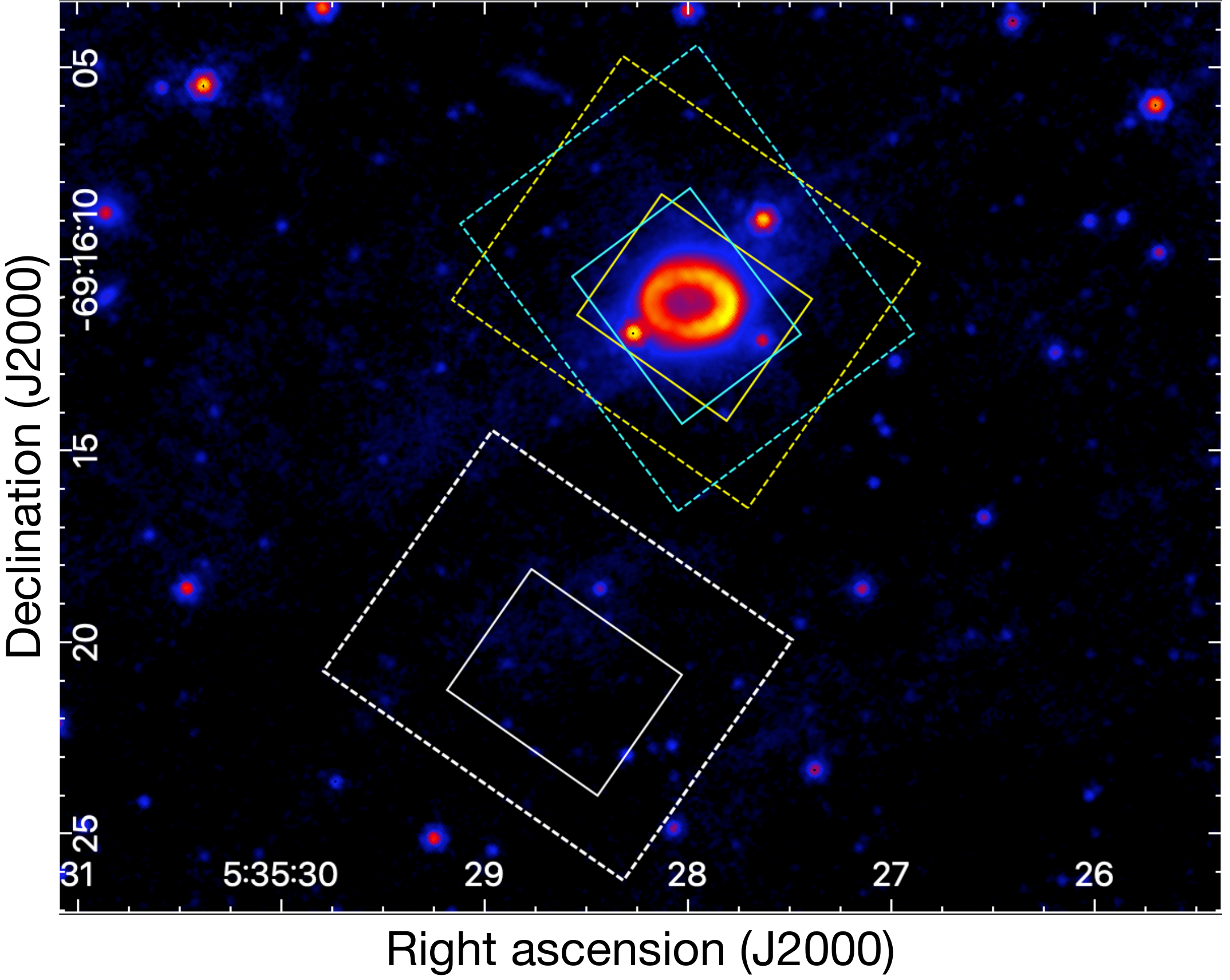}
\caption{The location of the MIRI/MRS fields-of-view (FOVs) shown on a MIRI imager F560W image of \sn\ \citep{Bouchet2024}. The Cycle 1 bands 1A and 4C FOVs are shown by the solid cyan and dashed cyan rectangles, respectively. Similarly, the Cycle 2 science FOVs are shown in yellow and background FOVs are shown in white.} 
\label{fig:fovs}
\end{figure}
%-------------------------------------------

Unlike Cycle~1, an additional dedicated background observation, located to the south of \sn\ (Fig.~\ref{fig:fovs}), was taken to negate the contribution of the detector features discussed in Sect.~2 of J23. The dedicated background comprised 4 dithers (extended source-negative pattern) with 1 integration of 94 groups per dither using the {\sc FASTR1} readout pattern for a total exposure time of 1043.4~s per MIRI/MRS sub-band. No target acquisition was used for any of the observations given the excellent blind pointing accuracy of $\sim0.1\arcsec$. The MRS fields of view (FOVs) range from 3.2$\arcsec$ $\times$ 3.7$\arcsec$ in channel 1 to 6.6$\arcsec$ $\times$ 7.7$\arcsec$ in channel 4 \citep{Wells2015, Argyriou2023}, which is illustrated in Fig.~\ref{fig:fovs}.

During both Cycle~1 and Cycle~2 MIRI/MRS observations, simultaneous MIRI imaging was performed using the F560W, F770W, and F1000W filters. Each covered a region to the north of \sn, with significant overlap between the Cycles 1 and 2 fields. The imaging observations were taken using the {\sc SLOWR1} readout pattern to reduce the data excess, and were comprised of 4 dithers with 3 integrations of 10 groups for a total exposure time of 2057.9~s per filter. We used these imager observations to correct the absolute astrometry of the MIRI/MRS observations (see Sect.~\ref{sect:pipe_proc}).

%----- DATA REDUCTION ------%
\subsection{Data reduction}\label{sect:data_red}
\subsubsection{Pipeline processing}\label{sect:pipe_proc}
\begin{figure*}
\centering
\includegraphics[width=\hsize]{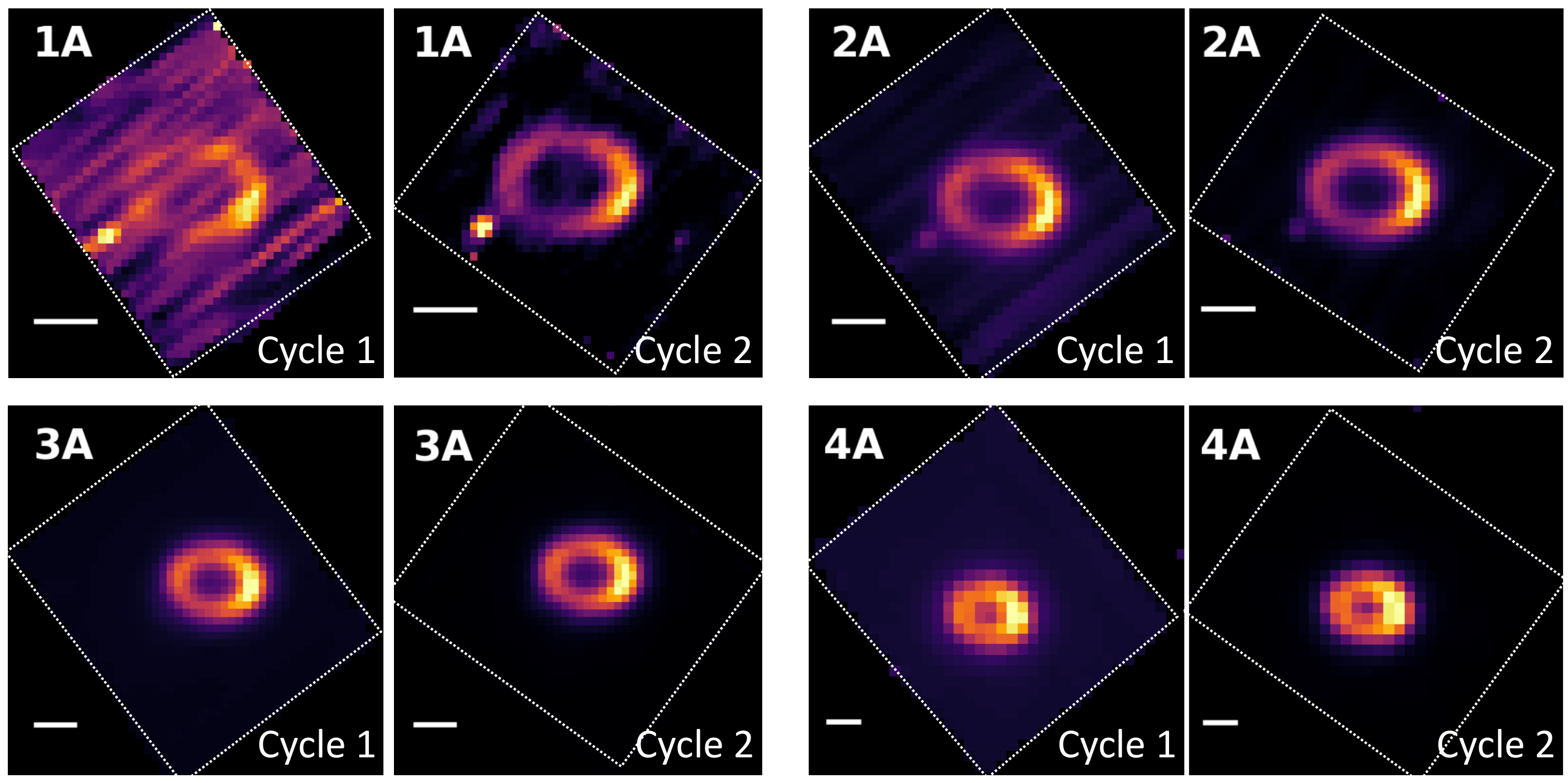}
\caption{Sample slices from the MIRI/MRS sub-band cubes from Cycles 1 (Day~12927) and 2 (Day~13311), with the band labels shown in the top-left. The white lines in each pane represent $1.0\arcsec$ to highlight the increasing size of the FOV (shown by the white dashed lines), as well as the decreasing spatial resolution from channels 1 to 4. `Star 3' is visible to the lower left of the ER in bands 1A--2A. The benefit of having a dedicated background field is clear with the removal of residual dark current at short wavelengths, particularly noticeable in band 1A, and background at long wavelengths for Cycle 2. The images are in the same orientation as Fig.~\ref{fig:fovs}. North is up, east is left.} 
\label{fig:cyc12_cubes}
\end{figure*}

To reduce our MIRI/MRS data we used version 1.14 of the JWST Calibration Pipeline \citep{2023zndo...7692609B} with versions 11.17.16 and ``jwst\_1223.pmap'' of the Calibration Reference Data System (CRDS) and CRDS context, respectively. We followed the same procedure as J23: all level 1b (ramp) files were processed through the \texttt{Detector1Pipeline} to produce level 2a (rate) images; absolute astrometry was corrected by applying an offset to the rate image world coordinate system (WCS) solution determined from the matching sources in the F560W simultaneous imaging fields to their Gaia Data Release 3\footnote{\url{https://www.cosmos.esa.int/web/gaia/dr3}} counterparts; in the case of Cycle~2, master background images per MIRI/MRS sub-band were created and subtracted from the science exposures; the level 2a files were processed through \texttt{Spec2Pipeline} with the optional \texttt{residual\_fringe} step switched on to produce flux level 2b calibrated rate (cal) images; the level 2b files were processed through \texttt{Spec3Pipeline} to produce spectral cubes for the 12 MIRI/MRS sub-bands by setting the \texttt{cube\_build} parameter `output\_type' set to `band'. Examples of the cubes are shown in Fig. \ref{fig:cyc12_cubes}.

%-------------------------------------------
%-------------------------------------------

%----- CUBES in ALPHA, BETA ------%
\subsubsection{Cubes in the MIRI/MRS IFU local coordinate system}\label{sect:alpha_beta_cubes}
\begin{figure}
\centering
\includegraphics[width=\hsize]{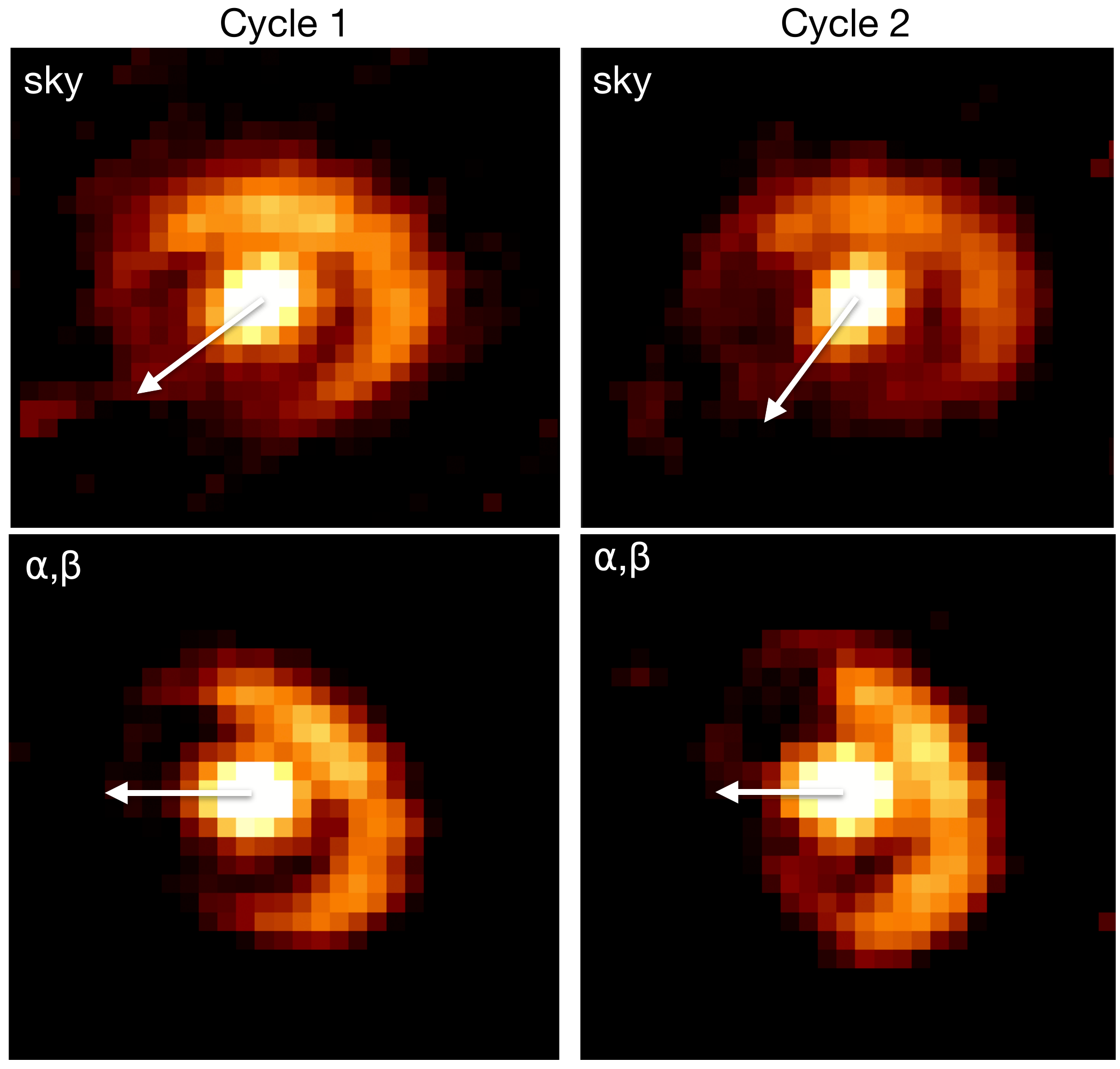}
\caption{Example of the effect of PSF broadening in channel 1 using the [Ar~{\sc ii}]~6.99~\micron\ line. The images in the left column are from Cycle 1 and the right from Cycle 2. The top panel in each is the cube constructed in sky coordinates, i.e., RA and Dec. The bottom panels in each show the cubes constructed in the MIRI/MRS IFU local coordinate system ($\alpha, \beta$). The arrows are included to highlight the direction of broadening. Note that the broadening in the sky coordinate cubes means a spectral extraction region defined on the sky that encompasses the ejecta emission in Cycle 1 will not be appropriate for Cycle 2. Building the cubes in $\alpha, \beta$ negates this issue.} 
\label{fig:psf_broad}
\end{figure}

The telescope position angle in Cycle 2 differs from that of Cycle 1 by approximately $18^{\circ}$ east-of-north (see Fig.~\ref{fig:fovs}). As discussed in \citet[][their Fig.~2]{Law2023}, the MIRI/MRS point spread function (PSF) recovered from the cube-building algorithm is more extended in the along-slice direction in channel 1. This has implications for spectral extraction and comparison between the MIRI/MRS epochs. In particular, the bright [Ar~{\sc ii}]~6.99~\micron\ feature located in the ejecta region associated with the compact object \citep{Fransson2024} is broadened at a different angle in Cycle 1 than in Cycle 2, which is illustrated in Fig.~\ref{fig:psf_broad}. This means that a spectral extraction region defined for the ejecta region in Cycle 1 is not appropriate for Cycle 2. The PSF broadening also has implications for PSF deconvolution (see Sect.~\ref{sect:psf_decon}) since we must ensure that the angle of the broadening of the model PSF matches that of the observations. Both of these issues can be accounted for by constructing the data cubes in the MIRI/MRS IFU local coordinate system with axes $\alpha, \beta$. This was achieved by running \texttt{Spec3Pipeline} with the \texttt{cube\_build} parameter `coord\_system' set to `ifualign' to produce cubes in this coordinate system for each of the MIRI/MRS sub-bands.

%-------------------------------------------
%-------------------------------------------

%----- SPECTRAL EXTRACTION ------%
\subsubsection{Spectral extraction}\label{sect:spec_extraction}
\begin{figure}[!ht]
\centering
\includegraphics[width=\hsize]{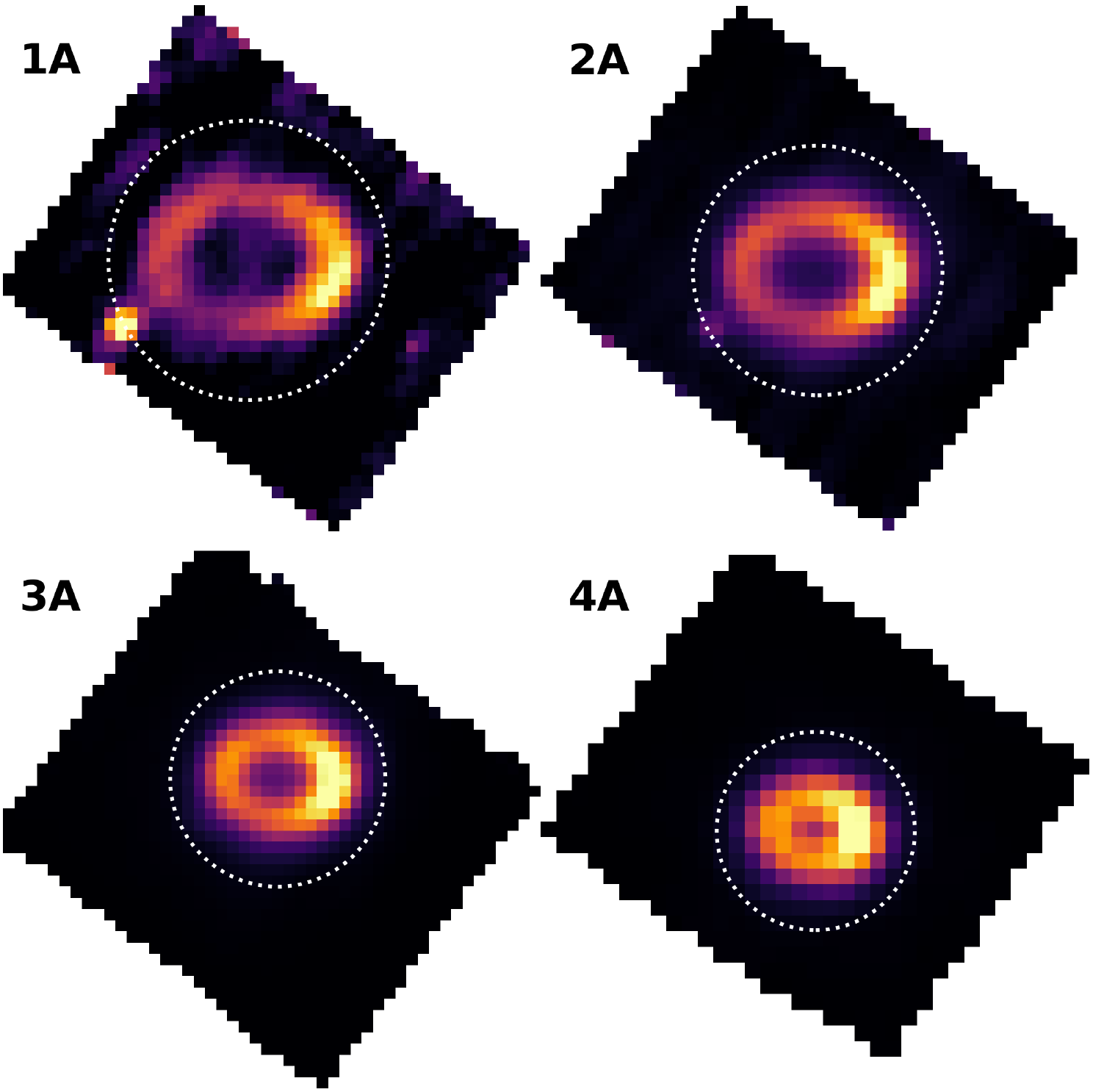}
\caption{Example of increasing aperture with wavelength for the `total' spectrum of \sn. The aperture size at the last wavelength frame of an individual sub-band data cube (white dashed circles), with the sub-band name given in the top-left of each panel. The location of `Star 3', visible just outside the eastern ER in band 1A, was masked during extraction. }
\label{fig:cone_aperture}
\end{figure}

We extracted spectra from several regions of \sn. To enable comparison with mid-IR spectra obtained from \spitzer, we produced a `total' spectrum for both MIRI/MRS epochs. Unlike in J23, these spectra were not extracted from a fixed aperture across the MIRI/MRS wavelength range. Instead we defined a circular aperture below 5~\micron\ in channel 1A, centered as in J23, of radius $1.5\arcsec$ that enclosed the outer edge of the ER. To this aperture radius we added an additional angle equivalent to the Rayleigh criterion ($\theta\approx1.22\lambda/D$) where $D=6.5$~m, resulting in conservative aperture sizes that grow with increasing wavelength, as shown in Fig.~\ref{fig:cone_aperture}. This minimizes the background contribution and optimizes signal-to-noise in the lower wavelength range. The location of `Star 3' was masked during extraction.

%-------------------------------------------
%-------------------------------------------

\begin{figure*}
\centering
\includegraphics[width=\hsize]{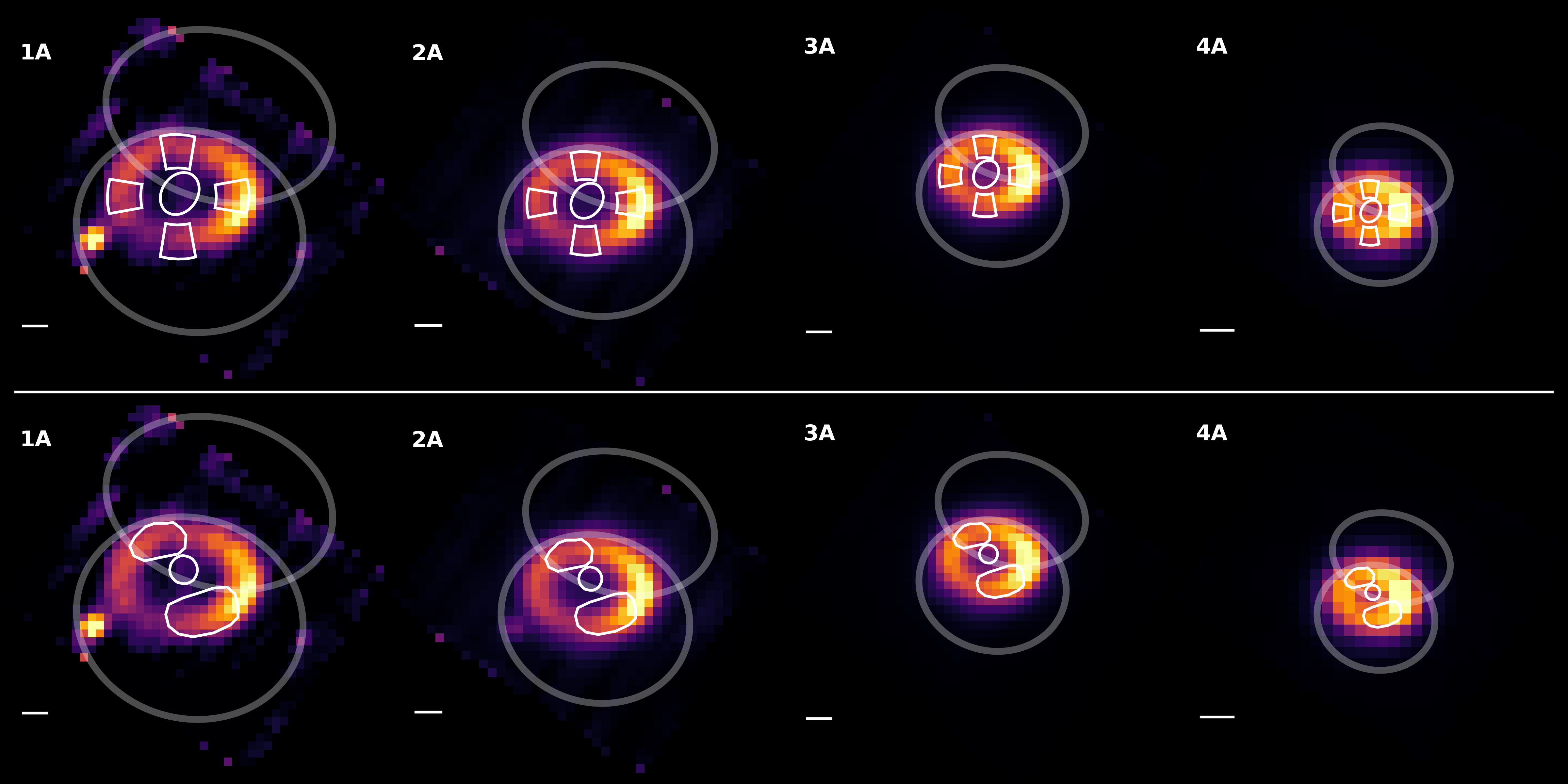}
\caption{Top row: The spectral extraction regions for the cardinal point and ejecta spectra plotted on the Cycle 2 (Day~13311) cubes. Bottom row: Same as top but for the interaction regions. The polygons at the top and bottom of the ER are the north and south interaction regions, respectively, while the circle represents the mid-north region. In both rows, the location of the outer rings is shown by the two faded ellipses. The bar in each panel indicates a 1$\arcsec$ scale. North is up, east is left.} 
\label{fig:spectral_extraction_regions}
\end{figure*}

In addition to the `total' spectra, we extracted spectra from the four cardinal point regions from the N, S, E and W portions of the ER and central ejecta region, shown in Fig~\ref{fig:spectral_extraction_regions}. The cardinal point regions are similar to those defined in J23 (see their Fig.~2) though larger. They also have slightly different areas with respect to each other to reduce contamination from the ejecta region (see Tables \ref{tab:cardinal_pointsNS} and \ref{tab:cardinal_pointsEW}). In addition, the ejecta region is modified for Cycle 2 due to the PSF broadening issue described in Sect.~\ref{sect:alpha_beta_cubes}. 

%-------------------------------------------
%-------------------------------------------

As in J23, spectral regions were defined using SAO~Image DS9 \citep{Joye2003} and extracted using the \texttt{aperture} module in the Astropy Photutils package\footnote{\url{https://photutils.readthedocs.io/en/stable/}} \citep{Bradley2022}. The DS9 regions files were parsed using the Astropy Regions package\footnote{\url{https://astropy-regions.readthedocs.io/en/stable/}} \citep{Bradley2022b}. We applied the additional post-pipeline residual fringe correction to all our spectra, which is included in the JWST Calibration Pipeline package under the \texttt{extract\_1d} step. To produce full-band MRS spectra from the `total' region, we stitched the 12 sub-band spectra using the \texttt{combine\_1d} tool in the JWST Calibration Pipeline package. No scaling of the sub-bands was performed.

%----- PSF DECON ------%
\subsection{PSF deconvolution}\label{sect:psf_decon}
To enable spatially resolved studies of the dust continuum we spatially deconvolved our MIRI/MRS datacubes using PSFs obtained from a development version of WebbPSF\footnote{https://github.com/spacetelescope/webbpsf/pull/691}. Using the \texttt{calc\_psf} method we produced model PSFs monochromatic (Patapis et al., in prep.) at each wavelength element of our datacubes at the same pixel scale. We then deconvolved our cubes at each wavelength using the Richardson-Lucy algorithm in the \texttt{restoration.richardson\_lucy} function \textit{scikit-image}\footnote{https://scikit-image.org/} Python package \citep{pedregosa2011}. The deconvolution algorithm is very sensitive to residual warm pixels and outliers from the processing and can produce large positive and negative spikes in the deconvolved cubes. For this reason we first smoothed out such imperfections using a sliding-box median with a width of 5 wavelength elements. While this does affect the shape of emission lines, the continuum is unaffected and emission lines are masked in our continuum analysis. We tuned the number of iterations in the algorithm based on testing of each sub-band. Prior to spectral extraction, we reprojected all datacubes from both MIRI/MRS epochs to the spatial grid of the Cycle 1 band 4C cube using the package Reproject\footnote{\url{https://reproject.readthedocs.io/en/stable/}}. This ensures that the spatial sampling of the ER is consistent across the entire MIRI/MRS range for all cubes.

%--------------- RESULTS ----------------%
%--------------- RESULTS ----------------%
\section{The mid-IR dust emission from the Equatorial Ring}\label{sect:results}

%----- DUST EVOLUTION ------%
\subsection{Dust continuum evolution}\label{sect:dust_evol}
\begin{figure*}
\centering
\includegraphics[width=\hsize]{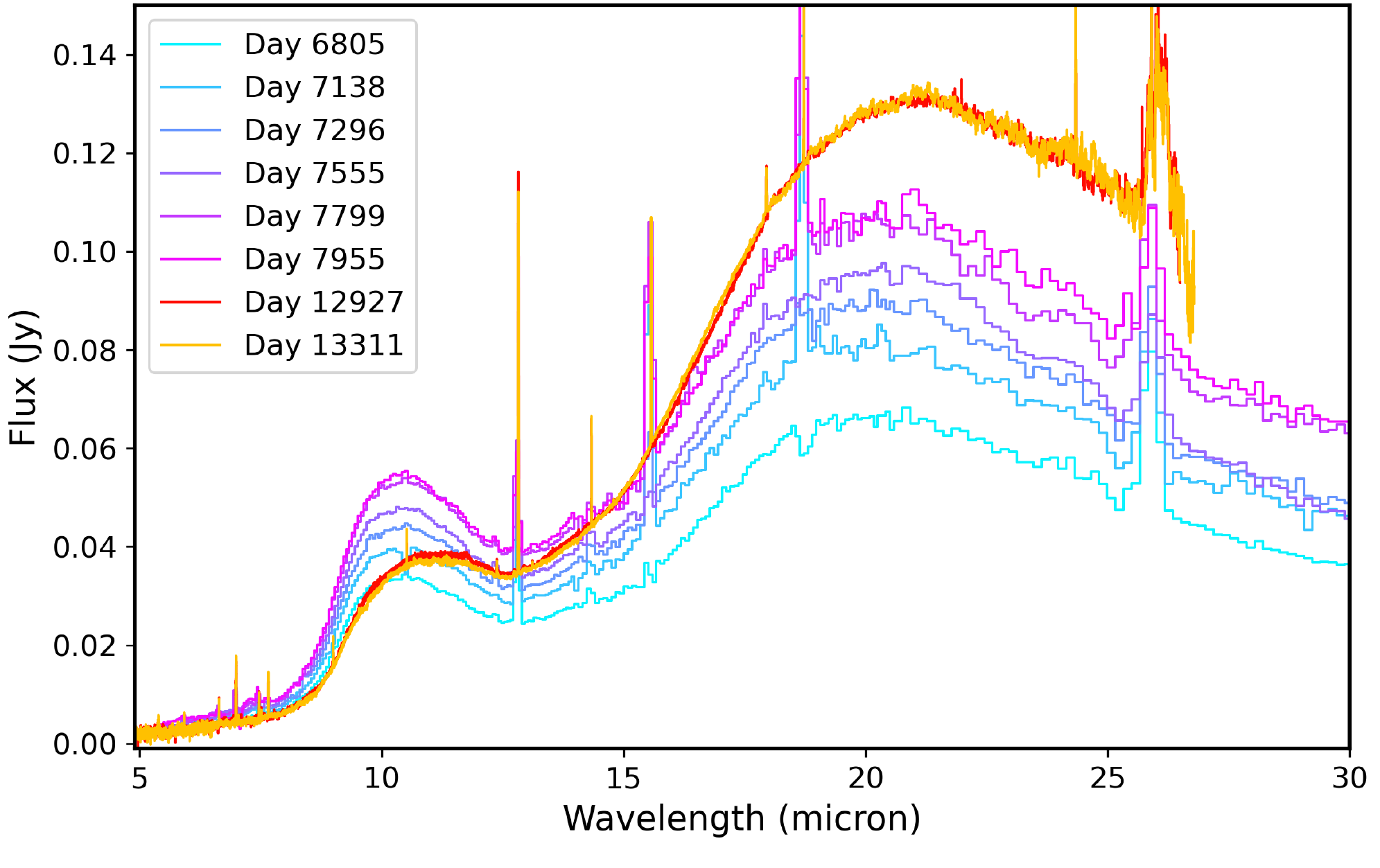}
\caption{The evolution of the MIR spectrum of \sn. The MIRI/MRS Day 12927 and 13311 spectra are shown in red and yellow, respectively. All other spectra are from \spitzer/IRS with the corresponding days since explosion shown in the legend. The MIRI/MRS spectra are cut off at 27~\micron\ as the quality of the flux calibration begins to degrade beyond this limit \citep{Law2024}.} 
\label{fig:spitzer_jwst_evol}
\end{figure*}

The Day~12927 and Day~13311 MIRI/MRS `total’ spectra are plotted in Fig.~\ref{fig:spitzer_jwst_evol} in red and orange, respectively. The broadband continuum shows little evolution between the two epochs. The 10~\micron\ silicate feature is slightly fainter at Day~13311. Otherwise, the spectra are remarkably similar in both shape and flux. The Spitzer/IRS spectra (see J23 for a description of these data), also plotted in Fig.~\ref{fig:spitzer_jwst_evol}, typically showed much more evolution on yearly timescales. At these epochs, the smallest change in the spectrum of \sn\ is observed between Days~7799 and 7955, a separation of 156 days, though the brightening trend due to the passage of the blast wave through the ER is still clear across the wavelength range. By comparison the MIRI/MRS epochs are separated by 384 days. Since the blast wave passed the ER around Day~8300, between the \spitzer/IRS and MIRI/MRS epochs, the ER was observed to fade at most wavelengths and this was also evident in the mid-IR spectral evolution below $\sim15~$\micron\ \citep{Jones2023}. Above this, the Days~12927 and 13311 MIRI/MRS spectra show a higher flux than the \spitzer\ epochs. This was interpreted by J23 as resulting from large grains in the ER preferentially surviving processing by the forward shock, with the total mass likely increasing as the  forward shock sweeps up circumstellar material. However, a further increase in the total spectrum longward of $\sim15~$\micron\ is not observed in the 384 days between the MIRI/MRS observations.

\begin{figure*}
\centering
\includegraphics[width=5in]{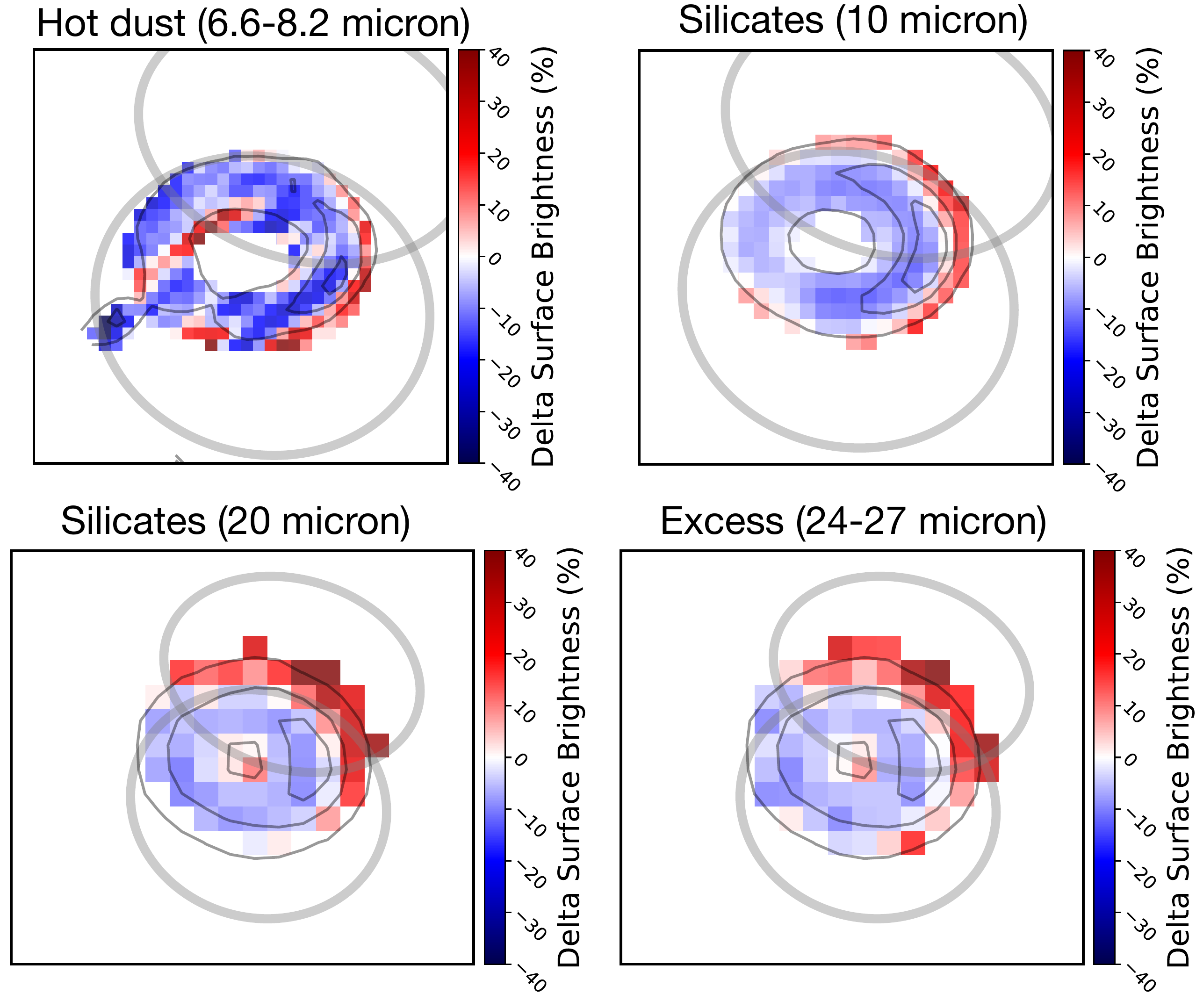}
\caption{The continuum brightness evolution maps between Days 12927 and 13311 for four continuum regions, chosen to highlight specific spectral features (see Sect.~\ref{sect:dust_evol} for details). The hot dust component is shown on top-left with the 10~\micron\ and 20~\micron\ silicate features on the top-right and bottom-left, respectively, with the 30--70~\micron\ excess, traced in the 24--27~\micron\ wavelength range, shown in bottom-right. Dust continuum contours are shown by the dark gray lines in each panel at 1$\sigma$, 3$\sigma$, and 5$\sigma$ above the median background level. The location of the outer rings is indicated by the light gray circles. Spatial regions with fluxes less than 10\% of the maximum have been masked which roughly corresponds to the 1$\sigma$ contour. Emission lines were masked when producing these maps. The northern OR appears to cross the ER in a slightly different region for the hot dust component than the other three. We note that the inner contour is not the same in each panel but determined from the continuum in their respective wavelength ranges. The inner contour position is dependent on the brightness distribution in the ring and the size of the PSF, which is large at longer wavelengths. This makes a comparison between the OR crossing location in the panels difficult.}
\label{fig:cont_evol}
\end{figure*}

With the second MIRI/MRS epoch we can investigate the brightness evolution of the broadband continuum in the \sn\ spectrum. To enable direct comparison of between the cycles we resampled the continuum maps of one onto the pixel grid of the other, in this case the Day~12927 onto Day~13311. We did this using the Reproject package. We defined four wavelength ranges which were selected to best show the evolution of the 10~$\mu$m and $20$~$\mu$m silicate features, which peak in the 10.25--11.5~$\mu$m and 19.5--20.75~$\mu$m ranges, respectively, the hot dust component traced in the 6.6--8.2~\micron\ range (MIRI/MRS band 1C), and the 30--70~\micron\ excess, probed in the $24-27$~\micron\ range (see Sect.~\ref{sect:excess}). For the hot dust component we chose the band 1C wavelength range as the continuum emission below $\sim6.5$~$\mu$m is faint and noisy. We masked the contributions of emission lines in each of the wavelength ranges. The resulting brightness evolution maps are shown in Fig.~\ref{fig:cont_evol}.

We find that in all cases, the main body of the ER has faded between Days~12927 and 13311. This appears most significant for the hot dust component though we caution that this is likely overstated given there was no dedicated background observation for Day~12927 so the difference image includes the contribution of the residual dark current signal (see Fig.~\ref{fig:cyc12_cubes}, top-left for example). In each case, the emission is increasing on the outer edge of the ER, indicating that more of the progenitor CSM is being swept-up and heated in the post-shock region. Interestingly, there appears to be a slight brightening of $\sim10$\% on the inner part of the south-western ER in the 20~\micron\ silicate and excess component bands, which is consistent with the south-western interaction region and could potentially indicate that dust in the ejecta is being heated by X-rays from the ER as the distance between them narrows.

%-------------------------------------------
%-------------------------------------------

%-------------------------------------------
%-------------------------------------------

%----- DUST FITTING ------%
\subsection{Dust continuum fitting}\label{sect:dust_fits}
J23 presented fits to the Day~12927 total spectrum using the
5.3--28~\micron\ MRS and the 0.9--5.3~\micron\ NIRSpec/IFU data \citep{Larsson2023}
with two-temperature dust models of varying grain
compositions and accounting for the synchrotron contribution at Day~12927 following
the descriptions of \citet{Cendes2018} and \citet{Cigan2019}.
An additional component representing continuum and lines from H and He
was also included in J23, which are described in \citet{Larsson2023}. Since we did not have
a contemporaneous NIRSpec/IFU observation at Day 13311, we limited our analysis to the MIRI/MRS range only. For this
reason, we do not include the H and He component in the fits but note that its contribution
to the MIRI/MRS range is an order of magnitude smaller than the combined dust and synchrotron 
emission at the shortest MIRI/MRS wavelength, and decreasing longward (see J23, their Fig.~9).

To fit the updated Day~12927 and new Day~13311 total spectra we adopted the same approach
as J23, i.e. using a modified blackbody analysis \citep[see, e.g.][]{Micelotta2018} and fixing the synchrotron component at each epoch using the relations given by \citet{Cendes2018} and \citet{Cigan2019}. 

\begin{figure*}
\centering
\includegraphics[width=\hsize,trim= 0cm 0cm 0cm 0cm,clip]{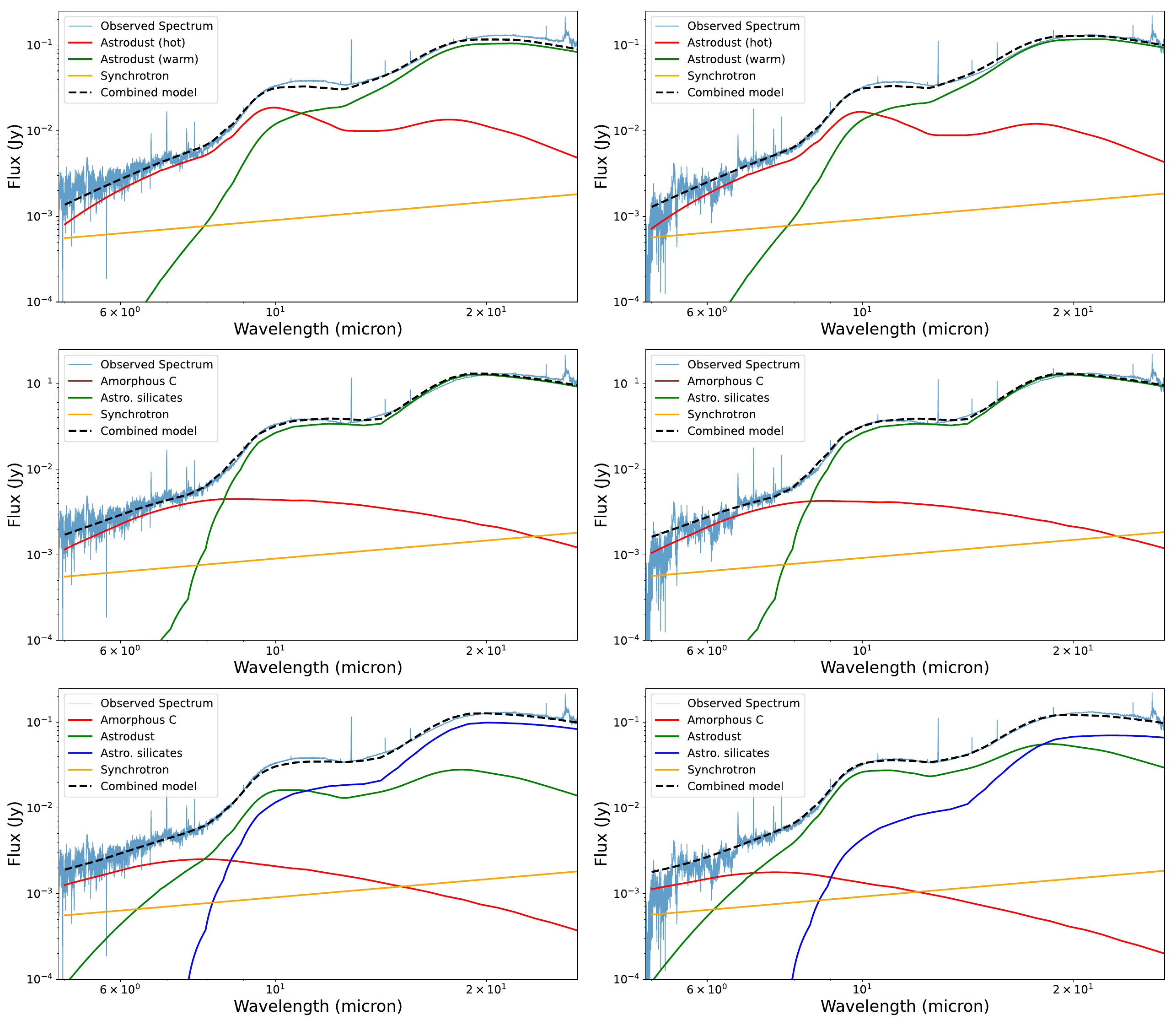}
\caption{Broadband dust continuum fits to the total spectrum of \sn\ at Days 12927 (left column) and 13311 (right column). The top row shows the fits for Case 1 (two-component astrodust), the middle row shows Case 2 (astronomical silicates+amorphous C), and the bottom row shows Case 3 which includes the astronomical silicates component to represent the 30-70~\micron\ excess (amorphous C+astrodust+astronomical silicates). The color of each component is indicated in the legends. The synchrotron component was fixed in all cases as described in the text. Best fit model parameters are given in Table~\ref{tab:dust_fit_params}}.
\label{fig:jwst_total_dust_cont_fits}
\end{figure*}

We fitted the models with the astrodust composition from \citet{Draine2021} and \citet{Hensley2023} (Case 1 in Table~\ref{tab:dust_fit_params}), and also the combination of a silicate grain composition from \citet{Draine1984} and an additional hot, featureless 
amorphous carbon component \citep{Draine1984} to account for emission below $\sim7$ \micron\ (Case 2 in Table~\ref{tab:dust_fit_params}). We use amorphous carbon as it has been used to account for this hot component by previous authors \citep[e.g.][]{Dwek2010,Matsuura2022}, though may not be the actual composition given the progenitor wind should be oxygen rich. We forego fitting the hot dust component with pure Fe, as in previous works, since J23 found that the expected dust mass is unrealistically high. Our best fits are plotted in Fig.~\ref{fig:jwst_total_dust_cont_fits}.

For Cases 1 and 2, our results are largely consistent with those of J23. Minor differences in derived temperature and mass parameters are observed. However, this is to be expected given the improvement in the spectro-photometric calibration in channel 4. This results in shallower decrease in the continuum above 20~\micron, resulting in the fitted parameter differences.

We find that over the 384 days between the Day~12927 and 13311 observations, the derived dust parameters for the `total' spectra are largely consistent indicating that a longer baseline is required to see any significant change in the overall spectral shape. However, as described in Sect.~\ref{sect:dust_evol}, the main body of the ER appears to fade between the observations with the extremities of the ER increasing in brightness. The western segment dominates the other parts of the ER, which is also where the largest increases in brightness are seen (Fig.~\ref{fig:cont_evol}). This, taken together with our fit results, suggests that the fading in some parts of the ER is balanced by the brightness increase in others leading to minor change in the `total' spectrum. 

%-------------------------------------------
%-------------------------------------------

\subsubsection{30--70~\micron\ excess}
\label{sect:excess}
\citet{Matsuura2019} reported the detection of excess emission at 31.5~\micron\ using SOFIA observations at Day~10732 post-explosion, which could be related to the 70~\micron\ excess detected by Herschel at Day~9090 \citep{Matsuura2015}. Two scenarios for this 30--70~\micron\ excess were proposed: an additional dust component prominent between the warm dust continuum detected by \spitzer\ and the cold dust components detected by ALMA, or a modified warm component with a flatter grain size distribution resulting in more large dust grains than typical ISM grain size distributions. \citet{Matsuura2019} suggested dust re-formation in the ER or a fraction of the ejecta dust being heated to higher temperatures by X-rays from the ER. J23 did not detect a warm dust component from the ejecta region and argued against grain growth in the ring given the consistency between the long-wavelength MIRI/MRS and Spitzer/IRS data, instead suggesting that large dust grains survive the passage of the shocks.

\begin{figure}
\centering
\includegraphics[width=\hsize]{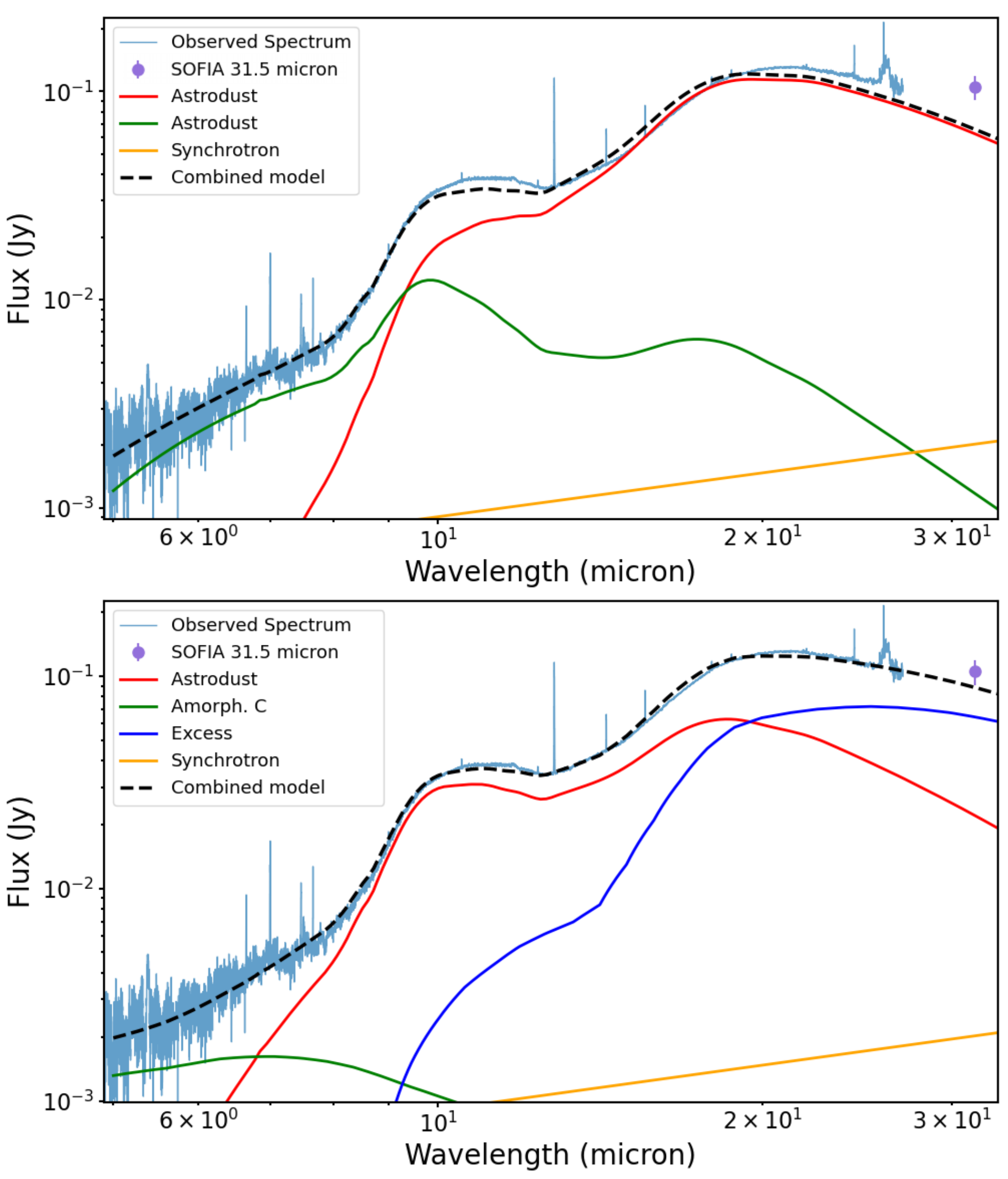}
\caption{Comparison of our model fits to the Day 12927 MIRI/MRS total spectra to the SOFIA 31.5~\micron\ flux measured at Day~10732, denoted by the purple filled circle with associated error bar. The best fit for our Case~1 (two-component astrodust), and Case~3 (amorphous C+astrodust+astronomical silicates) are shown on the top and bottom, respectively. The color of each component is indicated in the legends. Best fit model parameters are given in Table~\ref{tab:dust_fit_params}.} 
\label{fig:sofia_comp}
\end{figure}

Given our second epoch of MIRI/MRS observations with improved flux calibrations and that the low 
wavelength portion of the excess emission falls within the MIRI/MRS range, we assessed the possible 
contribution of the 30--70\micron\ excess to the MIRI/MRS spectra and attempted to identify the 
excess emission region to determine where the excess component could be located. We first fitted 
the total spectra from Days~12927 and 13311 using a model, which we refer to as Case 3, comprising the fixed synchrotron contribution, an astrodust component representing the warm dust in the ER, an amorphous C component to account for 
the hot featureless emission, and an astronomical silicates \citep{Draine1984} component representing 
the excess emission. The results of our fits for Case 3 are given in Table~\ref{tab:dust_fit_params} and shown in Fig.~\ref{fig:jwst_total_dust_cont_fits}. We found that Case 3 fits the spectra well, with the 
introduction of the excess component accounting for the $>20$~\micron\ region better than Cases 1 and 2. 
The temperature of the excess emission is consistent between both epochs at approximately 120~K which 
is hotter than the excess determined by \citet{Matsuura2019} at Day~10732. It is difficult to compare 
the excess component masses between our results and those of \citet{Matsuura2019} given that they are 
poorly constrained in some cases and also very sensitive to how well the ER and hot dust components 
fit the lower wavelength regions. In Fig.~\ref{fig:sofia_comp}, we compare our fits to the SOFIA 31.5~\micron\ 
flux at Day~10732, cautioning that the excess component has likely changed in the intervening time period. 
The excess component in our Case~3 fit is consistent with the excess measured by \citet{Matsuura2019}. 
Interestingly, our Case~1, which does not include an excess component, cannot account for the SOFIA 
measurement. We stress the caveat with this comparison that the SOFIA data point was taken much earlier 
in the evolution of \sn. In addition, there are uncertain contributions to the data point from 
the [S~{\sc iii}]~33.5~\micron\ and [Si~{\sc ii}]~34.8~\micron\ lines, which are observed to be 
significant in other young SNRs. We assessed their possible contributions as follows.

For a purely photoionised gas, the [S~{\sc iii}] 33.5/18.7~\micron\ line intensity
ratio is about 1.6 at electron densities $n_{e}<<1000$~cm$^{-3}$, and drops to
about 0.25 at $n_{e}=1\times10^{4}$~cm$^{-3}$. At the density of $n_{e}\sim2\times10^{3}$~cm$^{-3}$ 
(see Sect.~\ref{sect:fe-line-diag}), the ratio should be about 0.55. From the line fits to our `total'
\sn\ spectra, the flux of the [S~{\sc iii}]~18.7~\micron\ line is $605\times10^{-24}$~W~cm$^{-2}$ for Day 12927 and $255\times10^{-24}$~W~cm$^{-2}$ for Day 13311 (see Table~\ref{tab:total_spectra}). Assuming that the 
flux of the [S~{\sc iii}]~33.5~\micron~line is half of this and taking the SOFIA FORCAST F315 filter 
FWHM of 4.6~\micron, the expected flux density of [S~{\sc iii}]~33.5~\micron\ in the filter is no more than 0.5 mJy. The filter transmission at 33.5~\micron\ is 0.41 so the line contribution would be $\sim0.3$~mJy, compared to the observed 
flux of $105~(\pm14)$~mJy. To assess the contribution of the [Si~{\sc ii}]~34.8~\micron\ line, 
we used the results of an analysis of the {\it Spitzer} IRS spectrum of Cas~A from \citep{Smith2009}. 
For Cas~A, these authors found the [Si~{\sc ii}] line to be about three times stronger than the [S~{\sc iii}]~33.5~\micron\ line. The filter transmission is 0.27 at 34.8~\micron, which would correspond to a contribution of $\sim0.2$~mJy, if the ratio were to be similar in SN1987A.  Based on these estimates, we suggest the contribution of these two lines to the SOFIA F315 flux to be negligible, though this does not account for any temporal evolution.

%-------------------------------------------
%-------------------------------------------

We attempted to localize the excess emission component by creating continuum maps from the $>24$~\micron\ where the excess component is the primary contributor to the continuum. Similar to J23, we found no clear evidence of emission from the inner ejecta, with the distribution  $>24$~\micron\ following that of the ER. We assessed the evolution of the excess component between Days~12927 and 13311 as described in Sect.~\ref{sect:dust_evol}, which is shown in Fig.~\ref{fig:cont_evol}, bottom-right. The evolution map is remarkably similar to that of the 20~\micron\ silicate feature, indicating that the excess is likely related to the ER dust, possibly resulting from a top-heavy grain size distribution as suggested in \citet{Matsuura2015} and J23.

\subsubsection{Composition and properties}
As described in J23, the appropriateness of the astrodust ISM dust model of \citet{Draine2021} for the ER dust is uncertain given that the progenitor was a B supergiant which probably underwent a red supergiant (RSG) phase. As shown in their Fig.~10, the astrodust opacity is higher than the astronomical silicates models of \cite{Draine1984,Weingartner2001} below 7~\micron\ and in the 10-15~\micron\ range. In particular, the astrodust opacity curve appears to align well with the location of the `dip' between the 10~\micron\ and 20~\micron\ silicate features. J23 note that astrodust contains contributions from various components such as SiO$_{2}$, Al$_{2}$O$_{3}$, etc., at the few percent level which could account for the difference in opacity at these wavelengths. Additionally, the astrodust mass absorption coefficients used in J23 and this paper are computed for size distributions of ellipsoidal grains. 

\begin{figure}
\centering
\includegraphics[width=\hsize]{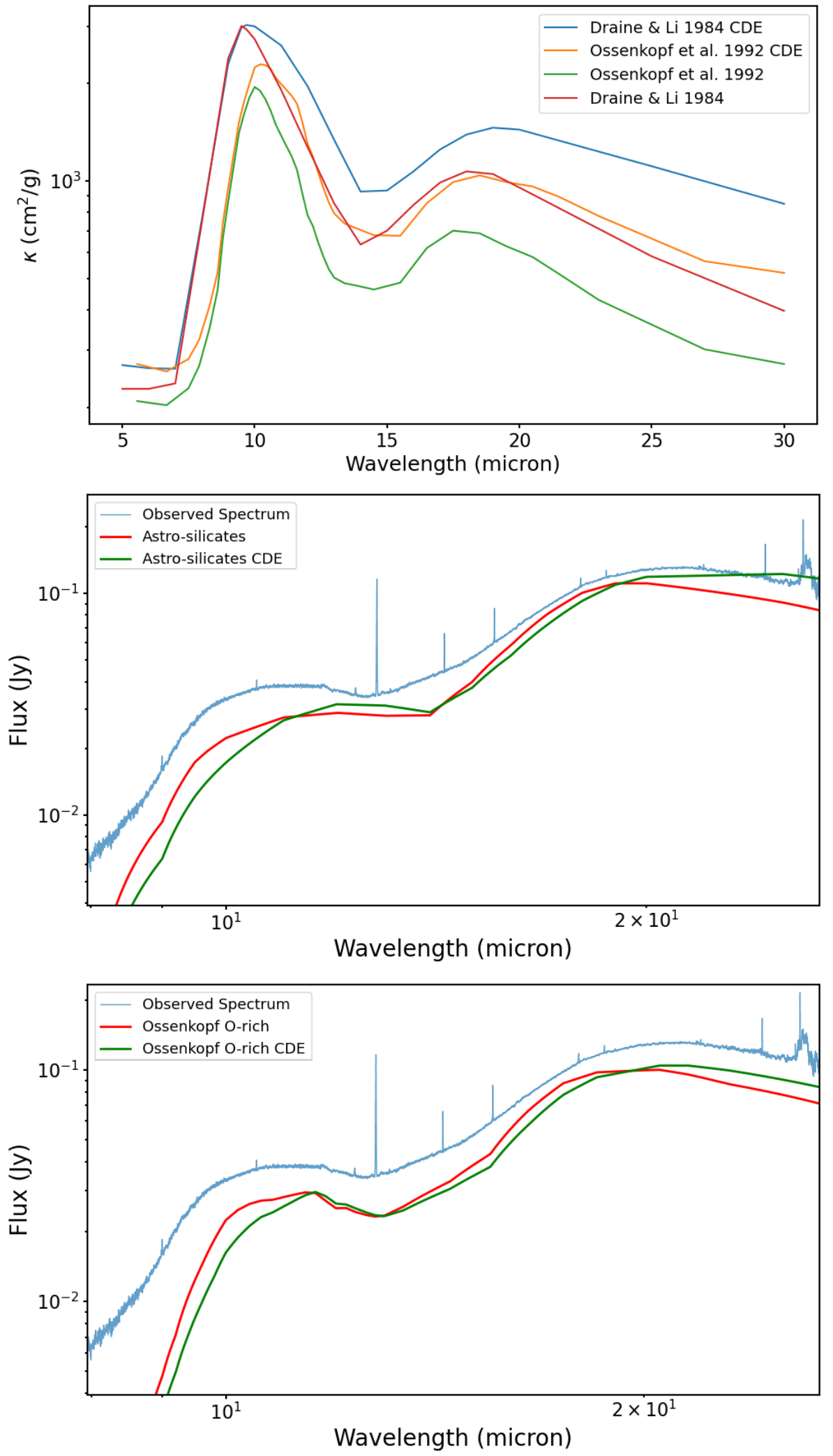}
\caption{Assessment of composition and shape distributions. Top: The calculated mass absorption coefficients ($\kappa$) for astronomical silicates (Mie in red, CDE in blue) of \citet{Draine1984} and O-rich CSM (Mie in green, CDE in orange) of \citet{Ossenkopf1992}. Middle: The calculated $\kappa$ values for astronomical silicates applied to a blackbody of temperature 150~K (Mie in red, CDE in green), and the \sn\ Day 12927 spectrum in blue. Bottom: Same as middle for the O-rich CSM. For middle and bottom, the normalizations of the models have been adjusted to enable qualitative comparison to the \sn\ spectrum. } 
\label{fig:mie_cde_comparison}
\end{figure}

We therefore investigated whether changing the size and shape distributions of the \cite{Draine1984} astronomical silicates model, as well as the O-rich composition of \citet{Ossenkopf1992} more representative of an O-rich circumstellar medium (CSM), could account for the shape of the \sn\ ER dust continuum. We experimented with how elongated dust could affect the spectral features by calculating the dust mass absorption coefficients ($\kappa$) of silicates using (a) Mie theory \citep{Mie1908} and (b) a continuous distribution of ellipsoids (CDE) shape distribution, assuming a grain density of 3.3~g~cm$^{-3}$ for both. For the Mie calculations, we used a fixed grain size of 0.1~\micron\ for this comparison. The resulting $\kappa$ values for the compositions and distributions are shown in Fig.~\ref{fig:mie_cde_comparison}, top. Overall, the spectral shape does not vary much between the Mie and CDE distributions, but the magnitude of the absorption coefficients does, in part because of the fixed grain size in the Mie calculations. The difference is not as prominent as that for crystalline silicates \citep{Min2003}.

To assess the effect of the compositions and distributions on the spectra of \sn\, we applied the mass absorption coefficients to a blackbody of temperature 150~K, as found for the warm component in Cases 1 and 2, scaling the normalization so that the models are approximately the same flux as the \sn\ spectra to enable a qualitative comparison. These models are shown in Fig.~\ref{fig:mie_cde_comparison}, middle and bottom for the astronomical silicates and the O-rich CSM, respectively. It is clear that the astronomical silicates models cannot account for the location of the `dip' between the 10~\micron\ and 20~\micron\ silicate features for either shape distribution. However, the O-rich CSM models of \citet{Ossenkopf1992} both match the observed dip, and the general shape of the CDE model spectrum longward of 12~\micron\ agrees very well with the \sn\ spectrum. This indicates that the O-rich CSM model of \citep{Ossenkopf1992} offers a better representation of the dust composition of the ER. We attempted to fit a more complete model incorporating the O-rich CDE composition but our attempts still failed to account for the shorter wavelength part of the spectrum, i.e. the hot dust component. It is likely that the O-rich CSM composition lacks a species present in the astrodust composition that increases opacity at these wavelengths, such as Fe oxides. A more complete analysis of dust composition, size, shape distributions will be presented in a future work.

%-------------------------------------------
%-------------------------------------------

\begin{table*}
\begin{tabular}{llllllll}
\hline
& & \multicolumn{2}{c}{Hot} & \multicolumn{2}{c}{Warm} & \multicolumn{2}{c}{Excess}\\
\cmidrule(lr){3-4}\cmidrule(lr){5-6}\cmidrule(lr){7-8}

Case & Day & T & M & T & M & T & M \\
 &  & (K) & ($10^{-8}$ M$_{\odot}$) & (K) & ($10^{-6}$ M$_{\odot}$) & (K) & ($10^{-5}$ M$_{\odot}$) \\

\hline
% 2 astrodust
1 & 12927 & 320 ($\pm8$) & 17.79 ($\pm0.82$) & 144 ($\pm2$) & 29.65 ($\pm0.90$) & -- & -- \\
1 & 13311 & 319 ($\pm3$) & 15.74 ($\pm0.42$) & 143 ($\pm1$) & 32.93 ($\pm0.65$) & -- & -- \\
% C+sil
2 & 12927 & 347 ($\pm1$) & 12.23 ($\pm0.23$) & 153 ($\pm1$) & 27.64 ($\pm0.51$) & -- & -- \\
2 & 13311 & 343 ($\pm1$) & 12.34 ($\pm0.27$) & 152 ($\pm1$) & 29.04 ($\pm0.53$) & -- & -- \\
% C+astrodust+excess
3 & 12927 & 435 ($\pm10$) & 2.51 ($\pm0.17$) & 218 ($\pm1$) & 1.26 ($\pm0.10$) & 137 ($\pm2$) & 3.98 ($\pm0.11$) \\
3 & 13311 & 488 ($\pm20$) & 1.12 ($\pm0.19$) & 205 ($\pm2$) & 2.96 ($\pm0.30$) & 124 ($\pm3$) & 5.14 ($\pm0.62$) \\
\hline
\end{tabular}
\caption{Temperatures and masses for our fitted multi-component dust models. Case 1 is the two-component astrodust model, Case 2 is astronomical silicates+amorphous C, and Case 3 is amorphous C+astrodust+astronomical silicates. Further details of the models are given in Sect.~\ref{sect:dust_fits}. Uncertainties are the $3\sigma$ uncertainties on the fit parameters.}
\label{tab:dust_fit_params}
\end{table*}

%----- SPATIALLY RESOLVED DUST ------%
\subsection{Spatially resolved dust continua fitting}\label{sect:spatial_res_dust}
J23 compared the general shape of the east and west segments of the ER and found that the west segment is approximately 50\% brighter than the east. They also reported that the eastern part is relatively stronger at longer wavelengths with the 10~\micron\ silicate band being weaker and its peak slightly shifted to the red in comparison to the west. It was not possible to perform a more spatially resolved spectroscopic study of the dust around the ER for two reasons. First, the MIRI/MRS datacubes produced by the calibration pipeline cannot be used to perform broadband spatially resolved spectroscopy on an object the size of the ER. While the MIRI/MRS spatial resolution is small in comparison to the ER at the shortest wavelengths, it increases in size towards the longest wavelengths becoming comparable to the ER. Therefore, a spectrum extracted from a fixed aperture on part of the ER would contain increasing contamination from nearby ER regions towards longer wavelengths. Second, modeling of the spectra would require constraints on the synchrotron continuum at the sub-ER level.

\begin{figure}
\centering
\includegraphics[width=\hsize,trim= 0cm 0cm 0cm 0cm,clip]{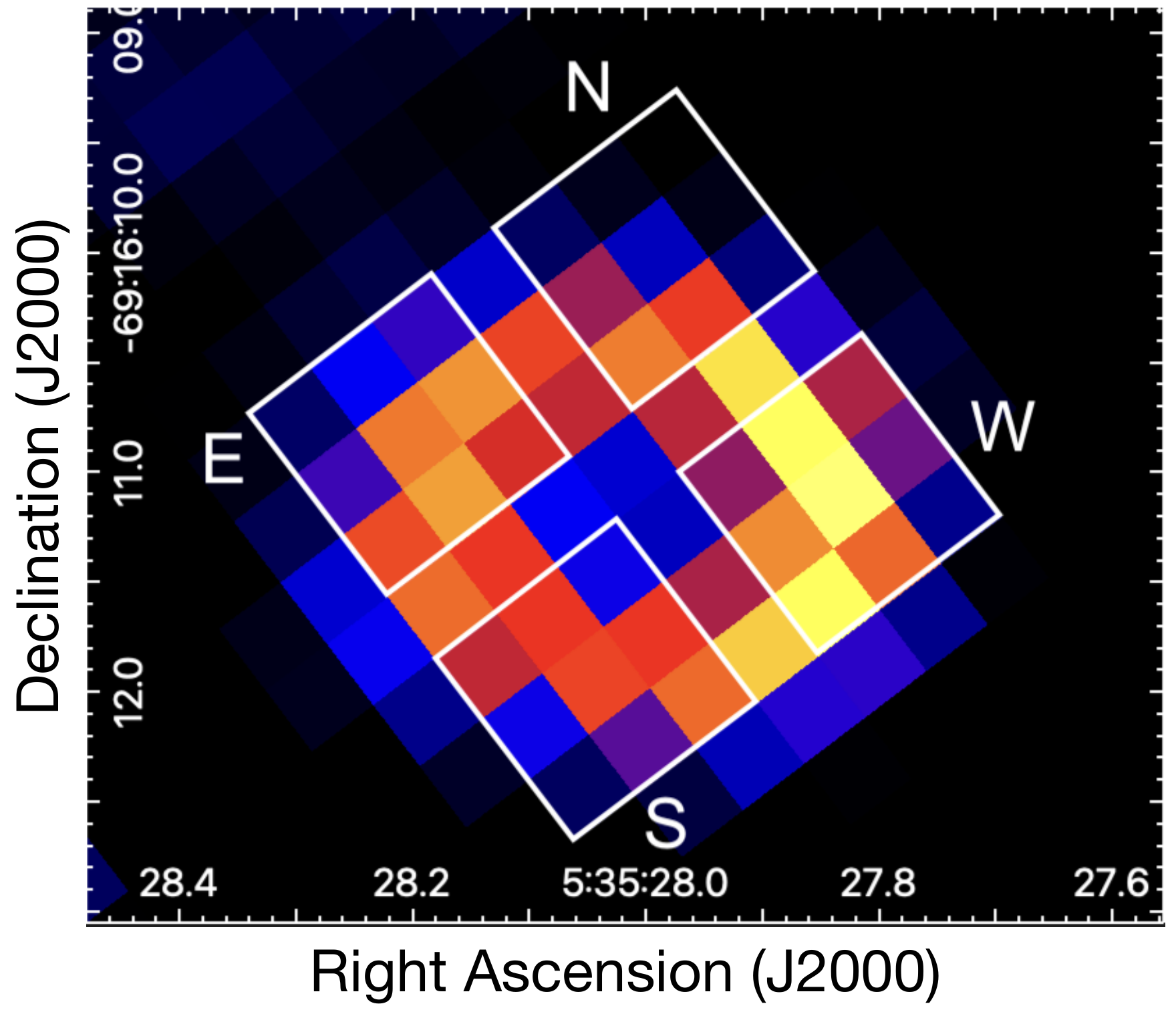}
\caption{The north, south, east and west extraction regions for the deconvolved and reprojected sub-band data cubes. } 
\label{fig:er_regions}
\end{figure}

\begin{figure*}
\centering
\includegraphics[width=\hsize,trim= 0cm 0cm 0cm 0cm,clip]{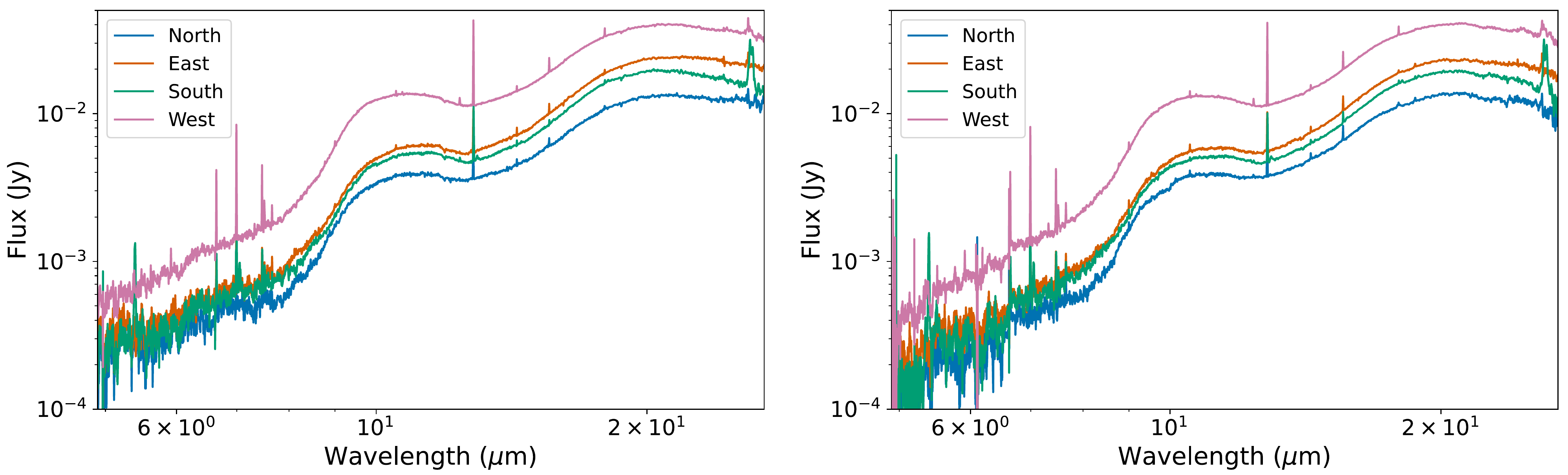}
\caption{Extracted spectra from sub-regions of the ER on Days~12927 (left) and 13311 (right), with north in blue, east in orange, south in green, and west in magenta. The location of the spectral extraction regions are shown in Fig.~\ref{fig:er_regions}.} 
\label{fig:er_regions_spectra}
\end{figure*}

To address the first issue, we used spatially deconvolved datacubes (see Sect.~\ref{sect:psf_decon}). We defined four extraction regions $3\times3$ pixels in size roughly corresponding to the north, south, east, and west of the ER (see Fig.~\ref{fig:er_regions}). The separation between the regions was set to 1 pixel to ensure that there is no cross-talk between the regions. We extracted spectra for each of the regions from each of the MIRI/MRS sub-bands. We inspected any offsets between the sub-bands and found that in the majority of cases, the offsets were $\lesssim5$\% of the continuum level. The exception to this was band 4C in the northern regions at both epochs with offsets of $\sim10$\%, though the shape appeared consistent with the lower bands. We therefore stitched the sub-band spectra to remove offsets $>5\%$, anchoring to the flux level of band 2C. The extracted spectra are shown in Fig.~\ref{fig:er_regions_spectra}.  To constrain the synchrotron component in each of the ER regions, we used the results of \citet{Matsuura2024} and the NIRSpec/IFU data reported in \citet{Larsson2023}. We first identified a synchrotron dominated spectral region free of emission lines. We then used aperture photometry to estimate the flux from the entire ER in this range, as well as each of the ER regions. We scaled the synchrotron component fitted to the total spectra in each of Days~12927 and 13311 described in Sect.~\ref{sect:dust_fits} using a scaling factor defined as the ratio of the ER region fluxes to the total ER flux.

\begin{figure*}
\centering
\includegraphics[width=\hsize,trim= 0cm 0cm 0cm 0cm,clip]{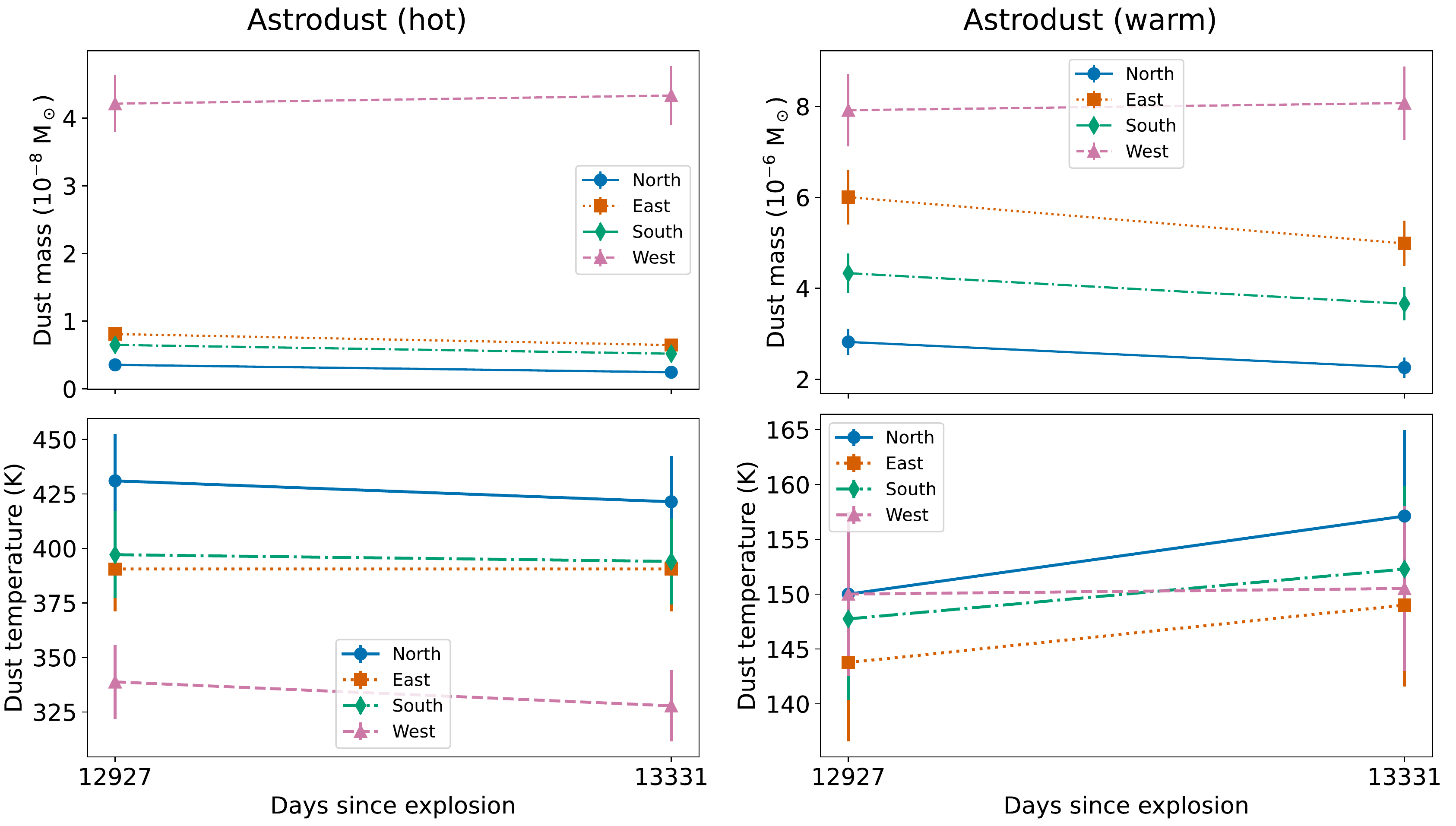}
\caption{Evolution of dust components derived from Case 1 model fits in the sub-regions of the ER between Days~12927 and 13311. The hot component is on the left and the warm component on the right. Dust masses and temperatures are shown on the top and bottom rows, respectively. In each panel, north in shown in blue, east in orange, south in green, and west in magenta.} 
\label{fig:er_regions_dust_evol}
\end{figure*}

We fitted the spectra of each of the ER regions using the two-temperature astrodust model (Case 1) with a fixed synchrotron component, the results of which are given in Table~\ref{tab:dust_fit_params_regions}. We used only Case~1 as it allowed us to assess the variation in the warm and hot components using a single composition. Given the offsets in the spectral sub-bands mentioned above, we applied a conservative signal-to-noise on the fitted spectra of 15\%. The evolution of the fitted dust properties is shown in Fig.~\ref{fig:er_regions_dust_evol}.

The temperatures in the individual regions mostly differ from that of the `total' spectrum. Since higher dust temperature leads to lower derived dust masses, the sum of the component masses from the four regions is not directly comparable to the mass estimates in Table~\ref{tab:dust_fit_params}. This is most evident for the hot component. The fits to the total spectrum Table~\ref{tab:dust_fit_params} and the western ER in Table~\ref{tab:dust_fit_params_regions} show approximately consistent temperatures, which is unsurprising given the western ER dominates the total continuum emission. The fits to the other regions show much higher ($\sim70-100~K$) temperatures and, therefore, much lower masses, highlighting the importance of a spatially resolved analysis to probe the temperature and mass distribution around the ER.

The spectra from each of the four regions are consistent in terms of contributions from warm and hot dust components. The hot component appears to be more concentrated in the west ER, the mass being $\gtrsim2$ times the mass of the other three regions combined. The warm dust mass is also largest in the west region, though the difference is not as stark as for the hot component. The temperature of the warm dust is consistent within the errors in all four regions. The hot dust temperature is consistent in the north, south, and east regions, though somewhat lower in the west region. As shown in Fig.~\ref{fig:er_regions_dust_evol}, we find negligible evolution of the hot dust component between Days~12927 and 13311. The warm dust component appears to be trending downward in dust mass in all apart from the west region which shows little evolution. The temperature of the warm component is trending higher in the north, south, and east. We note however that for the dust components, there is a dependence on fitted dust mass and temperatures for a given component flux, i.e. a higher dust temperature will lead to a lower dust mass. Future observations will be required to confirm these trends.

%-------------------------------------------
%-------------------------------------------

%-------------------------------------------
%-------------------------------------------

%-------------------------------------------
%-------------------------------------------

\begin{table}[h!]
\centering
\begin{tabular}{lllll}
\hline
 & \multicolumn{2}{c}{Hot} & \multicolumn{2}{c}{Warm}\\
 \cmidrule(lr){2-3}\cmidrule(lr){4-5}

Day & T & M & T & M \\
  & (K) & ($10^{-8}$ M$_{\odot}$) & (K) & ($10^{-6}$ M$_{\odot}$)  \\

\hline
\multicolumn{5}{c}{North}\\
 12927 & 431 ($\pm8$) & 0.35 ($\pm0.05$) & 150 ($\pm2$) & 2.82 ($\pm0.05$)  \\
 13311 & 421 ($\pm5$) & 0.25 ($\pm0.07$) & 157 ($\pm2$) & 2.26 ($\pm0.07$)  \\
 
\multicolumn{5}{c}{East}\\
 12927 & 391 ($\pm5$) & 0.81 ($\pm0.03$) & 144 ($\pm2$) & 6.00 ($\pm0.08$)  \\
 13311 & 391 ($\pm4$) & 0.65 ($\pm0.08$) & 149 ($\pm1$) & 4.99 ($\pm0.10$)  \\

\multicolumn{5}{c}{South}\\
 12927 & 397 ($\pm6$) & 0.65 ($\pm0.02$) & 148 ($\pm1$) & 4.33 ($\pm0.13$)  \\
 13311 & 394 ($\pm4$) & 0.52 ($\pm0.18$) & 153 ($\pm2$) & 3.66 ($\pm0.19$)  \\

\multicolumn{5}{c}{West}\\
 12927 & 339 ($\pm4$) & 4.21 ($\pm0.13$) & 150 ($\pm2$) & 7.91 ($\pm0.31$)  \\
 13311 & 328 ($\pm3$) & 4.33 ($\pm0.09$) & 151 ($\pm2$) & 8.07 ($\pm0.31$)  \\
 
\hline
\end{tabular}
\caption{Temperatures and masses for the Case 1 model fitted to the ER regions. Uncertainties are the $3\sigma$ uncertainties on the fit parameters.}
\label{tab:dust_fit_params_regions}
\end{table}

%----- LINE EVOLUTION ------%
\section{Emission line properties and evolution}\label{sect:line_evol}

%----- LINE FITS ------%
\subsection{Spectral line fits}\label{sect:line_fits}
As in J23, but for both the Day~12927 and 13311 observations,
we have analysed spectra extracted from the ejecta region and from the four cardinal point sub-regions whose boundaries are shown in Fig.~\ref{fig:spectral_extraction_regions}. In addition, a `total' spectrum, whose boundaries are shown in Fig.~\ref{fig:cone_aperture}, was also extracted for each epoch.

The software used to perform the line fits within the {\sc dipso} analysis package
\citep{Howarth2014} are the Gaussian emission line fitting ({\sc elf}) routines written by P. J. Storey. The software returns an error estimate for the best-fitting central wavelength, simultaneously with errors on the fitted full width at half maximum (FWHM) line width and total line flux, by tracing the behavior of the $\chi^{2}$ surface around an iteratively determined best-fitting minimum. The particular method used to
achieve this is the parabolic expansion of $\chi^{2}$ as described by \citet[][their Sect. 8]{Bevington2003}. For some ejecta lines, which are not necessarily Gaussian, we approximate their shape using single or multi-component Gaussians.
Our line-fitting results, for both Days~12927 and 13311, are presented in 
Tables~\ref{tab:cardinal_pointsNS} and 
\ref{tab:cardinal_pointsEW}
for the four cardinal point regions, in Table~\ref{tab:total_spectra} for the `total' spectra and in Table~\ref{tab:ejecta} for the ejecta
spectra. 

\subsection{Emission line properties in the different extraction regions}\label{sect:line_properties}

In J23 it was noted that ER cardinal point lines from singly ionized species (e.g. H~I, [Ni~{\sc ii}], [Ar~{\sc ii}] and [Ne~{\sc ii}], plus the [Fe~{\sc ii}] 17.93~$\mu$m line) displayed much larger FWHMs than lines from more highly ionized species (e.g. [S~{\sc iv}], [O~{\sc iv}], [Ne~{\sc v}] and [Ne~{\sc vi}]), with the FWHMs of these more highly ionized species being effectively unresolved in the MIRI/MRS spectra. From our Day~12927 and 13311 spectra, fine-structure (FS) lines from singly ionized species have a mean FWHM of 239$\pm$20~km~s$^{-1}$, while those from the above more highly ionized species have a mean FWHM of 110$\pm$12~km~s$^{-1}$. The broader lines from the singly ionized species were attributed to an origin in cooling post-shock gas in the ER. On the other hand, the much narrower lines from the highly ionized species were found to be spatially extended well beyond the ER and were attributed to much lower density diffuse gas that had been flash ionized by the UV pulse from the original supernova explosion and which has still to recombine due to its low density. This was also reported in the recent NIRSpec study of \citet{Larsson2025}, who identify clear narrow components at 0~km~s$^{-1}$ from an extended region in [Ar~{\sc iv}], [Mg~{\sc iv}] and [Ca~{\sc v}].
Confirming these very different origins for the singly and multiply ionized emitting species, the mean radial velocity {\em differences} between the north and south cardinal point regions (Table~\ref{tab:cardinal_pointsNS})
are large (275$\pm$16~km~s$^{-1}$) for the singly ionized FS lines while negligible (19$\pm$20~km~s$^{-1}$) for the highly ionized species, consistent with an ER origin for the singly ionized species. Amongst the singly ionized species, the H~I recombination lines show a larger radial velocity difference between the north and south cardinal point regions than do the singly ionized FS lines (341$\pm$25~km~s$^{-1}$ vs. 275$\pm$16~km~s$^{-1}$), indicating origins from somewhat different regions of the post-shock ER gas 
(their respective FWHMs are 292$\pm$32 vs. 239$\pm$20~km~s$^{-1}$). 

Of the three doubly ionized species observed, the radial velocities and line widths of the [S~{\sc iii}] 18.71~$\mu$m line from all four cardinal point regions are similar to those of the more highly ionized species, while the [Ar~{\sc iii}] 8.99~$\mu$m line is narrow in all but the west cardinal point, suggesting an ER contribution to its emission from the latter region. The [Ne~{\sc iii}] 15.56~$\mu$m line required a two-component fit for all four cardinal point regions: one narrow and one broad component, whose relative contributions to the total line flux varied from region to region. For the north and south regions, whose radial velocity patterns show the largest differences between singly ionized and highly ionized species, the broad [Ne~{\sc iii}] components show similar radial velocities to those of the singly ionized species, indicating 
an origin in the ER, while the narrow [Ne~{\sc iii}] components have similar
radial velocities to those of the more highly ionized species, indicating an
origin from the same extended diffuse emitting region.

The line width patterns with degree of ionization measured for the Days~12927 and 13311 `total' spectra
(Table~\ref{tab:total_spectra}) are
similar to those discussed above for the cardinal point regions, with the obvious qualification that the singly ionized species line widths are even broader than those measured for the individual cardinal point spectra, their additional broadening being attributable to the large radial velocity differences between the cardinal point regions included in the `total' spectral extraction.

The Day~12927 and 13311 ejecta spectra (Table~\ref{tab:ejecta}) show the usual narrow lines ($\sim$120~km~s$^{-1}$ FWHM) from the three-, four- and five-times ionized species, with similar narrow line widths measured for the [Ar~{\sc iii}] and [S~{\sc iii}] lines. These lines are therefore interpreted as coming from the diffuse background rather than the ejecta. Of the singly ionized species, [Ni~{\sc ii}] 6.64~$\mu$m from the ejecta region is noticeably broader than in the north cardinal point spectrum (383$\pm$45 vs. 194$\pm$12~km~s$^{-1}$ FWHM), as is [Ne~{\sc ii}] 12.81~$\mu$m  (380$\pm$15 vs.
256$\pm$5~km~s$^{-1}$ FWHM).  These two lines from the ejecta region are also broader than in the `total' spectral extraction, where the FWHM line widths were 340$\pm$30 and 296$\pm$6~km~s$^{-1}$ for the [Ni~{\sc ii}] and [Ne~{\sc ii}] lines respectively.

For the ejecta lines, we note that except for the lines associated with the compact object, we expect very broad lines from the inner ejecta ($\gtrsim10^{3}$~km~s$^{-1}$, as seen in optical and NIR spectra). In our MIRI/MRS spectra, the only such broad emission lines are [Fe~{\sc ii}]~17.9~\micron\ and 26~\micron\ lines. The other lines in Table~\ref{tab:ejecta} that have widths of $\sim300-400$~km~s$^{-1}$ are likely due to contamination of emission from the ER. 
Therefore, the lines listed in Table~\ref{tab:ejecta} likely have four different origins: diffuse background, ER contamination, the central region/compact object, and bulk ejecta. Multi-component lines such as [S~{\sc iv}] 10.51~$\mu$m and [Fe~{\sc ii}] 17.94~$\mu$m have contributions from two of these different origins.

As discussed by \citet{Fransson2024}, the ejecta [Ar~{\sc ii}] 6.99~$\mu$m line in the Day~12927 spectrum shows two blue-shifted components, at -262 and -177~km~s$^{-1}$, respectively. The [S~{\sc iv}] 10.51~$\mu$m line also shows a weak component with a similar blue-shift to that of the [Ar~{\sc ii}] line
(see Table~\ref{tab:ejecta}).
\citet{Fransson2024} showed that the
[Ar~{\sc ii}] emission originated from the very center of the ejecta region and was spatially unresolved. From photoionization modeling, they showed that the blue-shifted lines in the MIRI/MRS ejecta spectrum, along with several blue-shifted lines in their NIRSpec/IFU spectrum of the ejecta, were most plausibly powered by a bare neutron star or pulsar wind nebula located at the center of the ejecta. The MIRI/MRS and NIRSpec/IFU data for the ejecta region from both Days~12927 and 13311 for MIRI/MRS and Days 12927 and 13511/13512 for NIRSpec/IFU, along with their implications for the compact source model, are discussed in \citet{Larsson2025}.

\subsection{Comparison of Cycle 1 and 2 line fluxes}\label{sect:line_fluxes}
The final columns of Tables~\ref{tab:cardinal_pointsNS} to
\ref{tab:ejecta}, for the ER and ejecta regions, list for each line and velocity component the ratio of the Day~13311 line flux divided by the Day~12927 flux. The vast majority of these ratios are consistent within the uncertainties with a ratio close to unity, i.e. no change during the intervening 384 days. Apart from [Ne~{\sc ii}]~12.81~$\mu$m and H~{\sc i}+He~{\sc i} 6-5 7.45~$\mu$m (see next subsection),
the relatively small number of exceptions to this are found to be lines from high-ionization species, an effect which we attribute to the differing amounts of emission from these species in the two spatially distinct regions from where the background spectra were extracted at each epoch. \\

As shown by Table~\ref {tab:background_spectra} in the Appendix, the dedicated background region to the south of \sn\ (see Fig.~\ref{fig:fovs}) which was used for Day~13311 background estimation exhibits no line emission from highly ionized species, consistent with emission originating from LMC diffuse ionized gas. On the other hand, Table~\ref{tab:background_spectra} shows that the region used for Day~12927 background estimation, immediately adjacent to the aperture used to extract the `total' spectrum, shows significant line emission from highly ionized species.\footnote{We note that all of the emission lines in the Days~12927 and 13311 background spectra (Table~\ref{tab:background_spectra}), including those from singly ionized species, are narrow and spectrally unresolved by the MIRI/MRS.} The differences between Days~12927 and 13311 in the amount of background line emission being subtracted for the case of the highly ionized species, together with spatial variations in their line emission between the Day~12927 background region and the various target aperture extractions, appears to account for those Day (13311/12927) line flux ratios in Tables~\ref{tab:cardinal_pointsNS} to \ref{tab:ejecta} which differ significantly from unity.

\renewcommand{\tabcolsep}{2pt}
\begin{table*}
\caption{Detected emission lines, their measured radial velocities\tablenotemark{a}, full-width half maxima, and line fluxes, for the North and South cardinal points regions of the ER, for Days~12927 and 13311. Note that some lines required multi-component fits. Fine-structure lines from singly ionized species {(mean FWHM~$\sim240$~km~s$^{-1}$) are attributed to emission from the ER. Highly ionized species (mean FWHM$\sim110$~km~s$^{-1}$) originate in much lower density diffuse gas surrounding \sn. Very broad components (FWHM$>$1000~km~s$^{-1}$) result from the rapidly expanding inner ejecta. See Sect.~\ref{sect:line_properties} for details.}}
\label{tab:cardinal_pointsNS}
\centering
\begin{tabular}{lllllllll}
\hline
\hline
 & & \multicolumn{3}{c}{Day 12927} & \multicolumn{3}{c}{Day 13311} &  \\
\cmidrule(lr){3-5}\cmidrule(lr){6-8}
Species & $\lambda_{\rm lab}$\tablenotemark{b} 
& Velocity & FWHM & Flux & Velocity & FWHM & Flux & Flux ratio\tablenotemark{c} \\ 
 & $\mu$m & km~s$^{-1}$ & km~s$^{-1}$ & $10^{-24}$~W~cm$^{-2}$ & km~s$^{-1}$ & km~s$^{-1}$ & $10^{-24}$~W~cm$^{-2}$ &  \\
\hline
% -----------------------------------------------
\multicolumn{9}{c}{North (6.05882$\times10^{-12}$ sr)}\\
\cmidrule(lr){3-8}
{[}Ni~{\sc ii}{]} & 6.6360 & -143.2$\pm$2.8 & 206.2$\pm$6.8 & 17.9$\pm$6.8 & -157.1$\pm$3.1 & 181.9$\pm$7.4 & 13.5$\pm$0.5 & 0.75$\pm$0.03 \\
{[}Ar~{\sc ii}{]} & 6.985274 & -151.4$\pm$2.0 & 235.5$\pm$4.6 & 38.1$\pm$0.7 & -160.6$\pm$2.8 & 259.2$\pm$6.3 & 32.6$\pm$0.7 & 0.86$\pm$0.02 \\
'' & '' & -2290 & FWZI=3660 & 19.6$\pm$1.8 & -2290 & FWZI=3510 & 16.4$\pm$1.7 & 0.84$\pm$0.12 \\
H~{\sc i}+He~{\sc i} 6-5 & 7.459858 & -189.0$\pm$4.3 & 259.8$\pm$10.6 & 17.2$\pm$0.6 & -209.6$\pm$5.7 & 294.7$\pm$13.9 & 14.8$\pm$0.6 & 0.86$\pm$0.05 \\
{[}Ne~{\sc vi}{]} & 7.6524 & -15.2$\pm$3.6 & 106.8$\pm$9.1 & 4.4$\pm$0.3 & -20.1$\pm$3.4 & 112.2$\pm$8.5 & 4.9$\pm$0.3 & 1.09$\pm$0.10 \\
H~{\sc i}+He~{\sc i} 10-7 & 8.760064 & -174.4$\pm$31.3 & 343.1$\pm$70.3 & 2.0$\pm$0.4 & -200.2$\pm$22.4 & 335.8$\pm$66.0 & 2.7$\pm$0.4 & 1.39$\pm$0.36 \\
{[}Ar~{\sc iii}{]} & 8.99138 & -117.0$\pm$41.8 & 261.3$\pm$72.6 & 1.8$\pm$0.4 & -34.7$\pm$4.3 & 112.1$\pm$10.4 & 2.7$\pm$0.2 & 1.50$\pm$0.35 \\
{[}S~{\sc iv}{]} & 10.51049 & -0.9$\pm$7.2 & 100.5$\pm$11.8 & 1.1$\pm$0.2 & -19.2$\pm$3.2 & 88.9$\pm$7.2 & 2.0$\pm$0.1 & 1.79$\pm$0.28 \\
{[}Ni~{\sc ii}{]} & 10.682308 & -159.8$\pm$15.7 & 256.2$\pm$43.1 & 1.8$\pm$0.2 & -157.0$\pm$14.8 & 227.0$\pm$31.2 & 1.6$\pm$0.2 & 0.92$\pm$0.18 \\
H~{\sc i}+He~{\sc i} 7-6 & 12.371898 & -183.7$\pm$7.4 & 255.3$\pm$18.5 & 4.1$\pm$0.3 & -165.6$\pm$7.2 & 295.2$\pm$18.7 & 4.0$\pm$0.2 & 0.98$\pm$0.08 \\
{[}Ne~{\sc ii}{]} & 12.813548 & -141.8$\pm$0.6 & 251.8$\pm$1.6 & 97.2$\pm$0.5 & -142.2$\pm$0.6 & 260.9$\pm$1.4 & 85.4$\pm$0.4 & 0.88$\pm$0.01 \\
 '' & '' & -2140.1$\pm$58.9 & 1032.0$\pm$148.4 & 8.8$\pm$1.4 & -2140.9$\pm$37.3 & 990.1$\pm$83.9 & 9.9$\pm$1.0 & 1.14$\pm$0.22 \\
{[}Ne~{\sc v}{]} & 14.32168 & -17.3$\pm$1.2 & 94.4$\pm$2.9 & 4.4$\pm$0.1 & -16.6$\pm$1.3 & 98.4$\pm$3.3 & 4.6$\pm$0.1 & 1.06$\pm$0.04 \\
{[}Ne~{\sc iii}{]} & 15.5551 & 3.1$\pm$3.0 & 113.0$\pm$10.1 & 3.8$\pm$0.5 & 8.9$\pm$0.9 & 120.4$\pm$2.8 & 9.9$\pm$0.4 & 2.60$\pm$0.37 \\
 '' & '' & -85.6$\pm$12.1 & 323.2$\pm$14.8 &7.6$\pm$0.6 & -134.6$\pm$12.9 & 344.4$\pm$20.1 & 6.0$\pm$0.5 & 0.80$\pm$0.09 \\
{[}Fe~{\sc ii}{]} & 17.936026 & -130.4$\pm$8.7 & 247.1$\pm$21.3 & 6.8$\pm$0.5 & -134.4$\pm$7.2 & 217.8$\pm$19.6 & 5.5$\pm$0.4 & 0.80$\pm$0.09 \\
{[}S~{\sc iii}{]} & 18.71303 & -21.3$\pm$12.6 & 137.4$\pm$19.7 & 2.3$\pm$0.4 & -17.5$\pm$4.0 & 128.1$\pm$6.6 & 5.6$\pm$0.3 & 2.4$\pm$0.4 \\ 
H~{\sc i}+He~{\sc i} 8-7 & 19.061898 & -205.0$\pm$45.2 & 348.8$\pm$80.7 & 2.2$\pm$0.5 & -187.4$\pm$25.2 & 236.0$\pm$56.2 & 1.7$\pm$0.4 & 0.77$\pm$0.25 \\
{[}Ne~{\sc v}{]} & 24.3175 & 2.9$\pm$6.6 & 115.4$\pm$11.0 & 4.7$\pm$0.5 & -7.0$\pm$9.6 & 115.1$\pm$18.2 & 4.6$\pm$0.7 & 0.96$\pm$0.19 \\ 
{[}O~{\sc iv}{]} & 25.8903 & -3.5$\pm$5.5 & 165.7$\pm$15.7 & 10.6$\pm$1.1 & -4.1$\pm$7.2 & 120.8$\pm$15.2 & 8.5$\pm$1.3 & 0.81$\pm$0.15 \\
{[}Fe~{\sc ii}{]} & 25.98839 & -2800 -- 3100 & FWZI$\sim$5900 & 178.2$\pm$15.8 & -2800 -- 2800 & FWZI$\sim$5600 & 177.2$\pm$16.4 & 0.99$\pm$0.13 \\
% -----------------------------------------------
\hline
\multicolumn{9}{c}{South (6.31920$\times10^{-12}$ sr)}\\
\cmidrule(lr){3-8}
{[}Fe~{\sc ii}{]} & 5.3401693 & 2696.4$\pm$15.9 & 1035.0$\pm$39.9 & 64.5$\pm$2.2 & 2756.0$\pm$12.7 & 820.7$\pm$31.6 & 63.0$\pm$2.2 & 0.98$\pm$0.05 \\
{[}Ni~{\sc ii}{]} & 6.6360 & 130.9$\pm$6.3 & 210.1$\pm$13.5 & 7.8$\pm$0.5 & 145.6$\pm$6.8 & 207.3$\pm$18.7 & 6.7$\pm$0.5 & 0.85$\pm$0.08 \\
{[}Ar~{\sc ii}{]} & 6.985274 & 136.9$\pm$3.3 & 224.0$\pm$9.0 & 15.0$\pm$0.6 & 133.7$\pm$4.7 & 266.5$\pm$11.3 & 14.0$\pm$0.6 & 0.94$\pm$0.03 \\
'' & '' & 2700 & FWZI$\sim$6020 & 16.3 & 2700 & FWZI$\sim$5760 & 16.4 & 1.0 \\
H~{\sc i}+He~{\sc i} 6-5 & 7.459858 & 138.5$\pm$8.6 & 288.4$\pm$22.2 & 7.6$\pm$0.5 & 164.7$\pm$11.4 & 270.6$\pm$26.1 & 6.8$\pm$0.6 & 0.89$\pm$0.10 \\
{[}Ne~{\sc vi}{]} & 7.6524 & 9.6$\pm$7.8 & 92.5$\pm$15.7 & 1.5$\pm$0.2 & -0.3$\pm$6.1 & 113.7$\pm$14.2 & 2.1$\pm$0.2 & 1.41$\pm$0.28 \\
{[}Ar~{\sc iii}{]} & 8.99138 & -- & -- & -- & -32.9$\pm$5.5 & 108.3$\pm$12.7 & 2.0$\pm$0.2 & -- \\
{[}S~{\sc iv}{]} & 10.51049 & -- & -- & -- & -16.8$\pm$10.5 & 124.9$\pm$21.6 & 1.0$\pm$0.2 & --
\\
H~{\sc i}+He~{\sc i} 7-6 & 12.371898 & 115.9$\pm$14.2 & 346.2$\pm$43.5 & 3.3$\pm$0.3 & 186.7$\pm$31.4 & 592.7$\pm$96.1 & 3.6$\pm$0.5 & 1.11$\pm$0.19 \\
{[}Ne~{\sc ii}{]} & 12.813548 & 114.4$\pm$0.9 & 260.2$\pm$2.2 & 55.6$\pm$0.4 & 114.5$\pm$0.9 & 280.2$\pm$2.2 & 46.7$\pm$0.3 & 0.84$\pm$0.01 \\
 '' & '' & 2822.0$\pm$48.8 & 651.7$\pm$92.3 & 4.2$\pm$1.3 & 2892.3$\pm$53.6 & 958.1$\pm$183.8 & 5.2$\pm$1.3 & 1.25$\pm$0.34 \\
{[}Ne~{\sc v}{]} & 14.32168 & -3.8$\pm$2.5 & 103.2$\pm$6.2 & 2.7$\pm$0.1 & -2.3$\pm$2.6 & 99.7$\pm$6.7 & 2.5$\pm$0.1 & 0.91$\pm$0.07 \\
{[}Ne~{\sc iii}{]} & 15.5551 & 44.2$\pm$3.7 & 45.6$\pm$18.2 & 1.0$\pm$0.2 & 15.3$\pm$1.7 & 128.9$\pm$5.5 & 6.4$\pm$0.6 & 6.3$\pm$1.5 \\
 '' & '' & 147.7$\pm$6.5 & 218.5$\pm$13.8 & 3.6$\pm$0.7 & 128.5$\pm$22.9 & 280.3$\pm$29.3 & 3.6$\pm$0.7 & 0.99$\pm$0.18 \\
{[}Fe~{\sc ii}{]} & 17.936026 & 98.5$\pm$8.8 & 234.2$\pm$22.2 & 4.7$\pm$0.4 & 128.0$\pm$9.9 & 276.1$\pm$21.4 & 4.2$\pm$0.3 & 0.91$\pm$0.10 \\
{[}S~{\sc iii}{]} & 18.71303 & -- & -- & -- & -15.8$\pm$4.9 & 126.8$\pm$8.5 & 4.7$\pm$0.3 & -- \\
{[}Ne~{\sc v}{]} & 24.3175 & 17.1$\pm$9.4 & 130.0$\pm$18.8 & 3.4$\pm$0.5 & 51.6$\pm$19.6 & 164.2$\pm$46.3 & 3.3$\pm$0.8 & 0.98$\pm$0.27 \\
{[}O~{\sc iv}{]} & 25.8903 & 23.1$\pm$15.7 & 184.7$\pm$47.5 & 6.6$\pm$1.8 & -1.9$\pm$19.0 & 126.5$\pm$36.1 & 4.2$\pm$1.3 & 0.63$\pm$0.26 \\
{[}Fe~{\sc ii}{]} & 25.98839 & -2200 -- 4300 & FWZI$\sim$6500 & 270.9$\pm$20.2 & -2100 -- 4100 & FWZI$\sim$6000 & 343.9$\pm$22.3 & 1.27$\pm$0.13 \\
% -----------------------------------------------

%
\hline
\end{tabular}
\flushleft
\tablenotetext{a}{Radial velocities are with respect to the SN~1987A frame, defined by \citet{Groningsson2008} as corresponding to +286.7 km~s$^{-1}$ heliocentric.}
\tablenotetext{b}{Vacuum wavelengths from the compilation by \cite{vanHoof2018}.}
\tablenotetext{c}{Ratio of Day~13311 to 12927 line fluxes.}
\end{table*}

\begin{table*}
\caption{Detected emission lines, their measured radial velocities\tablenotemark{a}, full-width half maxima, and line fluxes, for the East and West cardinal points regions of the ER, for Days~12927 and 13311. See caption from Table~\ref{tab:cardinal_pointsNS} for note on line origins.} 
\label{tab:cardinal_pointsEW}
\centering
\begin{tabular}{lllllllll}
\hline
\hline
 & & \multicolumn{3}{c}{Day 12927} & \multicolumn{3}{c}{Day 13311} &  \\
\cmidrule(lr){3-5}\cmidrule(lr){6-8}
Species & $\lambda_{\rm lab}$\tablenotemark{b}
& Velocity & FWHM & Flux & Velocity & FWHM & Flux & Flux ratio\tablenotemark{c}\\ 
 & $\mu$m & km~s$^{-1}$ & km~s$^{-1}$ & $10^{-24}$~W~cm$^{-2}$ & km~s$^{-1}$ & km~s$^{-1}$ & $10^{-24}$~W~cm$^{-2}$ &  \\

% -----------------------------------------------
\hline
\multicolumn{9}{c}{East (5.98718$\times10^{-12}$ sr)}\\
\cmidrule(lr){3-8}
{[}Ni~{\sc ii}{]} & 6.6360 & 39.5$\pm$10.4 & 206.0$\pm$29.3 & 4.9$\pm$0.7 & -2.5$\pm$15.2 & 277.6$\pm$33.8 & 4.9$\pm$0.6 & 1.00$\pm$0.18 \\
{[}Ar~{\sc ii}{]} & 6.985274 & 33.3$\pm$6.7 & 248.2$\pm$18.9 & 9.1$\pm$0.6 & 32.7$\pm$5.6 & 233.9$\pm$17.7 & 10.7$\pm$0.7 & 1.17$\pm$0.11 \\
H~{\sc i}+He~{\sc i} 6-5 & 7.459858 & 13.5$\pm$8.3 & 196.4$\pm$18.3 & 4.0$\pm$0.4 & 24.2$\pm$14.0 & 269.1$\pm$39.7 & 5.0$\pm$0.6 & 1.25$\pm$0.18 \\
{[}Ne~{\sc vi}{]} & 7.6524 & -9.7$\pm$3.6 & 113.3$\pm$8.4 & 4.1$\pm$0.3 & -13.6$\pm$2.5 & 85.6$\pm$8.3 & 3.6$\pm$0.3 & 0.94$\pm$0.09 \\
{[}Ar~{\sc iii}{]} & 8.99138 & -45.7$\pm$6.2 & 109.2$\pm$17.9 & 1.4$\pm$0.2 & -36.2$\pm$3.6 & 100.8$\pm$8.6 & 2.5$\pm$0.2 & 1.88$\pm$0.28 \\
{[}S~{\sc iv}{]} & 10.51049 & -8.8$\pm$7.6 & 89.6$\pm$15.2 & 0.9$\pm$0.1 & -20.4$\pm$3.2 & 101.4$\pm$8.0 & 2.4$\pm$0.2 & 2.7$\pm$0.5 \\
H~{\sc i}+He~{\sc i} 7-6 & 12.371898 & -- & -- & -- & -30.6$\pm$17.0 & 140.6$\pm$38.7 & 0.6$\pm$0.2 & -- \\
 {[}Ne~{\sc ii}{]} & 12.813548 & 14.9$\pm$1.6 & 275.4$\pm$4.2 & 35.3$\pm$0.5 & 13.8$\pm$1.7 & 248.2$\pm$5.2 & 29.6$\pm$0.6 & 0.84$\pm$0.02 \\
 {[}Ne~{\sc v}{]} & 14.32168 & -16.6$\pm$1.1 & 105.8$\pm$2.8 & 4.9$\pm$0.1 & -15.2$\pm$1.2 & 94.5$\pm$3.2 & 4.8$\pm$0.1 & 0.98$\pm$0.04 \\
{[}Ne~{\sc iii}{]} & 15.5551 & 17.2$\pm$3.2 & 164.4$\pm$9.2 & 5.0$\pm$0.5 & 6.8$\pm$0.9 & 131.6$\pm$2.6 & 10.0$\pm$2.1 & 2.0$\pm$0.2 \\
 '' & '' & -45.6$\pm$47.2 & 462.5$\pm$91.6 & 1.8$\pm$0.6 & -116.6$\pm$49.7 & 579.3$\pm$79.9 & 2.1$\pm$0.4 & 1.16$\pm$0.44 \\
{[}Fe~{\sc ii}{]} & 17.936026 & 3.3$\pm$13.7 & 253.5$\pm$31.0 & 3.2$\pm$0.4 & 13.9$\pm$20.0 & 254.7$\pm$51.0 & 2.2$\pm$0.4 & 0.71$\pm$0.15 \\
{[}S~{\sc iii}{]} & 18.71303 & -27.0$\pm$9.8 & 148.9$\pm$14.9 & 2.3$\pm$0.3 & -19.4$\pm$3.5 & 135.0$\pm$5.8 & 6.3$\pm$0.3 & 2.7$\pm$0.3 \\
{[}Ne~{\sc v}{]} & 24.3175 & 7.8$\pm$6.0 & 134.9$\pm$11.8 & 5.6$\pm$0.5 & 12.6$\pm$7.4 & 127.9$\pm$14,4 & 5.7$\pm$0.7 & 1.03$\pm$0.15 \\
{[}O~{\sc iv}{]} & 25.8903 & 3.0$\pm$7.0 & 174.2$\pm$18.2 & 11.5$\pm$1.0 & 7.4$\pm$6.0 & 185.7$\pm$15.5 & 15.4$\pm$1.1 & 1.34$\pm$0.14 \\
{[}Fe~{\sc ii}{]} & 25.98839 & -2700 -- 2800 & FWZI$\sim$5500 & 71.1$\pm$8.8 & -2700 -- 2800 & FWZI$\sim$5500 & 77.3$\pm$9.3 & 1.09$\pm$0.17 \\

% -----------------------------------------------
\hline
\multicolumn{9}{c}{West (5.80030$\times10^{-12}$ sr)}\\
\cmidrule(lr){3-8}
{[}Fe~{\sc ii}{]} & 5.3401693 & 1809.6$\pm$46.4 & 1174.4$\pm$117.3 & 22.7$\pm$2.2 & 1582.4$\pm$95.2 & 2480.4$\pm$273.0 & 31.8$\pm$3.8 & 1.40$\pm$0.17 \\
H~{\sc i}+He~{\sc i} 9-6 & 5.908213 & -29.5$\pm$10.5 & 300.0$\pm$22.6 & 10.3$\pm$0.8 & -58.5$\pm$9.3 & 235.1$\pm$24.4 & 9.5$\pm$0.7 & 0.93$\pm$0.10 \\
{[}Ni~{\sc ii}{]} & 6.6360 & -20.8$\pm$1.0 & 238.3$\pm$2.5 & 56.0$\pm$0.5 & -20.7$\pm$1.0 & 238.1$\pm$2.6 & 52.2$\pm$0.5 & 0.93$\pm$0.01 \\
{[}Ar~{\sc ii}{]} & 6.985274 & -11.9$\pm$0.6 & 229.0$\pm$1.4 & 117.1$\pm$0.6 & -8.0$\pm$0.7 & 231.4$\pm$1.7 & 107.6$\pm$0.7 & 0.92$\pm$0.01 \\
H~{\sc i}+He~{\sc i} 6-5 & 7.459858 & -29.0$\pm$1.5 & 284.0$\pm$3.9 & 50.8$\pm$0.7 & -24.7$\pm$2.2 & 289.3$\pm$6.1 & 45.9$\pm$0.9 & 0.90$\pm$0.02 \\
{[}Ne~{\sc vi}{]} & 7.6524 & -12.2$\pm$2.2 & 127.5$\pm$5.9 & 7.9$\pm$0.3 & -9.7$\pm$2.5 & 131.8$\pm$7.1 & 8.2$\pm$0.4 & 1.04$\pm$0.06 \\
H~{\sc i}+He~{\sc i} 10-7 & 8.760064 & -32.6$\pm$14.3 & 358.4$\pm$36.1 & 4.6$\pm$0.4 & -72.3$\pm$15.5 & 417.7$\pm$37.7 & 5.4$\pm$0.5 & 1.16$\pm$0.15 \\
{[}Ar~{\sc iii}{]} & 8.99138 & -17.8$\pm$9.9 & 211.1$\pm$30.3 & 4.8$\pm$5.3 & -17.0$\pm$7.5 & 219.9$\pm$23.0 & 5.1$\pm$0.5 & 1.07$\pm$0.15 \\
{[}S~{\sc iv}{]} & 10.51049 & 11.1$\pm$3.7 & 83.3$\pm$8.3 & 2.4$\pm$0.2 & 1.0$\pm$3.3 & 102.6$\pm$7.9 & 3.8$\pm$0.3 & 1.63$\pm$0.19 \\
{[}Ni~{\sc ii}{]} & 10.682308 & -23.5$\pm$6.8 & 232.0$\pm$16.9 & 5.2$\pm$0.3 & -42.4$\pm$9.8 & 242.2$\pm$25.8 & 4.6$\pm$0.4 & 0.89$\pm$0.10 \\
H~{\sc i}+He~{\sc i} 7-6 & 12.371898 & -18.5$\pm$4.7 & 307.7$\pm$13.2 & 14.5$\pm$0.5 & -17.8$\pm$4.2 & 302.6$\pm$11.6 & 13.7$\pm$0.5 & 0.94$\pm$0.05 \\
 {[}Ne~{\sc ii}{]} & 12.813548 & -4.4$\pm$0.5 & 238.5$\pm$1.1 & 291.5$\pm$0.7 & -0.4$\pm$0.5 & 238.1$\pm$1.4 & 265.4$\pm$1.4 & 0.91$\pm$0.02 \\
{[}Ne~{\sc v}{]} & 14.32168 & -14.3$\pm$1.0 & 97.9$\pm$2.7 & 7.3$\pm$0.2 & -12.6$\pm$1.2 & 93.6$\pm$3.2 & 6.8$\pm$0.2 & 0.93$\pm$0.03 \\
{[}Ne~{\sc iii}{]} & 15.5551 & 21.2$\pm$2.5 & 169.4$\pm$14.2 & 15.0$\pm$4.8 & 23.0$\pm$1.8 & 127.1$\pm$9.6 & 12.2$\pm$2.7 & 0.81$\pm$0.32 \\
'' & '' & 22.0$\pm$8.3 & 287.4$\pm$48.1 & 9.6$\pm$4.8 & 21.0$\pm$4.3 & 255.5$\pm$22.4 & 15.1$\pm$2.6 & 1.57$\pm$0.82 
 \\
{[}Fe~{\sc ii}{]} & 17.936026 & -28.1$\pm$4.7 & 266.7$\pm$11.5 & 18.0$\pm$0.7 & -11.0$\pm$6.3 & 262.4$\pm$14.8 & 16.2$\pm$0.8 & 0.90$\pm$0.06 \\
{[}S~{\sc iii}{]} & 18.71303 & -4.2$\pm$9.8 & 120.4$\pm$27.1 & 3.5$\pm$0.4 & 7.3$\pm$25.8 & 76.5$\pm$99.1 & 7.2$\pm$3.7 & 2.1$\pm$1.1 \\
H~{\sc i}+He~{\sc i} 8-7 & 19.061898 & -29.2$\pm$33.6 & 440.6$\pm$80.9 & 7.0$\pm$1.2 & -39.7$\pm$22.0 & 264.4$\pm$71.1 & 4.2$\pm$0.8 & 0.59$\pm$0.16 \\
{[}Ne~{\sc v}{]} & 24.3175 & -0.4$\pm$4.0 & 113.7$\pm$6.8 & 7.6$\pm$0.5 & 4.7$\pm$5.6 & 115.6$\pm$9.4 & 7.2$\pm$0.7 & 0.95$\pm$0.11 \\
{[}O~{\sc iv}{]} & 25.8903 & -2.7.0$\pm$2.9 & 132.0$\pm$6.6 & 15.1$\pm$0.8 & -4.7$\pm$4.3 & 133.6$\pm$11.3 & 14.0$\pm$1.3 & 0.93$\pm$0.10 \\
{[}Fe~{\sc ii}{]} & 25.98839 & -2700 -- 3500 & FWZI$\sim$6200 & 195.3$\pm$18.6 & -2400 -- 3500 & FWZI$\sim$5900 & 197.0$\pm$18.1 & 1.01$\pm$0.13 \\
\hline
\end{tabular}
\flushleft
\tablenotetext{a}{Radial velocities are with respect to the SN~1987A frame, defined by \citet{Groningsson2008} as corresponding to +286.7 km~s$^{-1}$ heliocentric.}
\tablenotetext{b}{Vacuum wavelengths from the compilation by \cite{vanHoof2018}.}
\tablenotetext{c}{Ratio of Day~13311 to 12927 line fluxes.}
\end{table*}

%-------------------------------------------

\begin{table*}
\caption{Detected emission lines, their measured radial velocities\tablenotemark{a}, full-width half maxima, and line fluxes, for the `total' spectral extraction for Days~12927 and 13311. See caption from Table~\ref{tab:cardinal_pointsNS} for note on line origins.} 
\label{tab:total_spectra}
\centering
\begin{tabular}{lllllllll}
\hline
\hline
 & & \multicolumn{3}{c}{Day 12927} & \multicolumn{3}{c}{Day 13311} &  \\
\cmidrule(lr){3-5}\cmidrule(lr){6-8}
Species & $\lambda_{\rm lab}$\tablenotemark{b} 
& Velocity & FWHM & Flux & Velocity & FWHM & Flux & Flux ratio\tablenotemark{c}
\\ 
 & $\mu$m & km~s$^{-1}$ & km~s$^{-1}$ & $10^{-24}$~W~cm$^{-2}$ & km~s$^{-1}$ & km~s$^{-1}$ & $10^{-24}$~W~cm$^{-2}$ &  \\
\hline
{[}Fe~{\sc ii}{]} & 5.3401693 & $\sim$2180 & FWZI$\sim$2100 & 530.6$\pm$33.8 & $\sim$2000 & FWZI$\sim$2300 & 727.4$\pm$50.9 & 1.37$\pm$0.13 \\
{[}Ni~{\sc ii}{]} & 6.6360 & -24.7$\pm$3.3 & 310.3$\pm$7.8 & 269.6$\pm$6.2 & -37.6$\pm$4.8 & 373.2$\pm$12.8 & 306.0$\pm$9.0 & 1.14$\pm$0.04\\
{[}Ar~{\sc ii}{]} & 6.985274 & -10.0$\pm$2.0 & 253.6$\pm$5.2 & 560.1$\pm$9.7 & -6.4$\pm$2.8 & 250.9$\pm$7.6 & 489.5$\pm$12.6 & 0.87$\pm$0.03 \\
'' & '' & -256.5$\pm$3.0 & 148.3$\pm$6.6 & 154.6$\pm$8.2 & -254.7$\pm$3.9 & 148.5$\pm$8.4 & 154.4$\pm$10.3 & 1.00$\pm$0.09 \\
H~{\sc i}+He~{\sc i} 6-5 & 7.459858 & -36.2$\pm$5.0 & 338.5$\pm$12.1 & 243.3$\pm$7.7 & -35.5$\pm$5.5 & 380.2$\pm$14.7 & 269.7$\pm$10.4 & 1.11$\pm$0.06 \\
H~{\sc i}+He~{\sc i} 8-6 & 7.502493\tablenotemark{d} & -20.5$\pm$20.9 & 562.5$\pm$49.9 & 137.2$\pm$11.2 & 87.4$\pm$19.2 & 519.2$\pm$57.9 & 111.2$\pm$11.5 & 0.81$\pm$0.11 \\
{[}Ne~{\sc vi}{]} & 7.6524 & -8.2$\pm$0.8 & 103.0$\pm$2.1 & 151.9$\pm$2.6 & -7.9$\pm$1.0 & 110.1$\pm$2.6 & 158.1$\pm$3.1 & 1.04$\pm$0.03 \\
{[}Ar~{\sc iii}{]} & 8.99138 & -26.7$\pm$0.9 & 105.5$\pm$2.2 & 146.0$\pm$2.5 & -29.4$\pm$1.4 & 112.0$\pm$3.5 & 86.1$\pm$2.3 & 0.59$\pm$0.02
\\
{[}S~{\sc iv}{]} & 10.51049 & -16.2$\pm$0.6 & 98.9$\pm$1.5 & 171.7$\pm$2.2 & -12.0$\pm$1.5 & 99.1$\pm$3.5 & 76.4$\pm$2.4 & 0.45$\pm$0.11 \\
H~{\sc i}+He~{\sc i} 7-6 & 12.371898 & -21.9$\pm$9.0 & 381.9$\pm$25.0 & 89.1$\pm$5.1 & -22.4$\pm$6.4 & 328.4$\pm$17.2 & 80.1$\pm$3.6 & 0.90$\pm$0.07 \\
{[}Ne~{\sc ii}{]} & 12.813548 & -10.9$\pm$1.2 & 302.3$\pm$3.2 & 1902.0$\pm$17.0 & -10.1$\pm$1.4 & 289.9$\pm$3.7 & 1731.0$\pm$18.7 & 0.91$\pm$0.01
%\footnote{also two comp fit for cycle 2} 
\\
{[}Ar~{\sc v}{]} & 13.10219 & 3.9$\pm$3.9 & 81.6$\pm$19.9 & 10.1$\pm$1.2 & 10.3$\pm$11.1 & 148.8$\pm$27.9 & 14.9$\pm$2.3 & 0.68$\pm$0.13 \\
{[}Ne~{\sc v}{]} & 14.32168 & -13.7$\pm$0.4 & 98.8$\pm$0.9 & 191.9$\pm$1.5 & -13.1$\pm$0.4 & 99.1$\pm$1.1 & 184.7$\pm$1.7 & 0.96$\pm$0.01 \\
{[}Ne~{\sc iii}{]} & 15.5551 & 4.6$\pm$0.3 & 114.5$\pm$1.2 & 678.4$\pm$17.9 & 8.8$\pm$0.6 & 121.9$\pm$2.0 & 375.5$\pm$11.5 & 0.54$\pm$0.02 \\
'' & '' & 17.3$\pm$2.6 & 256.9$\pm$14.1 & 209.2$\pm$16.6 & 19.0$\pm$9.8 & 367.9$\pm$38.4 & 109.5$\pm$10.9 & 0.52$\pm$0.07 \\
{[}Fe~{\sc ii}{]} & 17.936026 & -11.2$\pm$6.3 & 246.1$\pm$15.4 & 147.0$\pm$8.4 & -6.7$\pm$5.8 & 266.7$\pm$14.6 & 136.5$\pm$6.8 & 0.93$\pm$0.07 \\
{[}S~{\sc iii}{]} & 18.71303 & -21.7$\pm$0.7 & 140.1$\pm$1.2 & 605.1$\pm$5.6 & -14.9$\pm$1.7 & 136.4$\pm$3.1 & 255.0$\pm$5.5 & 0.42$\pm$0.01 \\
{[}Ne~{\sc v}{]} & 24.3175 & 3.1$\pm$1.6 & 122.0$\pm$2.9 & 298.2$\pm$7.7 & 4.9$\pm$3.0 & 122.0$\pm$5.4 & 267.8$\pm$12.4 & 0.90$\pm$0.05 \\
{[}O~{\sc iv}{]} & 25.8903 & -1.1$\pm$5.3 & 157.3$\pm$13.5 & 628.4$\pm$47.2 & -1.1$\pm$2.9 & 146.8$\pm$6.7 & 597.0$\pm$24.0 & 0.95$\pm$0.08 \\
{[}Fe~{\sc ii}{]} & 25.98839 & -2400 -- 4400 & FWZI$\sim$6000 & 4420$\pm$304 & -2500 -- 4600 & FWZI$\sim$6100  & 4586$\pm$72 & 1.04$\pm$0.07 \\
\hline
\end{tabular}
\flushleft
\tablenotetext{a}{Radial velocities are with respect to the SN~1987A frame, defined by \citet{Groningsson2008} as corresponding to +286.7 km~s$^{-1}$ heliocentric.}
\tablenotetext{b}{Vacuum wavelengths from the compilation by \cite{vanHoof2018}.}
\tablenotetext{c}{Ratio of Day~13311 to 12927 line fluxes.}
\tablenotetext{d}{Quoted wavelength is for H~{\sc i} 8-6; the blended 11-7 transition lies +224.3 km~s$^{-1}$ to the red of the 8-6 transition.}
\end{table*}

%-------------------------------------------

\begin{table*}
\caption{Detected emission lines, their measured radial velocities\tablenotemark{a}, full-width half maxima, and line fluxes, for Days~12927 and 13311. MIRI/MRS spectra of the ejecta region (7.762060$\times10^{-12}$ sr). We note that many of these lines are not associated with the bulk ejecta and have origins from other emission components in the \sn\ system (see Sect.~\ref{sect:line_properties} for details).} 
\label{tab:ejecta}
\centering
\begin{tabular}{lllllllll}
\hline
\hline
 & & \multicolumn{3}{c}{Day 12927} & \multicolumn{3}{c}{Day 13311} &  \\
\cmidrule(lr){3-5}\cmidrule(lr){6-8}
Species & $\lambda_{\rm lab}$\tablenotemark{b} & Velocity & FWHM & Flux & Velocity & FWHM & Flux & Flux ratio\tablenotemark{c} \\ 
 & $\mu$m & km~s$^{-1}$ & km~s$^{-1}$ & $10^{-24}$~W~cm$^{-2}$ & km~s$^{-1}$ & km~s$^{-1}$ & $10^{-24}$~W~cm$^{-2}$ &  \\
\hline
{[}Ni~{\sc ii}{]} & 6.6360 & -80.6$\pm$22.7 & 403.1$\pm$44.6 & 5.6$\pm$0.7 & -63.9$\pm$30.7 & 363.5$\pm$59.2 & 5.4$\pm$0.9 & 0.95$\pm$0.20 \\
{[}Ar~{\sc ii}{]} & 6.985274 & -261.7$\pm$0.4 & 122.3$\pm$1.4 & 67.3$\pm$1.3 & -260.1$\pm$0.7 & 115.3$\pm$2.1 & 70.7$\pm$2.0 & 1.05$\pm$0.05 \\

'' & '' & -176.6$\pm$6.0 & 362.2$\pm$9.1 & 40.0$\pm$1.5 & -193.1$\pm$7.8 & 383.3$\pm$17.3 & 47.3$\pm$2.4 & 1.18$\pm$0.04 \\
H~{\sc i}+He~{\sc i} 6-5 & 7.459858 & -29.7$\pm$12.0 & 207.7$\pm$60.3 & 3.5$\pm$0.6 & -52.1$\pm$13.6 & 389.0$\pm$47.5 & 7.4$\pm$0.7 & 2.11$\pm$0.42 \\
H~{\sc i}+He~{\sc i} 8-6 & 7.502493\tablenotemark{d} & 115.7$\pm$29.4 & 234.7$\pm$62.7 & 1.6$\pm$0.4 & 74.6$\pm$36.9 & 349.4$\pm$70.7 & 2.3$\pm$0.5 & 1.47$\pm$0.51 \\
{[}Ne~{\sc vi}{]} & 7.6524 & -55.5$\pm$4.9 & 114.6$\pm$12.0 & 3.7$\pm$0.3 & -11.5$\pm$5.4 & 98.0$\pm$13.0 & 3.0$\pm$0.3 & 0.83$\pm$0.12 \\
{[}Ar~{\sc iii}{]} & 8.99138 & -29.2$\pm$2.1 & 106.3$\pm$5.2 & 4.8$\pm$0.2 & -32.6$\pm$2.5 & 111.2$\pm$6.1 & 5.1$\pm$0.2 & 1.05$\pm$0.07 \\
{[}S~{\sc iv}{]} & 10.51049 & -16.8$\pm$1.3 & 98.5$\pm$3.4 & 4.8$\pm$0.2 & -18.9$\pm$1.8 & 100.3$\pm$4.6 & 4.9$\pm$0.2 & 1.03$\pm$0.06 \\
'' & '' & -268.5$\pm$28.1 & 336.4$\pm$107.0 & 1.6$\pm$0.4 & -254.0$\pm$16.8 & 177.4$\pm$56.8 & 1.2$\pm$0.3 & 0.78$\pm$0.26 \\
H~{\sc i}+He~{\sc i} 7-6\tablenotemark{e} & 12.371898 & -47.3$\pm$13.2 & 465.0$\pm$34.4 & 4.1$\pm$0.3 & 5.1$\pm$19.0 & 548.5$\pm$41.9 & 4.2$\pm$0.3 & 1.02$\pm$0.11 \\
{[}Ne~{\sc ii}{]} & 12.813548 & -24.9$\pm$2.1 & 364.8$\pm$5.3 & 77.8$\pm$1.0 & -23.1$\pm$2.5 & 394.1$\pm$6.5 & 66.5$\pm$1.1 & 0.86$\pm$0.18 \\
{[}Ne~{\sc v}{]} & 14.32168 & -15.1$\pm$1.3 & 110.1$\pm$3.1 & 4.2$\pm$0.1 & -13.1$\pm$1.4 & 107.9$\pm$3.5 & 4.0$\pm$0.1 & 0.97$\pm$0.04 \\
{[}Cl~{\sc ii}{]} & 14.3678 & -302.6$\pm$18.5 & 328.2$\pm$64.1 & 1.5$\pm$0.2 & -303.1$\pm$13.4 & 242.7$\pm$45.9 & 1.4$\pm$0.2 & 0.92$\pm$0.18 \\
{[}Ne~{\sc iii}{]} & 15.5551 & 2.2$\pm$0.5 & 112.1$\pm$2.0 & 19.4$\pm$0.6 & 1.5$\pm$0.6 & 116.1$\pm$2.0 & 20.4$\pm$0..5 & 1.05$\pm$0.04 \\
'' & '' & -20.7$\pm$6.4 & 334.6$\pm$24.0 & 8.3$\pm$0.6 & -37.4$\pm$12.7 & 413.3$\pm$45.7 & 6.5$\pm$0.6 & 0.78$\pm$0.09 \\ 
{[}Fe~{\sc ii}{]} & 17.936026 & -26.7$\pm$6.0 & 323.5$\pm$14.3 & 9.2$\pm$0.4 & -17.9$\pm$12.5 & 342.2$\pm$30.5 & 7.6$\pm$0.6 & 0.83$\pm$0.07 \\
'' & '' & 2426.6$\pm$81.3 & 1199.6$\pm$14.2 & 5.1$\pm$0.8 & 2511.8$\pm$82.6 & 1003.9$\pm$201.8 & 5.8$\pm$1.2 & 1.14$\pm$0.29 \\
{[}S~{\sc iii}{]} & 18.71303 & -21.8$\pm$1.3 & 145.5$\pm$2.3 & 17.2$\pm$0.3 & -15.0$\pm$7.8 & 139.8$\pm$15.2 & 16.4$\pm$1.4 & 0.95$\pm$0.09
\\
{[}Ne~{\sc v}{]} & 24.3175 & 4.3$\pm$6.8 & 118.8$\pm$11.6 & 5.7$\pm$0.6 & 1.3$\pm$7.2 & 128.2$\pm$13.2 & 6.4$\pm$0.7 & 1.13$\pm$0.17 \\
{[}O~{\sc iv}{]} & 25.8903 & 11.9$\pm$10.6 & 155.8$\pm$33.6 & 8.3$\pm$2.7 & 25.5$\pm$15.8 & 102.4$\pm$43.9 & 4.5$\pm$1.7 & 0.60$\pm$0.30 \\
{[}Fe~{\sc ii}{]} & 25.988390 & -2700 -- 4800 & FWZI$\sim$7500 & 645.9$\pm$12.7 & -2500 -- 5000 & FWZI$\sim$7500 & 654.4$\pm$12.3 & 1.01$\pm$0.03 \\
\hline
\end{tabular}
\flushleft
\tablenotetext{a}{Radial velocities are with respect to the SN~1987A frame, defined by \citet{Groningsson2008} as corresponding to +286.7 km~s$^{-1}$ heliocentric.}
\tablenotetext{b}{Vacuum wavelengths from the compilation by \cite{vanHoof2018}.}
\tablenotetext{c}{Ratio of Day~13311 to 12927 line fluxes.}
\tablenotetext{d}{Quoted wavelength is for H~{\sc i} 8-6; the blended 11-7 transition lies +224.3 km~s$^{-1}$ to the red of the 8-6 transition.}
\tablenotetext{e}{Possible contribution by redshifted H$_{2}$~0-0~S(2) line (see Sect.~\ref{sect:h2_lines}).}
\end{table*}

%-------------------------------------------

%----- LINE MAPS ------%
\subsection{Emission line maps}\label{sect:line_maps}
\begin{figure*}[!h]
\centering
\includegraphics[width=\hsize]{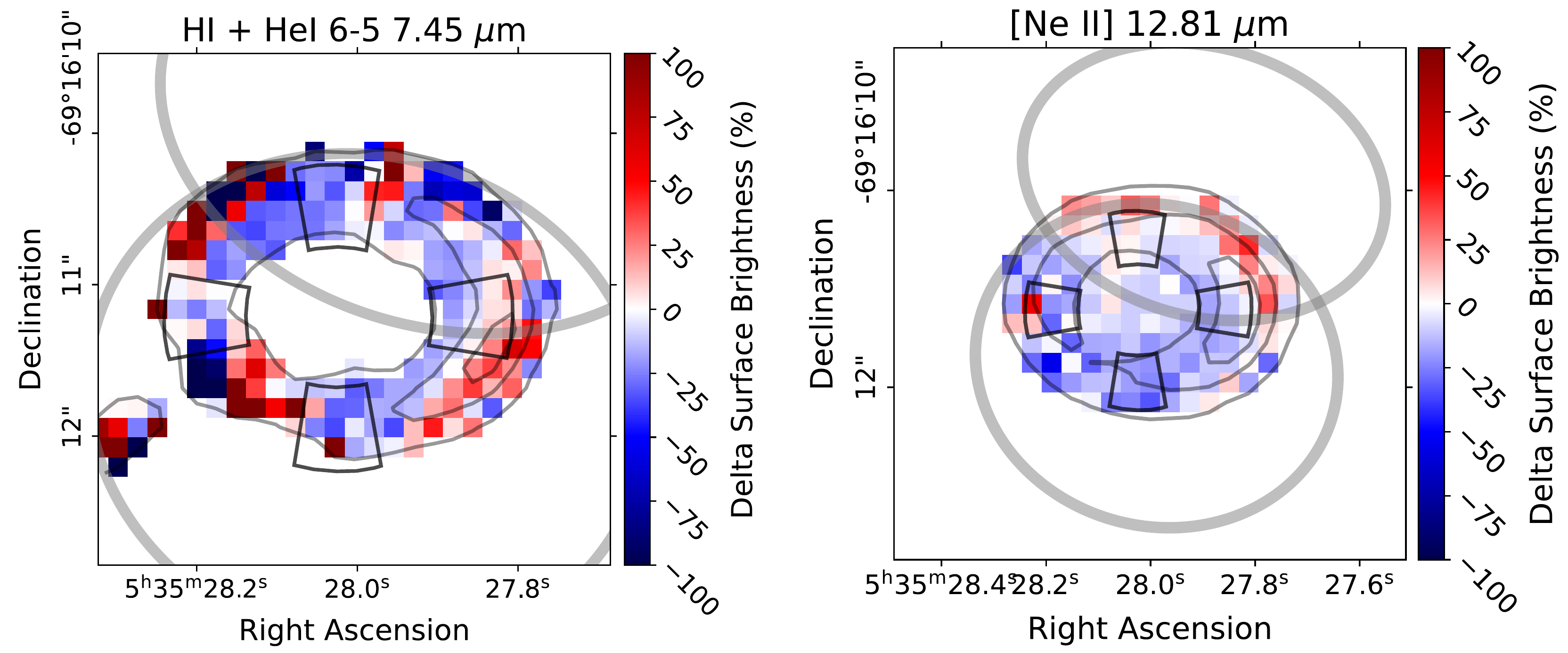}
\caption{Evolution in surface brightness between Days~12927 and 13311 for selected emission lines which show changes in surface brightness in some or all of the cardinal point regions. The velocity ranges have been restricted to only show the emission components associated with the ER. The H~{\sc i}+He~{\sc i} 6-5 7.45~\micron\ line is shown on the left and the {[}Ne~{\sc ii}{]} 12.81~\micron\ line on the right. Dust continuum contours are shown by the dark gray lines in each panel at 1$\sigma$, 3$\sigma$, and 5$\sigma$ above the median background level. The cardinal point extractions regions are plotted as dark gray polygons. The large gray circles indicate the position of the Outer Rings. The images have been masked so that only pixels above the 1$\sigma$ contour are shown.}
\label{fig:er_line_evol}
\end{figure*}

To assess variations in line brightness across \sn\ between the epochs we produced moment-0 maps of selected emission lines whose fluxes were measured to have changed between Days~12927 and 13311 in some or all of the cardinal point regions in order to map the percentage changes in surface brightness. We generated the moment-0 maps following the same procedure described in J23. The Day~13311 maps were reprojected onto the Day~12927 pixel grid as described in Sect.~\ref{sect:psf_decon}. Inspection of Tables 3-5 shows that the majority of the ER emission lines do not show flux changes between the two epochs that exceed the formal uncertainty limits. Two lines that do are the H~{\sc i}+He~{\sc i} 6-5 7.45~\micron\ and {[}Ne~{\sc ii}{]} 12.81~\micron\ lines. The surface brightness evolution maps for these are shown in Fig.~\ref{fig:er_line_evol}.

The day 13311 cardinal point region integrated fluxes for the H~{\sc i}+He~{\sc i} 6-5 7.45~\micron\ feature were measured to have decreased to $\sim80-90$\% of their day~12927 values in all regions except the east. This is reflected in the surface brightness evolution map in Fig.~\ref{fig:er_line_evol} left. The north and south regions are fading. The west appears to be fading overall, though the extraction regions are not capturing the full picture. On the western edge and to the south of this region, the emission has increased between the epochs. The location of this brightening is consistent with the outer edge of the ER which saw the most recent interaction with the blast wave. The H~{\sc i}+He~{\sc i} 6-5 7.45~\micron\ line was measured to brighten in the eastern region by $\sim25$\%, contrary to the other regions. The reason for this appears to be due to contamination of the region by residual dark current noise which affects channel 1 (see Fig.~\ref{fig:cyc12_cubes}). The flux in this region is low, making contamination by this noise more of an issue, causing large, erratic values in the brightness evolution map. 

The day 13311 {[}Ne~{\sc ii}{]} 12.81~\micron\ integrated line flux was measured to have decreased to $\sim80-90$\% of the day 12927 value in all cardinal point regions regions, which is also represented in the brightness evolution map in Fig.~\ref{fig:er_line_evol} right. In addition, there appears to be a brightening along the northern and western edge of the ER. Again, the increase in surface brightness here is most likely due to the recent passage of the blast wave.

%-------------------------------------------
%-------------------------------------------

\begin{table*}
\caption{Measured properties of \htwo\ lines detected in the mid-north region at Days~12927 and 13311.}
\label{tab:h2_lines}
\centering
\begin{tabular}{llllllll}
\hline
\hline
 & & \multicolumn{3}{c}{Day 12927} & \multicolumn{3}{c}{Day 13311}  \\
 \cmidrule(lr){3-5}\cmidrule(lr){6-8}
Transition & $\lambda_{\rm lab}$\tablenotemark{b} & Velocity & FWHM & Flux & Velocity & FWHM & Flux \\ 
 & $\mu$m & km~s$^{-1}$ & km~s$^{-1}$ & $10^{-23}$~W~cm$^{-2}$ & km~s$^{-1}$ & km~s$^{-1}$ & $10^{-23}$~W~cm$^{-2}$  \\\hline
\hline
0-0 S(7) & 5.5115 & -1819$\pm$613 & 937$\pm$297 & 1.3$\pm$1.1 & -1078$\pm$194 & 1558$\pm$190 & 3.0$\pm$0.7 \\
0-0 S(6) & 6.1088 & -2193$\pm$1375 & 1037$\pm$682 & 1.2$\pm$1.0 & -2230$\pm$1046 & 1851$\pm$684 & 2.0$\pm$1.4 \\
0-0 S(5) & 6.9091 & -1728$\pm$18 & 913$\pm$44 & 2.4$\pm$0.2 & -1731$\pm$24 & 773$\pm$56 & 1.9$\pm$0.2 \\
0-0 S(4) & 8.0258 & -1770$\pm$339 & 432$\pm$111 & 0.3$\pm$0.2 & -1729$\pm$24 & 474$\pm$54 & 0.6$\pm$0.1 \\
0-0 S(3) & 9.6649 & -1856$\pm$175 & 747$\pm$120 & 0.7$\pm$0.2 & -1820$\pm$112 & 702$\pm$81 & 0.8$\pm$0.1 \\
0-0 S(1) & 17.035 & -2138$\pm$90 & 2784$\pm$206 & 0.5$\pm$0.1 & -2355$\pm$323 & 2249$\pm$310 & 0.4$\pm$0.1 \\
\hline
\end{tabular}
\flushleft
\tablenotetext{a}{Radial velocities are with respect to the \sn\ frame, defined by \citet{Groningsson2008} as corresponding to +286.7 km~s$^{-1}$ heliocentric.}
\end{table*}

\subsection{Molecular hydrogen lines}\label{sect:h2_lines}
\begin{figure*}
\centering
\includegraphics[width=\hsize]{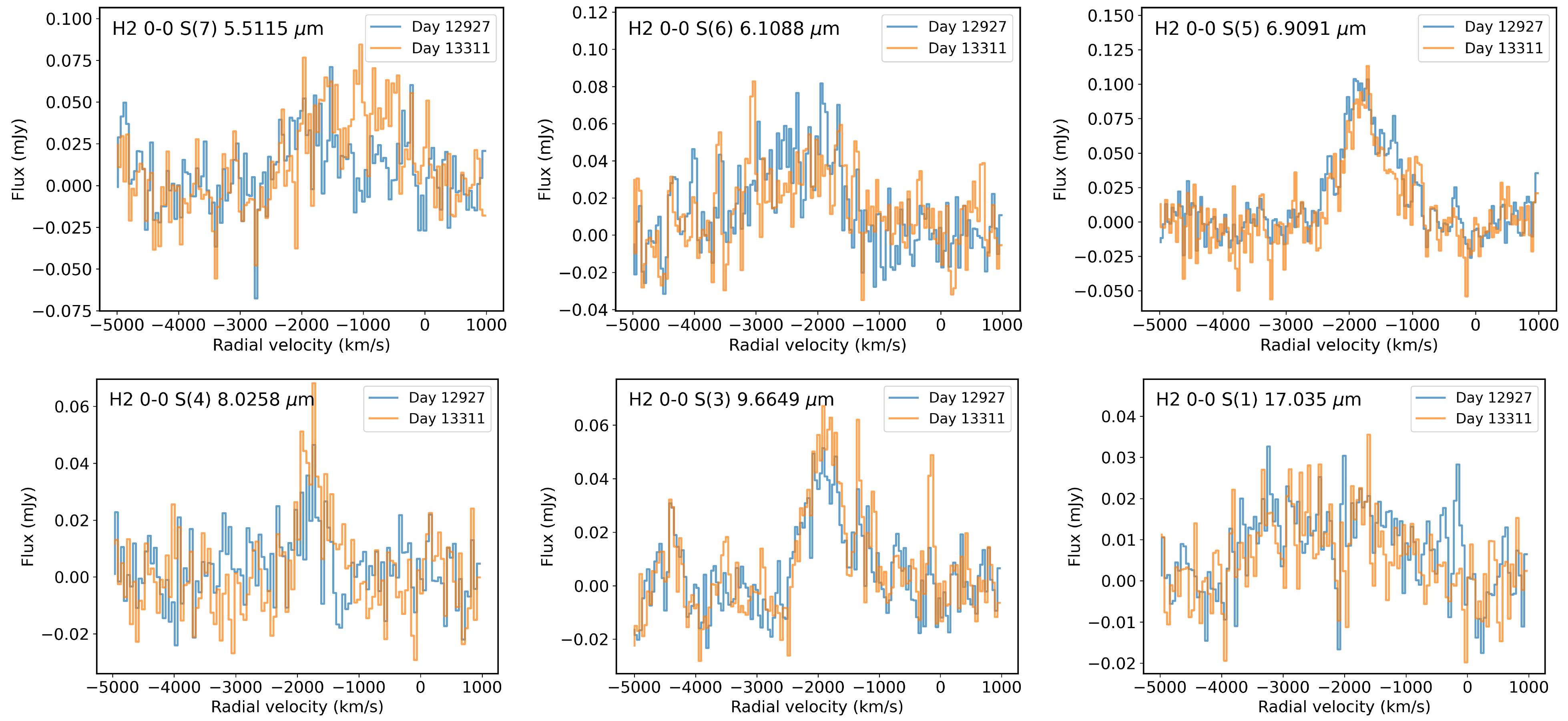}
\caption{Profiles of the six identified \htwo\ lines at both Days 12927 and 13311 extracted from the `mid north' region. Radial velocities are with respect to the SN~1987A frame, defined by \citet{Groningsson2008} as corresponding to +286.7 km~s$^{-1}$ heliocentric.} 
\label{fig:h2_profiles}
\end{figure*}

In addition to the atomic lines, discussed above, we also identified molecular lines from H$_2$. Lines of H$_2$ were previously discovered in ground based near-IR observations of \sn\ \citep{Fransson2016} and in more detail over the full NIRSpec/IFU range, $1-5 \ \mu$m, in \cite{Larsson2023}. These authors used synthetic spectra based on models for photodisssociation regions (PDRs) to compared to the observed spectrum and a good fit was obtained for PDR models having a high density and strong UV flux in the $912-1110$~ \AA\ range. As a source of this flux the UV emission from the ER/ejecta shocks were proposed. Based on the best fit PDR models in \cite{Draine1996} we searched the MIRI/MRS spectrum and found lines at 5.5115, 6.1088, 6.9091, 8.0258, 9.6649 and 17.035~$\mu$m (see Fig.~\ref{fig:h2_profiles}). These are among the strongest lines in the MIRI/MRS range in the PDR models, and all coming from pure rotational transitions from the ground vibrational level, namely the 0-0 S(1), S(3), S(5) and S(7) transitions of ortho-H$_2$, together with the 0-0 S(4) and S(6) transitions of para-H$_2$ (see Table \ref{tab:h2_lines}). We did not detect the 0-0~S(2) transition of para-H$_2$ at 12.28~\micron. From their relative statistical weights, para-H$_2$ transitions should be three times weaker than adjacent ortho-H$_2$ transitions. Interpolating between the fluxes of the 17.035~\micron\ S(1) and 9.665~\micron\ S(3) transitions, the 12.28~\micron\ S(2) transition would be expected to have a flux of $\sim0.2\times10^{-23}$~W~cm$^{-2}$ at both epochs. We suspect the line was not clearly detected due to the presence of the H~{\sc i}+He~{\sc i}~7-6~12.37~\micron\ line which is located at the same position where the redshifted 12.28~\micron\ S(2) transition is expected. 

\begin{figure*}
\centering
\includegraphics[width=\hsize]{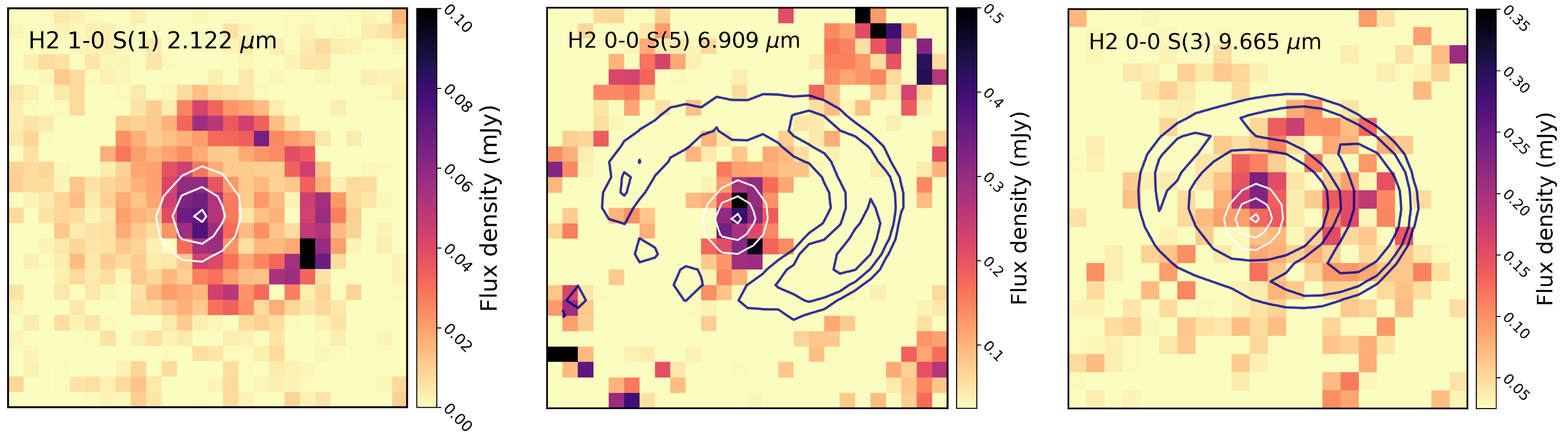}
\caption{Continuum subtracted maps from identified \htwo\ lines. Left: Map of the \htwo\ 1-0 S(1) 2.122~$\mu$m line extracted from NIRSpec/IFU data \citep{Larsson2025}. The contours of the [Ar~{\sc ii}]~6.99~\micron\ line associated with the compact object are shown in white for reference. Middle: Same as left, for \htwo\ 0-0 S(5) 6.9091~$\mu$m, with the adjacent continuum indicated by the dark blue contours at $2\sigma$, $3\sigma$, and $5\sigma$. The continuum was determined from 3000--3500~km~s$^{-1}$ on both the blue and red-shifted sides. Right: Same as middle, for \htwo\ 0-0 S(3).} 
\label{fig:h2_maps}
\end{figure*}

We investigated the spatial morphology of the \htwo\ emission lines and found that most were confined to a region between the center of the ejecta and the north ER, a region we refer to as mid-north. Only the \htwo\ 0-0 S(5) line was detected beyond the mid-north region, showing a red-shifted wing located south of the inner ejecta. This was also found in a previous VLT/SINFONI study by \citet{Larsson2019a}. We compared both the line profile and spatial map of the \htwo\ 0-0 S(5) line to lines observed in the NIRSpec/IFU range reported by \citet{Larsson2023}. Fig.~\ref{fig:h2_maps} shows a comparison of the surface brightness distribution of the \htwo\ 1-0 (S1) line at 2.122~\micron\ with those of \htwo\ 0-0 S(3) 9.665~$\mu$m and 0-0 S(5) 6.909~$\mu$m, the two brightest lines detected in the mid-infrared. The 0-0 S(5) 6.909~$\mu$m line shows good correlation with the 1-0 S(1) 2.122~$\mu$m line, extending in the north-south direction, aligning well with the general shape of the inner ejecta. The 0-0 S(3) 9.665~$\mu$m line is only detected in the mid-north region.

%-------------------------------------------
%-------------------------------------------

%-------------------------------------------
%-------------------------------------------

In Fig.~\ref{fig:h2_profiles}  we show the velocity profiles of the \htwo\ lines and in Table \ref{tab:h2_lines} we give velocities, line widths and fluxes determined from spectra extracted from the mid-north region. We note that Gaussian models do not fully capture the asymmetric profiles in Fig.~\ref{fig:h2_profiles} and so are approximations to the true profiles in this case. Many of the observed fluxes are among the strongest in the PDR models. Checking with the models in \cite{Draine1996}, we find that their Model Rw3o gives the best fit, followed by  Model Qm3o. Fluxes are for the former model within a factor of two, although the 17.035~$\mu$m line is overproduced by a factor five. Other models give a very poor fit. These models were also those which gave the best fits to the Day~12927 observations in \cite{Larsson2023}, characterized by a large UV flux and high density.

However, one has to keep in mind that the observed fluxes have considerable uncertainties, depending on the strong dust continuum above $\sim8~\mu$m and increasing noise for the shorter lines. In addition, the PDR models were calculated for PDR regions with UV spectra, intensities, temperatures and densities specific to these. Most models assumed UV spectra similar to that of a B0 star with a color temperature of 29,000 K, while the UV spectrum in this wavelength range from the shocks in \sn\ may be dominated by two-photon emission \citep[][]{France2011}. The temperature was varied between 300-1000 K, with an approximate adjustment for dust shielding. Although dust is present in the core of \sn, these assumptions are especially doubtful for the inner regions of the SN ejecta. In addition, the heating source in \sn\ is in the inner core radioactive decay of ${}^{44}$Ti and outside this X-ray heating from the shock interaction. 

%----- EJECTA/ER INTERACTION ------%
\section{Ejecta-ER interactions}\label{sect:ejecta_er_int_regions}
\subsection{Emission lines associated with the interaction}
Although the ejecta-ER interaction started already in $\sim 1995$ \citep{Lawrence2000}, up until recently, 35 years after the explosion, the inner ejecta had not yet encountered the reverse shock. Both \citet{Larsson2023} and J23 identified broad lines and highly blue- and red-shifted lines ($\sim3000-4000$~km~s$^{-1}$) from Fe, Ar, Ne, and S in the northeast (blue-shifted) and southwest (red-shifted) of the ER, interpreted as resulting from the inner ejecta being shock heated by the reverse shock near the ring. 

With our Day 13311 MIRI/MRS observations, we can assess the morphological and spectral evolution of the interaction regions in the $\sim1$~year since the initial observation at Day 12927. \citet{Larsson2023} identified several spectral lines associated with the interaction (see their Sect. 3.4 and Fig.~16). We used the [Fe~{\sc ii}]~5.34~\micron\ line, the brightest and best resolved at Day~12927, to define spectral extraction regions in the northern and southern interaction regions. We fixed the size of these regions across the MIRI/MRS wavelength range to optimize the search for new spectral features using the brightest part of the interaction region. We compiled a list from the spectra at each epoch which is presented in Table~\ref{tab:int_region_lines}, along with the results of our fits. Some of the fainter lines were identified with the aid of ionisation models (see Sect.~\ref{sect:fe-line-diag} below).

\renewcommand{\tabcolsep}{3pt}
\begin{table*}
\caption{Detected emission lines, their measured radial velocity ranges and peaks\tablenotemark{a}, and line fluxes, for Days~12927 and 13311 MIRI/MRS spectra of the interaction regions.}
\label{tab:int_region_lines}
\centering
\begin{tabular}{llllllll}
\hline
\hline
 & & \multicolumn{3}{c}{Day 12927} & \multicolumn{3}{c}{Day 13311}  \\
 \cmidrule(lr){3-5}\cmidrule(lr){6-8}
Species & $\lambda_{lab}$ & Flux & Range & Peaks & Flux & Range & Peaks \\
& $\mu$m & $10^{-23}$ W~cm$^{-2}$ & km~s$^{-1}$ & km~s$^{-1}$ & $10^{-23}$ W~cm$^{-2}$ & km~s$^{-1}$ & km~s$^{-1}$ \\
\hline
\hline
\multicolumn{8}{c}{North} \\
\hline
{[}\ion{Fe}{2}{]} & 5.340  & 3.8$\pm$0.4& -2675 to -1020 & -2000 & 3.3$\pm$0.4  & -2690 to -930 & -2100 \\
{[}\ion{Ar}{2}{]} & 6.985 & 2.8$\pm$0.1  & -2710 to -880 & -2070 & 3.2$\pm$0.1 & -2610 to -950 & -1990 \\
{[}\ion{Ne}{2}{]} & 12.814 & 1.8$\pm$0.1 & -3060 to -815 & -1945, -1530 & 1.8$\pm$0.1  & -3040 to -640 & -1910, -1520 \\
{[}\ion{Fe}{2}{]} & 25.988  & 51.1$\pm$12.2  & --$^{c}$ & --$^{c}$ & 58.4$\pm$7.5 & --$^{c}$ & --$^{c}$ \\
\hline
\multicolumn{8}{c}{South} \\
\hline
{[}\ion{Fe}{2}{]} & 5.340  & 50.6$\pm$0.1  & -1390 to +4010 & +2500 & 63.0$\pm$0.1  & +30 to +3875 & +2525 \\
{[}\ion{Fe}{2}{]} & 6.721  & 3.2$\pm$0.2 & +1000 to +3300 & -- & 3.6$\pm$0.2  & +1000 to +3500 & -- \\
{[}\ion{Ar}{2}{]} & 6.985  & 10.5$\pm$0.1 & 224 (FWHM) & +2550 & 13.0$\pm$0.1  & 260 (FWHM) & +2560 \\
{[}\ion{Ar}{3}{]} & 8.991  & 1.5$\pm$0.3 & +980 to +3220 & +2075 & 3.2$\pm$0.4 & +560 to +4580 & +2090 \\
{[}\ion{S}{4}{]} & 10.510  & 1.1$\pm$0.1 & +1355 to +3060 & +2070 & 1.7$\pm$0.1 & +1470 to +4100 & +1905 \\
{[}\ion{Ne}{2}{]} & 12.814  & 2.1$\pm$0.1 & 224 (FWHM) & +2550 & 2.5$\pm$0.1  & 260 (FWHM) & +2560 \\
{[}\ion{Cl}{2}{]} & 14.3678  & 1.9$\pm$0.1 & +1120 to +3515 & -2615 & 2.1$\pm$0.1 & +1410 to +3410 & -- \\
{[}\ion{Ne}{3}{]} & 15.5551 & 1.0$\pm$0.1 & +1000 to +4600 & --  & 1.7$\pm$0.1 & +1000 to +4600 & --  \\
{[}\ion{Fe}{2}{]}  & 17.936  & 2.4$\pm$0.1  & +1460 to +3430 & -- & 4.3$\pm$0.2 & +1150 to +3730 & -- \\
{[}\ion{S}{3}{]} & 18.713  & 3.0$\pm$0.3 & +710 to +3810 & -- & 4.5$\pm$0.3 & +935 to +3910 & -- \\
{[}\ion{Fe}{2}{]} & 25.988  & 83.7$\pm$14.6  & --$^{c}$ & --$^{c}$ & 91.2$\pm$8.9 & --$^{c}$ & --$^{c}$ \\
\hline
\end{tabular}
\flushleft
\tablenotetext{a}{Radial velocities are with respect to the SN~1987A frame, defined by \citet{Groningsson2008} as corresponding to +286.7 km~s$^{-1}$ heliocentric.}
\tablenotetext{b}{Vacuum wavelengths from the compilation by \cite{vanHoof2018}.}
\tablenotetext{c}{No ranges or peaks are quoted for the {[}\ion{Fe}{2}{]}~25.988~\micron\ line since they are assumed to be the same as the {[}\ion{Fe}{2}{]}~5.340~\micron\ line. See Sect.~\ref{sect:interaction_line_evol} for details.}
\end{table*}

\subsection{Morphological evolution}
\begin{figure*}[ht!]
\centering
\includegraphics[width=\hsize]{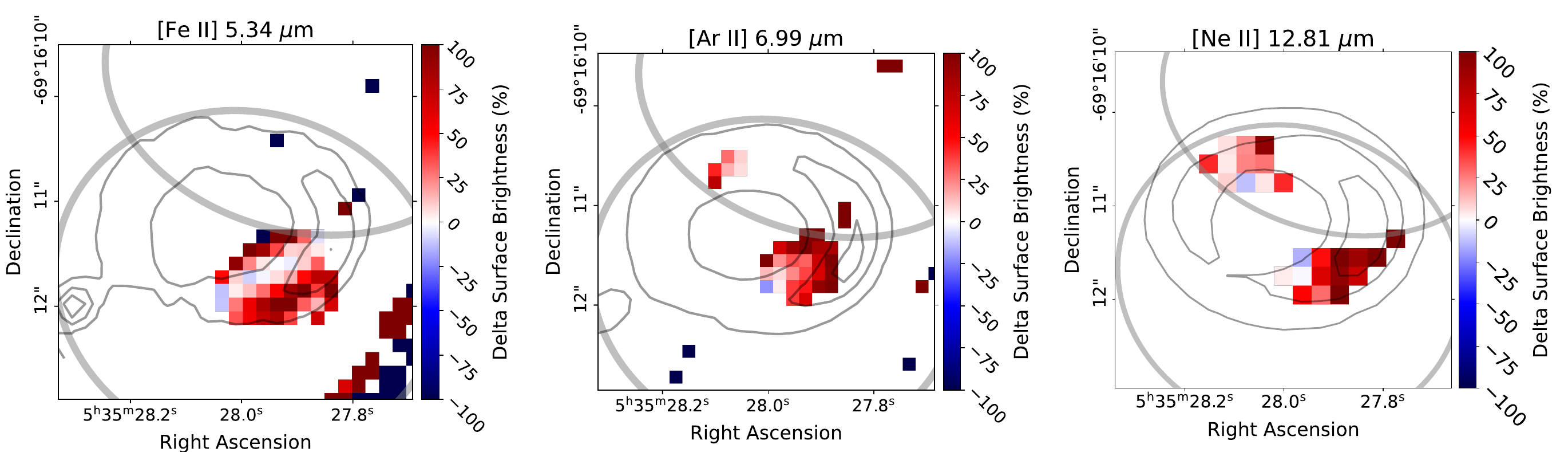}
\caption{Evolution in surface brightness between Days 12927 and 13311 for the three brightest lines in the interaction regions, {[}Fe~{\sc ii}{]}~5.34~\micron\ (left), {[}Ar~{\sc ii}{]}~6.99~\micron\ (middle), {[}Ne~{\sc ii}{]}~12.81~\micron\ (right). The velocity ranges have been restricted to only show the emission components associated with the interaction regions. Dust continuum contours are shown by the dark gray lines in each panel at 1$\sigma$, 3$\sigma$, and 5$\sigma$ above the median background level. The large gray circles indicate the position of the Outer Rings. The cluster of pixels in the lower right of the {[}Fe~{\sc ii}{]}~5.34~\micron\ image are due to residual dark current noise in band 1A from the Day 12927 observation.}
\label{fig:int_interaction}
\end{figure*}

To assess the morphological evolution of the emission lines, we constructed brightness evolution maps following the same procedure in Sect.~\ref{sect:line_maps} for lines in the interaction regions namely {[}\ion{Fe}{2}{]}~5.34~\micron, {[}\ion{Ar}{2}{]}~6.99~\micron, and {[}\ion{Ne}{2}{]}~12.81~\micron, shown in Fig.~\ref{fig:int_interaction}. We omitted the {[}\ion{Fe}{2}{]}~25.99~\micron\ line as the spatial resolution is much poorer in band 4C and significant blending between the interaction regions and high-velocity ejecta is present. Both regions have clearly evolved in the north-east/south-west direction, consistent with the shape of the expanding inner ejecta. The southern region shows the most significant evolution between the epochs.

%-------------------------------------------
%-------------------------------------------

\subsection{Spectral line evolution}
\label{sect:interaction_line_evol}
\begin{figure*}[!h]
\centering
\includegraphics[width=6in]{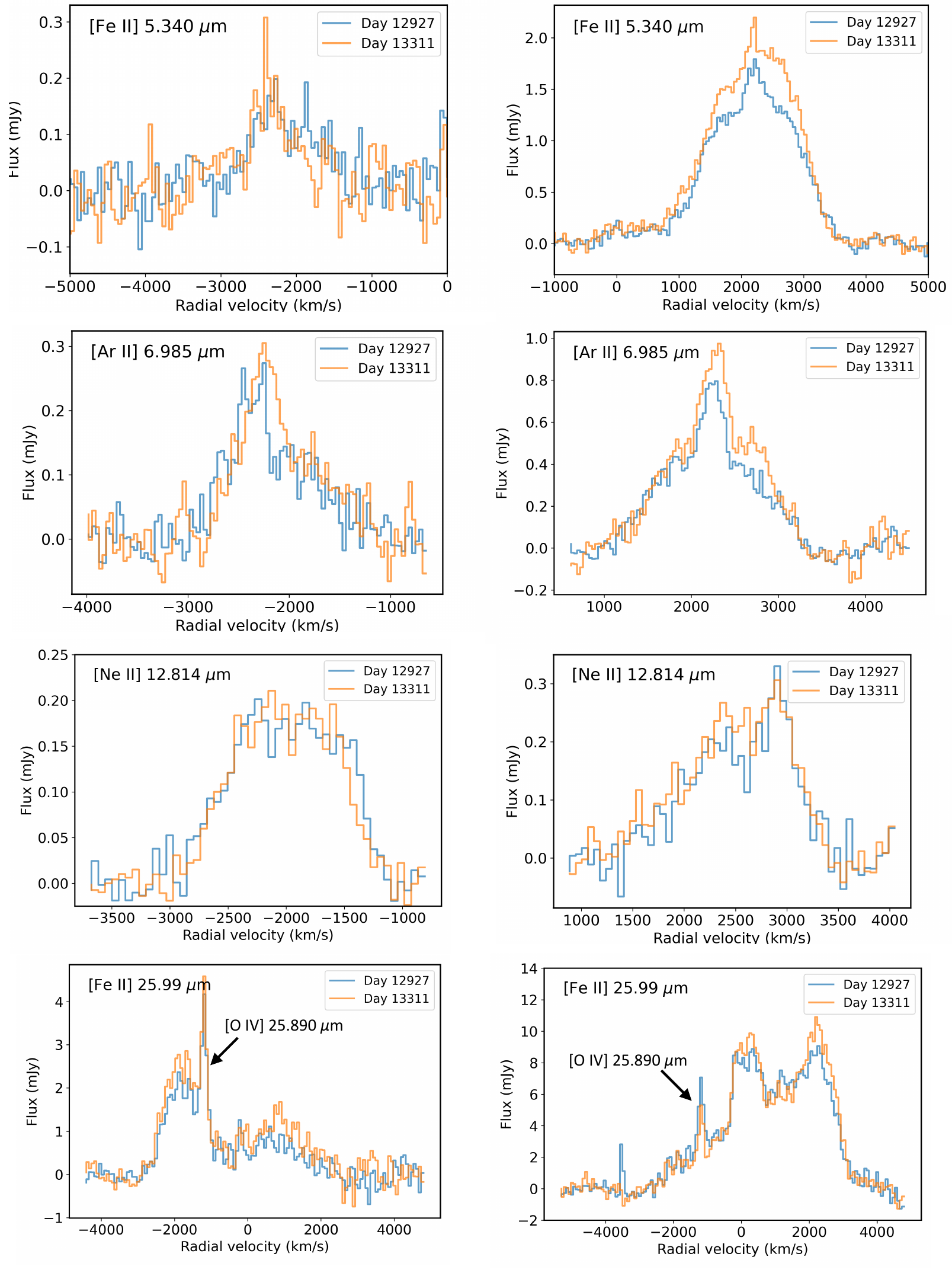}
\caption{Selected line profiles at Days 12927 (blue) and 13311 (orange) in the northern (left column) and southern (right column) interaction regions. The emission lines are identified in the top left of each panel. In the case of the {[}\ion{Fe}{2}{]}~25.99~\micron\ line, the {[}\ion{O}{4}{]}~25.89~\micron\ line which is unrelated to the interaction is also indicated. Radial velocities are with respect to the SN~1987A frame, defined by \citet{Groningsson2008} as corresponding to +286.7 km~s$^{-1}$ heliocentric.} 
\label{fig:int_profiles}
\end{figure*}

The emission line profiles from the interaction regions were found to be unstructured. In general, it was not possible to model the lines using a single or combination of Gaussian components. We therefore estimated the line fluxes by integrating a spline function fitted to the data. We estimated their error by determining the rms noise on the lines using the spline-subtracted data and the rms noise to generate and determine fluxes of 10000 realizations of the line profiles. We set our line error to be 1$\sigma$ of the flux estimates from the simulated lines. The results of our fits are given in Table~\ref{tab:int_region_lines}, also noting the `peaks' in the profiles. The flux estimate for the {[}\ion{Fe}{2}{]}~25.99~\micron\ line differed than the others. The {[}\ion{Fe}{2}{]}~25.99~\micron\ line complex is located in band 4C where the spatial resolution is poorest, with the PSF being comparable in size to the ER. For this reason, the spectral extraction regions defined for both the north and south will contain contributions from the central inner ejecta powered by radioactive decay, as well as the opposite interaction region. We therefore took the approach to scale the {[}\ion{Fe}{2}{]}~5.34~\micron\ profile to the {[}\ion{Fe}{2}{]}~25.99~\micron\ profile from the `total' spectrum (see Sect.~\ref{sect:spec_extraction}). We matched the red and blue sides of the profiles, in the north and south, respectively, and estimated the fluxes using the scaled {[}\ion{Fe}{2}{]}~5.34~\micron\ profiles, and errors using the {[}\ion{Fe}{2}{]}~5.34~\micron\ profiles with the noise properties of the {[}\ion{Fe}{2}{]}~25.99~\micron\ lines. We plotted the profiles at each epoch to assess their evolution, with examples of the brightest lines shown in Fig.~\ref{fig:int_profiles}. 

Both the {[}\ion{Ne}{2}{]}~12.81~\micron\ and {[}\ion{Fe}{2}{]}~25.99~\micron\ lines show minimal evolution between Days 12927 and 13311, exhibiting approximately the same profile shape with slight increases in flux. Similarly, the shapes of the {[}\ion{Fe}{2}{]}~5.34~\micron\ and {[}\ion{Ar}{2}{]}~6.99~\micron\ in the north region are unchanging, with minimal flux difference.

The emission line profiles of the strongest lines, {[}\ion{Fe}{2}{]}~5.34~\micron\ and {[}\ion{Ar}{2}{]}~6.99~\micron\ in the south, show significant increase in brightness between Days 12927 and 13311. Both comprise a broad (Full Width at Zero Intensity, FWZI$\sim3000$~\kmps) and narrow component centered at $\sim-2250$~\kmps. For both lines, the shoulder on the red side of the narrow line is more pronounced in the second epoch. This is also the case on the blue side for  {[}\ion{Fe}{2}{]}~5.34~\micron. The narrow component of the {[}\ion{Fe}{2}{]}~5.34~\micron\ line appears somewhat reduced at the second epoch, though this may be due to the increasing flux on the red side. 

%-------------------------------------------
%-------------------------------------------

\subsection{Line velocity structure comparison}
\begin{figure*}[!h]
\centering
\includegraphics[width=6in]{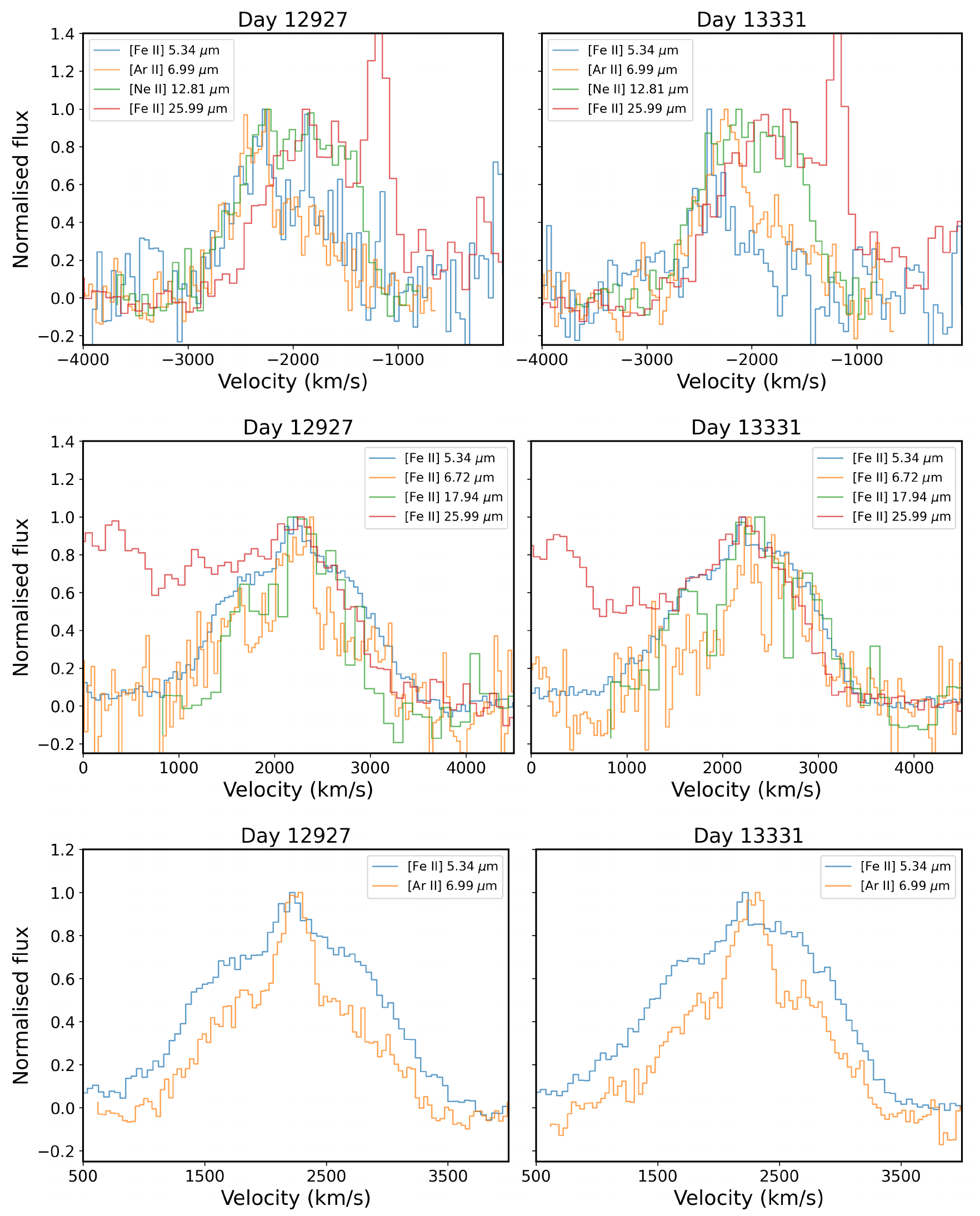}
\caption{Normalized profiles of broad, highly blue- or red-shifted emission lines. All lines from the northern interaction region are shown on the top row, [Fe~{\sc ii}] emission lines from the southern interaction region in the middle row, and bright {[}\ion{Fe}{2}{]}~5.34~\micron\ and bright {[}\ion{Ar}{2}{]}~6.99~\micron\ detected in the southern interaction region (bottom row) at Days 12927 (left column) and 13311 (right column). The species are indicated in the figure legend. The spectra have been normalized to their peak emission. The bright line in the {[}\ion{Fe}{2}{]}~25.99~\micron\ profile in the top row at $\sim-1200$~\kmps\ is {[}\ion{O}{4}{]}~25.89~\micron\ and is unrelated to the interaction. Radial velocities are with respect to the SN~1987A frame, defined by \citet{Groningsson2008} as corresponding to +286.7 km~s$^{-1}$ heliocentric.} 
\label{fig:lines_comp}
\end{figure*}

We assessed whether lines from one or more species exhibit similar profiles which may indicate whether they could be co-spatial in the ejecta structure. Four broad, highly blue-shifted lines were detected in the northern interaction region, namely {[}\ion{Fe}{2}{]}~5.34~\micron, {[}\ion{Ar}{2}{]}~6.99~\micron, {[}\ion{Ne}{2}{]}~12.81~\micron, {[}\ion{Fe}{2}{]}~25.99~\micron. The comparison of their profiles is shown in Fig.~\ref{fig:lines_comp}-left column. We note that the {[}\ion{Fe}{2}{]}~25.99~\micron\ profile comprises both the blue-shifted component from the interaction region as well as a contribution from the inner ejecta on the red side due to the large PSF in band 4C. For this reason, we only consider the blue side of the profile in the following. The {[}\ion{Fe}{2}{]}~5.34~\micron, {[}\ion{Ar}{2}{]}~6.99~\micron\ and {[}\ion{Ne}{2}{]}~12.81~\micron\ lines rise from $\sim-3000$~\kmps\ and peak at $\sim-2200$~\kmps\ on days 12927. The {[}\ion{Fe}{2}{]}~5.34~\micron\ and {[}\ion{Ar}{2}{]}~6.99~\micron\ are also quite similar on the red side at day 12927, both falling from their peaks to the continuum level at $\sim-1000$~\kmps. On Day 13311, {[}\ion{Ar}{2}{]}~6.99~\micron\ peaks at a slightly lower velocity and maintains a separation from {[}\ion{Ne}{2}{]}~12.81~\micron\ as it falls back to the continuum level. The {[}\ion{Ne}{2}{]}~12.81~\micron\ line is broader than both {[}\ion{Fe}{2}{]}~5.34~\micron\ and {[}\ion{Ar}{2}{]}~6.99~\micron, exhibiting a plateau near the peak level between $\sim-2200$ and $\sim1500$~\kmps\ with a sharper fall towards the continuum, also at $\sim-1000$~\kmps. The {[}\ion{Fe}{2}{]}~25.99~\micron\ profile is broad and similar to the {[}\ion{Ne}{2}{]}~12.81~\micron\ line, though extending to the red side likely due to the contribution of the inner ejecta. Given the profile of the {[}\ion{Ne}{2}{]}~12.81~\micron\ line is similar to both {[}\ion{Fe}{2}{]}~5.34~\micron\ and {[}\ion{Ar}{2}{]}~6.99~\micron\ on the blue side, and {[}\ion{Fe}{2}{]}~25.99~\micron\ on the red side, it is possible that it comprises {[}\ion{Fe}{2}{]}~5.34~\micron-like (or {[}\ion{Ar}{2}{]}~6.99~\micron-like) and {[}\ion{Fe}{2}{]}~25.99~\micron-like components.

Apart from the {[}\ion{Fe}{2}{]}~25.99~\micron\ line, the {[}\ion{Fe}{2}{]}~5.34~\micron\ and {[}\ion{Ar}{2}{]} lines are by far the brightest emission lines in the southern interaction region. These are plotted together in Fig.~\ref{fig:lines_comp}, bottom-row. Both lines exhibit a broad and narrow velocity component centered at $\sim2300$~km~s$^{-1}$, with the narrow component being more prominent in the {[}\ion{Ar}{2}{]} line. The broad components are quite similar at both epochs.

%-------------------------------------------
%-------------------------------------------

\subsection{{[}\ion{Fe}{2}{]} line diagnostics}
\label{sect:fe-line-diag}
To enable robust line diagnostics across the MIRI/MRS wavelength range it is essential that spectra are extracted using apertures that account for the decline in spatial resolution with increasing wavelength (see Sect.\ref{sect:spec_extraction} for example). The detected Fe lines in the southern interaction range from {[}\ion{Fe}{2}{]}~5.34~\micron\ to {[}\ion{Fe}{2}{]}~25.99~\micron, almost the full MRS range. We therefore extracted spectra from the southern interaction taking the same approach as in Sect.~\ref{sect:spec_extraction}. We used a circular aperture that enclosed the {[}\ion{Fe}{2}{]}~5.34~\micron\ line and allowed the aperture to grow with increasing wavelength, incrementally increasing by adding an angle equivalent to the Rayleigh criterion at each wavelength. This optimises the extraction apertures across the wavelength range so that the {[}\ion{Fe}{2}{]} emission from the southern interaction is captured correctly while also minimizing the noise contribution so that the fainter {[}\ion{Fe}{2}{]} lines are detectable. Upper limits to important diagnostic lines were determined as described in \citet{Larsson2025}. Our results are given in Table~\ref{tab:fe_lines_diagnostics}.

There are also numerous [\ion{Fe}{2}] lines from the interaction regions detected with NIRSpec at shorter wavelengths \citep{Larsson2023,Larsson2025}. We measured these using the NIRSpec observations from 13,500 days \citep{Larsson2025}, using an elliptical aperture that covers the full interaction region in the southwest. The most interesting lines for the nebular diagnostics are the [\ion{Fe}{2}]  1.534, 1.600, and 1.644-\micron\ lines, where the latter is by far the strongest. These lines all show broad profiles with a peak at $\sim 2200$~km~s$^{-1}$, similar to the [\ion{Fe}{2}]~5.34-\micron\ line in Fig.~\ref{fig:int_profiles}. We also analyzed the weak [\ion{Fe}{2}] 1.664- and 1.677-\micron\ lines, but could only obtain upper limits that did not provide meaningful constraints for the nebular diagnostics. 
%%%%%

\begin{table}[h!]
\caption{Fe line fluxes used for line diagnostics measured for Days~12927 and 13311 MIRI/MRS spectra of the interaction regions (see text for details). The upper limits for {[}Fe~{\sc iii}{]}~22.93~\micron\ and {[}Fe~{\sc ii}{]}~24.52~\micron\ were determined by combining both epochs.}
\label{tab:fe_lines_diagnostics}
\centering
\begin{tabular}{llll}
\hline
Species & $\lambda$ & Day~12927 flux & Day~13311 flux \\
 & ($\mu$m) & ($10^{-23}$ W~cm$^{-2}$) & ($10^{-23}$ W~cm$^{-2}$) \\
\hline
{[}Fe~{\sc ii}{]} & 5.34  & $51.7 \pm 5.4$     & $62.4 \pm 6.4$ \\
{[}Fe~{\sc ii}{]} & 6.72  & $3.77 \pm 0.65$    & $4.12 \pm 0.64$ \\
{[}Fe~{\sc ii}{]} & 17.94 & $2.40 \pm 0.72$    & $6.86 \pm 1.09$ \\
{[}Fe~{\sc iii}{]}  & 22.93 & \multicolumn{2}{l}{$<8.51$} \\
{[}Fe~{\sc ii}{]} & 24.52 & \multicolumn{2}{l}{$<6.75$} \\
{[}Fe~{\sc ii}{]}  & 25.99 & $131 \pm 14$       & $117 \pm 12$ \\
\hline
\end{tabular}
\end{table}

As is clear from 
Fig.~\ref{fig:int_interaction}, as well as from the 3D maps in \citet{Larsson2023} of the [\ion{Fe}{2}] 1.644~$\mu$m line, the high velocity emission close to the ER is coming from the expanding ejecta and not the ER. The emission is localized to the SW and NE regions, where the dense ejecta is closest to the ER and where the X-ray intensity is expected to be strongest. As discussed in \citet{Fransson2013}, the X-rays will ionize and heat especially the parts of the expanding ejecta closest to the ER. Most of the soft X-rays, for which the photoelectric absorption cross section is largest, will be deposited in the outer parts, while the harder X-rays can penetrate deeper. The absorption depends sensitively on the abundances of especially the heavier elements. In the spherically symmetric models of \citet{Fransson2013} the outer parts were dominated by the H/He envelope, while the observations clearly show that the Fe-rich dense ejecta is now close to the ER in the SW and NE directions, resulting in stronger X-ray absorption.

To estimate the physical parameters of the interaction region, assuming these to arise in the same region, we have calculated the ratios of the observed {[}\ion{Fe}{2}{]}  lines in the NIRSpec and MIRI/MRS ranges. For this we have used an extended version of the multilevel atom for \ion{Fe}{2} in \citet{Fransson2024}. This includes a full non-local thermodynamic equilibrium (NLTE) treatment of the statistical equilibrium equations for the lower 750  levels, including recombination from \ion{Fe}{3}. Line transfer is handled by the Sobolev approximation.

Transition rates are taken from \citet{Deb2011}, and  collisional rates from \citet{Smyth2019}. Recombination rates to individual levels are from \citet{Nahar1997}.  

Input parameters are the temperature, electron density, the total Fe abundance, $X(\rm{Fe})=n(\rm{Fe})/n(\rm total)$, and ionic fractions $X($\ion{Fe}{2}$)=n($\ion{Fe}{2})$/n(\rm Fe)$ and $X($\ion{Fe}{3})$=n($\ion{Fe}{3})$/n(\rm Fe)$.
\ion{Fe}{1} is likely to be nearly completely ionized away in the interaction region, supported by the absence of \ion{Si}{1} lines from this region \citep{Larsson2023}, while there may be a minor fraction of higher ionized stages, as reflected in the relatively weak doubly ionized lines relative to the singly ionized lines (Table \ref{tab:int_region_lines}). The \ion{Fe}{3}  fraction is important mainly as a source of recombination contributions to some of the \ion{Fe}{2} lines (see below). We can get an estimate or upper limit to the  \ion{Fe}{3}  fraction from the [\ion{Fe}{3}] 22.93 $\mu$m line relative to the [\ion{Fe}{2}] 25.99 $\mu$m line. 

If electron collisions are the only important excitation process, the ratio of the line fluxes is given by 
\begin{equation}
    \frac{F(\lambda 22.93)}{F(\lambda 25.99)} = \frac{25.99}{22.93} \frac{g_{\rm II}}{g_{\rm III}} \frac{\Upsilon_{\rm III}}{\Upsilon_{\rm II}}  
    \frac{X({\rm Fe~ III})}{X({\rm Fe ~II})} e^{-(630-553)/T_{\rm e}}
\label{eq:fe3_fe2}
\end{equation} 
Here $g_{II} = 10$ is the statistical weight of the \ion{Fe}{2} ${}^{6}$D$_{9/2}$ level and $g_{III} = 9$ is the statistical weight of the \ion{Fe}{3} ${}^{5}$D$_{4}$ level.
The collision strength for the [Fe~{\sc ii}]~25.99 $\mu$m transition is $\Upsilon_{II} \sim 5.0$ at $10^4$ K \citep{Ramsbottom2007} and the transition rate $2.56 \times 10^{-3}$ s$^{-1}$ and for [Fe~{\sc iii}]~22.93 $\mu$m it is $\Upsilon_{III} \sim 2.5$ at $10^4$ K \citep{Badnell2014} and the transition rate $2.8 \times 10^{-3}$ s$^{-1}$. The critical electron densities for collisional de-excitation are therefore $9.1 \times 10^4 (T_{\rm e}/10^4 {\rm K})^{1/2}$ cm$^{-3}$ for the [Fe~{\sc ii}] transition and $6.0 \times 10^4 (T_{\rm e}/10^4 {\rm K})^{1/2}$ cm$^{-3}$ for the [Fe~{\sc iii}] transition. We therefore assume that de-excitation is negligible at the densities here. Then 
\begin{equation}
\frac{X({\rm Fe~ III})}{X({\rm Fe ~II})} = 2.0 \ e^{77/T_{\rm e}} \frac{F(\lambda 22.93)}{F(\lambda 25.99)}
\label{eq:fe3_fe2_2}
\end{equation}
The ratio is therefore insensitive to temperature.
As an upper limit we find $F({\rm \lambda 22.93)} /{F(\lambda 25.99)} \lesssim 0.1$, so X(Fe~III)/X(Fe~II) $ \lesssim 0.2$. 

Optical depth effects may be important, depending on the [Fe~{\sc ii}] density, $n({\rm Fe \ II})$, in the emitting region. For homologous expansion the optical depth is given by
\begin{equation}
    \tau = \frac{\lambda^3 A_{u,l} g_u }{8 \pi g_l} \ n({\rm Fe \ II}) \ t
\end{equation}
where $t$ is the time since explosion. For the [Fe~{\sc ii}]~25.99~$\mu$m line at $13,000$ days we get 
\begin{equation}
    \tau = 1.34 \times 10^{-3} n({\rm Fe \ II}) \ .
\end{equation}
For $n({\rm Fe \ II}) \gtrsim 10^3 $ cm$^{-3}$ the line may therefore become optically thick. The optical depth of the [Fe~{\sc ii}]~5.34~\micron\ line is a factor $\sim 700$ smaller, reflecting that optical depth effects mainly affect the long wavelength lines.

\begin{figure*}[!h]
\centering
\includegraphics[width=13cm]{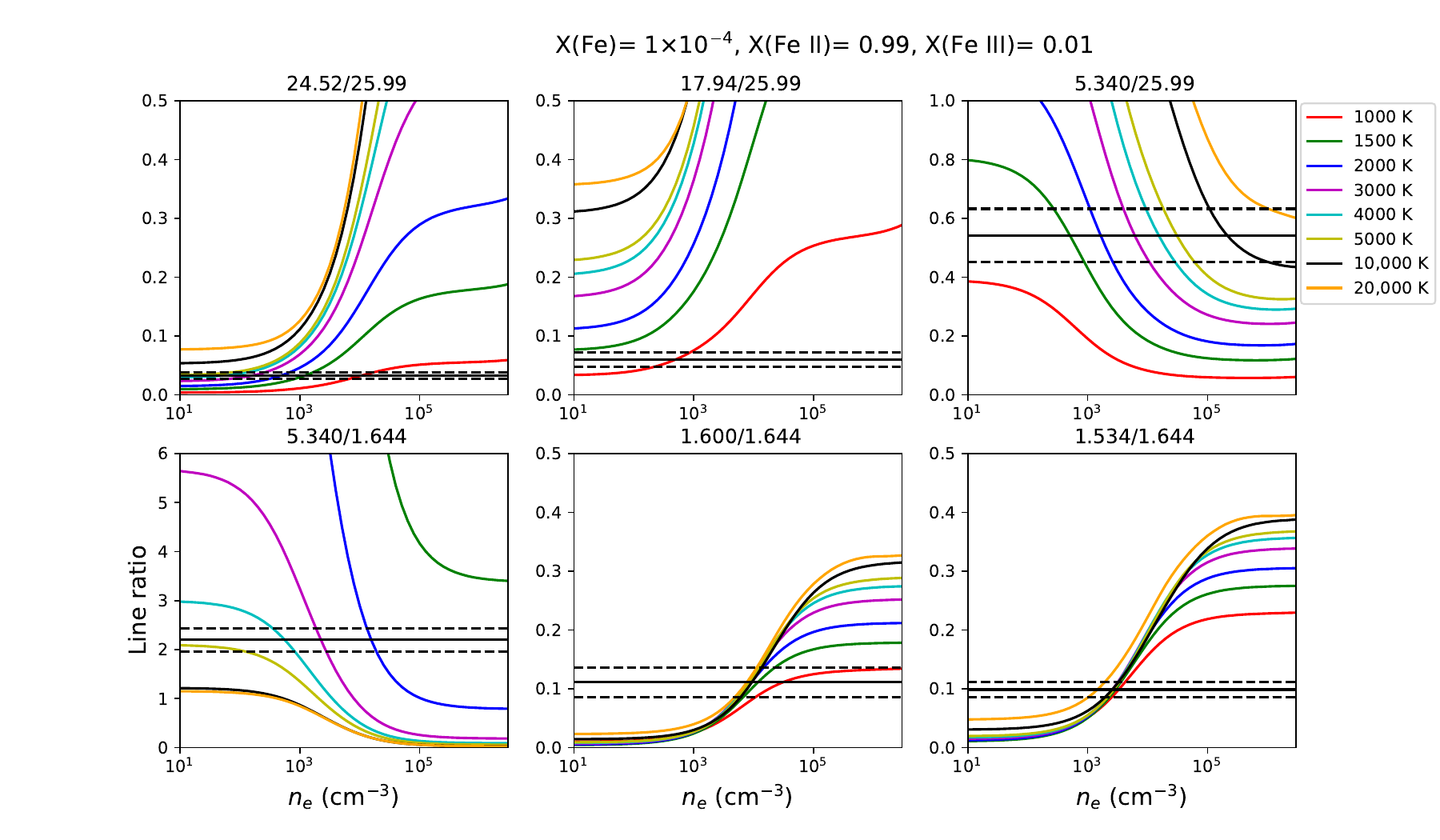}
\includegraphics[width=13cm]{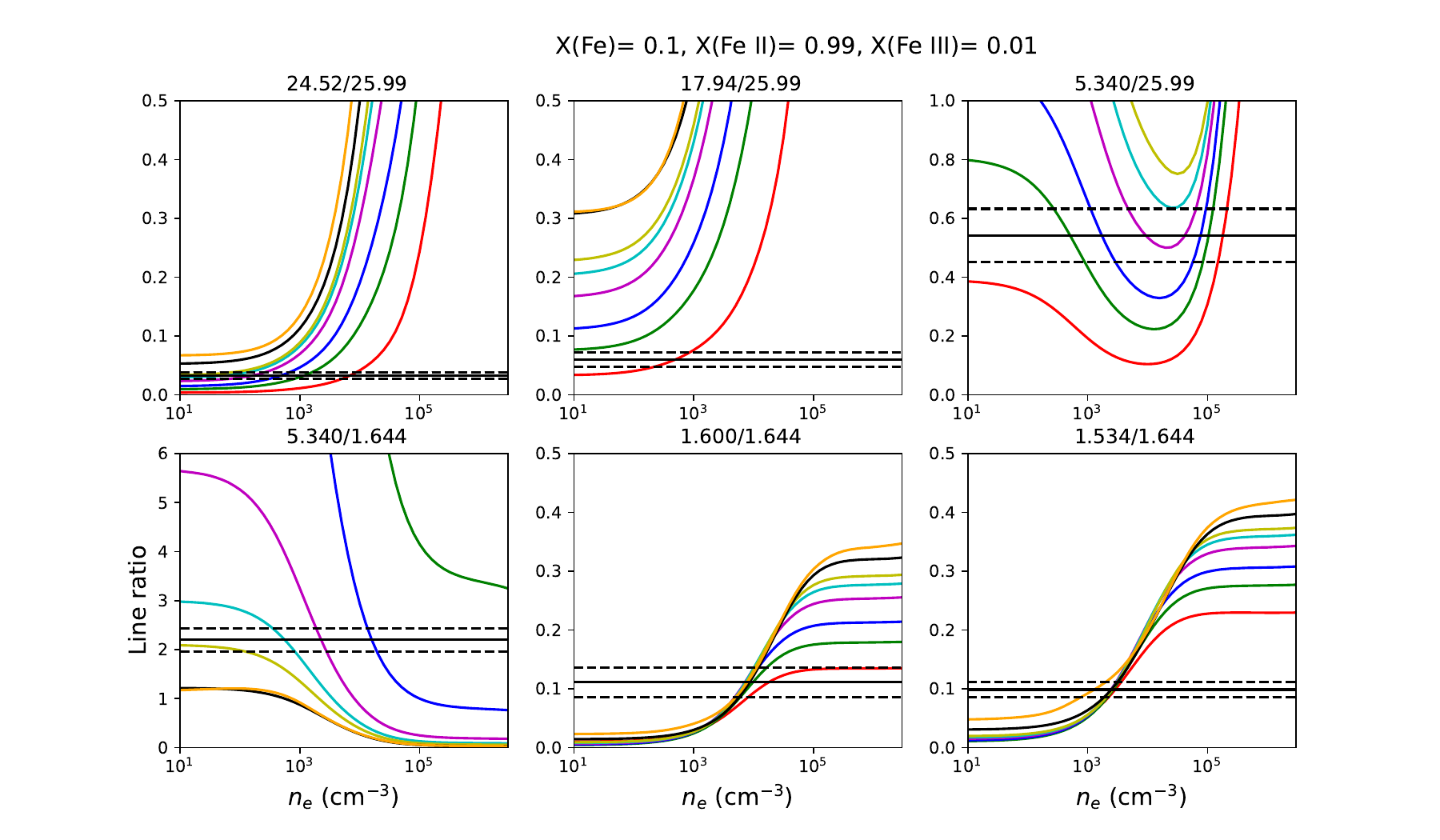}
\includegraphics[width=13cm]{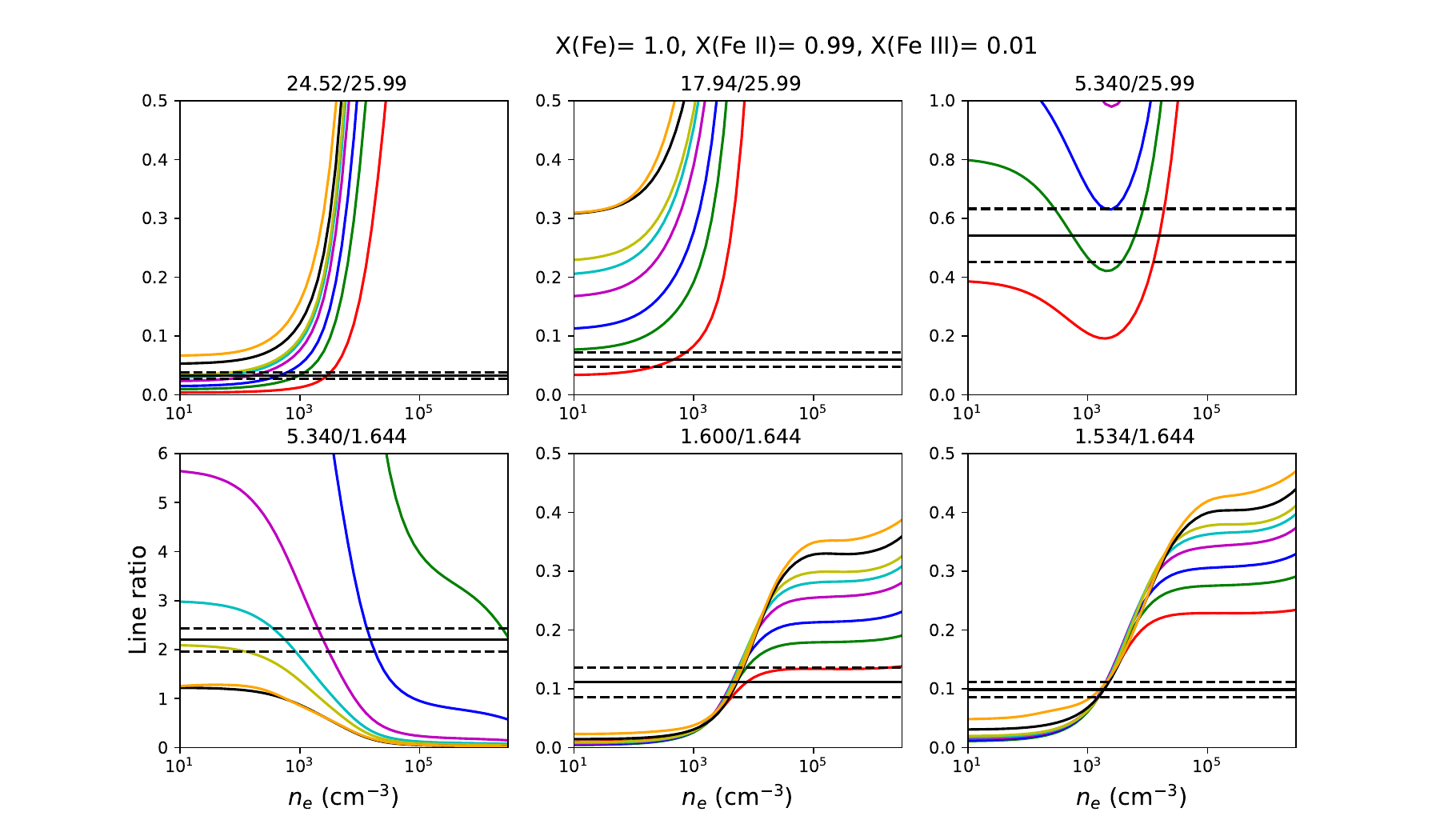}
\caption{Line ratios of the most important diagnostic [Fe~{\sc ii}] lines as a function of electron density, $n_{\rm e}$, and for temperatures of $10^3 - 2\times 10^4$ K. Observed line ratios are shown as solid horizontal lines, including errors (dashed lines). The three
different groups of figures show the effect of varying the total Fe abundance, X(Fe).}
\label{fig:feii_diagn}
\end{figure*}

In Fig. \ref{fig:feii_diagn} we show the line ratios of the most important diagnostic lines in the NIRSpec and MIRI/MRS ranges for different Fe abundances. We assume that once-ionized atoms dominate, as is consistent with the observed ions. From the discussion above we assume only a minor fraction of [Fe~{\sc iii}], $\sim 10^{-2}$.
To illustrate the sensitivity of the Fe abundances to the optical depth, we  show three cases with increasing Fe abundance, $X({\rm Fe})=1 \times 10^{-4}$. $X({\rm Fe})=0.1$ and $X({\rm Fe})=1.0$. 
The upper panel of Fig. \ref{fig:feii_diagn} shows the simplest case with only collisional excitation/de-excitation and radiative de-excitation important. The Fe abundance is here too low to explain the strong iron lines, but together with the other two cases illustrate the effects of the optical depth on the line ratios to $T_{\rm e}$ and $n_{\rm e}$. 

The 25.99 $\mu$m line of [Fe~{\sc ii}] arises from a transition between the lowest two fine-structure levels, while the 17.94 $\mu$m and 24.52 $\mu$m lines originate from the first excited multiplet. The 5.340 $\mu$m line connects the lowest levels of these two multiplets. The NIR lines at $1.53-1.75$ $\mu$m represent transitions between the third and second multiplets. The excitation temperature of the upper level of the 25.99 $\mu$m fine structure line is only 553 K, while it is  2694 K and  11445 K for the second and third multiplets, respectively. This explains the temperature sensitivity of both the $25.99 / 5.340$ $\mu$m and $5.340 / 1.664$ $\mu$m ratios at low densities. Transitions between fine-structure lines in the same multiplet are mainly sensitive to the electron density, where the levels at high density are in LTE, and the line ratios are nearly constant for a given temperature. 

For a high [Fe~{\sc ii}] abundance and high density, 
the $25.99$ $\mu$m, 17.94 $\mu$m and 24.52 $\mu$m lines, become optically thick, limiting their emissivity, while transitions between the NIR multiplets remain optically thin. This explains the decrease in the line ratios in the upper row in the lower two panels of 
Fig. \ref{fig:feii_diagn}.

Comparing to our observed line ratios, shown as horizontal lines in Fig \ref{fig:feii_diagn} , we find that for $X({\rm Fe})=1 \times 10^{-4} - 1.0$  the $1.600/1.644 \ \mu$m and $1.534/1.644$ $\mu$m ratios indicate an electron density $1.3 \times 10^3 \lesssim n_{\rm e} \lesssim  1.5 \times 10^4$ cm$^{-3}$. If we limit this to the range $X({\rm Fe})=0.1  - 1.0$ then $1.3 \times 10^3 \lesssim n_{\rm e} \lesssim  8 \times 10^3$ cm$^{-3}$, nearly independent of the temperature. As pointed out above, the $5.340 /25.99 / 5.340$ $\mu$m and $5.340 / 1.664$ $\mu$m ratios are most sensitive to the temperature. For the above electron density range the $5.340 / 1.664$ $\mu$m ratio results in a range of $1500 \lesssim T_{\rm e} \lesssim  5000$ K, while the $ 5.340/25.99$ $\mu$m ratio indicate a more limited range, $1500 \lesssim T_{\rm e} \lesssim  2000$ K. In this analysis one should keep in mind that the atomic data, especially the collision strengths, have an uncertainty, which in the absence of experimental data is difficult to assess.

\subsection{Implications for the nature of the ejecta - ER collision}
Our observations show not only lines from [Fe~{\sc ii}], but also from [Ne~{\sc ii}], [Ne~{\sc iii}], [S~{\sc iii}], [S~{\sc iv}], [Cl~{\sc ii}], [Ar~{\sc ii}], [Ar~{\sc iii}] and [Ni~{\sc ii}]. However, broad, redshifted lines from H~{\sc i} and He~{\sc i} are weak,  which indicates that the dense ejecta now interacting with the ER comes from the inner metal rich core. 

The ejecta temperature close to the ER found above, $1500 - 3000$ K, is much higher than in the core, where the energy input is only due to radioactive ${}^{44}$Ti heating and ionization. The temperature in the Fe-rich core region was estimated to $\sim 170$ K at 8 years after explosion \citep{Jerkstrand2011}.  Although the density in the core is a factor $\sim 20$ lower now, we do not expect the temperature to be much higher without the X-ray heating from the ER. Another observational indication of the effect of the X-ray input is the very low $5.340 / 25.99$ $\mu$m ratio from the central region, compared to that at high velocities, as seen in the line profiles \citep[middle panel in Fig. 15 of][]{Jones2023}. 

From spherically symmetric explosion models, having a steep outer ejecta profile $\rho \propto V^{-8.6}$, the density close to the reverse shock is estimated to be $\rho = 3.3 \times 10^{-21} k (t/36 \rm{yrs})^{5.6}$ g cm$^{-3}$, where $k=1-3$ \citep{Fransson2013}. To estimate the mass density from the electron density, derived above, one needs to determine the state of ionization, as well as the relative abundances of the different elements. This requires a complete shock model, including a self-consistent determination of the X-ray and ${}^{44}$Ti input \citep{Fransson2013}, which is out of scope for this paper. If we, however, assume an Fe abundance $X({\rm Fe})$ and that the electron density is dominated by singly ionized Fe, we can get an approximate estimate of the mass density $\rho \approx \sim 5 \times 10^3 \times 56 \times 1.67 \times 10^{-24} X({\rm Fe}) \approx 4.7 \times 10^{-19} X({\rm Fe})$ g cm$^{-3}$. For $X({\rm Fe}) \gtrsim 10^{-2}$ this is larger than that of the spherically symmetric ejecta model above, and consistent with the strong mixing and aspherical geometry seen from direct 3D imaging \citep{Larsson2023}.
Therefore, we find that, independent of the detailed parameters, the electron density shows that it is now likely that the dense inner Fe-rich ejecta has now reached the reverse shock.

%--------------- DISCUSSION ----------------%
%--------------- DISCUSSION ----------------%
%\section{Discussion}\label{sect:discussion}
%\textcolor{red}{TBA}

%--------------- CONCLUSIONS ----------------%
%--------------- CONCLUSIONS ----------------%
\section{Conclusions}\label{sect:conclusions}
We have presented the spatially-resolved spectroscopic evolution of SN 1987A in the mid-IR with two epochs of MIRI/MRS observations at Days 12927 and 13311. We find:

\begin{itemize}

    \item The `total' spectrum continuum shows little evolution between Days 12927 and 13311 in both shape and spectral fit parameters. However, brightness evolution maps revealed the inner ER structure to be fading while the outermost regions are brightening as more material is swept-up and heated by the blast wave. Both these effects appear to counteract each other resulting in a largely unchanged `total' spectrum between the epochs. 

    \item We assessed whether the 30--70~\micron\ excess reported by \citet{Matsuura2019} could be contributing to the continuum at the longer wavelengths by adding an additional dust component to our fits. We found that the introduction of an astronomical silicates component with $T\approx130$~K to represent the excess component accounts for the $>20$~\micron\ region better than our other models. In addition, a comparison of our fits to the 31.5~\micron\ SOFIA flux from \citet{Matsuura2019} indicated that our additional excess component could account for this measurement, though noting that the excess has likely evolved in the intervening time. We attempted to localize the excess component by creating continuum maps $>24$~\micron\ where the excess component is the primary contributor to the continuum. We found that the distribution  $>24$~\micron\ follows that of the ER. We assessed the evolution of the excess component which again suggests that the excess is likely related to the ER dust, possibly resulting from a top-heavy grain size distribution as suggested in \citet{Matsuura2015} and J23.

    \item We assessed the impact of changing the size and shape distribution for dust compositions expected to be more representative of the ER dust rather than the astrodust ISM model, given that the progenitor was a B supergiant which may have undergone a prior RSG phase. Our qualitative analysis found that the O-rich dust %circumstellar medium 
    composition of \citet{Ossenkopf1992} with a CDE shape distribution can account for the location of the `dip' between the 10-\micron\ and 20-\micron\ silicate features, while its continuum shape at longer wavelengths follows that of \sn\ well. However, there are problems in fitting the shorter wavelengths with this composition, likely due to the absence of species adding opacity in this wavelength range. This is often the case for mass losing evolved stars where also metallic Fe or amorphous C is added.

    \item We used spatially deconvolved data cubes to assess the dust continuum in four regions of the ER. We fitted the spectra from each region using a two-temperature astrodust model with a fixed synchrotron component to allow us to assess the variation in the warm and hot components using a single composition. We found the hot component to be concentrated in the western ER, with the derived mass for our west region being more than double that of the other three regions combined. The warm dust mass is also largest in the west region. The temperature of the warm dust is consistent around the ER, while the hot dust temperature is lower in the west than in the other regions. There was minimal evolution in the dust parameters between Days 12927 and 13311. Some trends are present. The warm dust mass component appears to be decreasing slightly everywhere apart from the west region, which shows little evolution. Similarly, the warm component temperature decreased in all regions apart from the west. Future observations will be required to confirm these trends.

    %\item \textcolor{red}{PJK: Mike, please check the following and edit as needed}. 
    \item We analysed emission lines from spectra extracted from the `total' region, from the ejecta region and from the four cardinal point sub-regions, for both Days 12927 and 13311. This included the more broadened lines from singly ionized species (e.g. H~I, [Ni~{\sc ii}], [Ar~{\sc ii}] and [Ne~{\sc ii}]), attributed to an origin in cooling post-shock gas in the ER, and the narrow-FWHM lines (consistent with being spectrally unresolved by the MRS) from more highly ionized species (e.g. [S~{\sc iv}], [O~{\sc iv}], [Ne~{\sc v}] and [Ne~{\sc vi}]), which are attributed to an origin in much lower density diffuse gas that had been flash ionized by the UV pulse from the supernova explosion. We assessed for variations in the line fluxes between the two epochs. The majority of the line fluxes are consistent within the uncertainties with a ratio close to unity. Exceptions to this were found to be lines from highly ionized species, which we determined to be related to the different background treatments at each epoch, with an in-field background used for Day 12927 and a dedicated background pointing for Day 13311. Other exceptions include lines due to the ejecta/ER interaction.

    \item We identify molecular lines from H$_2$ from the inner ejecta. Near-IR lines of H$_2$ from the ejecta were previously found by \citet{Fransson2016} and studied in more detail in \cite{Larsson2023}. We detected lines at 5.5115, 6.1088, 6.9091, 8.0258, 9.6649 and 17.035~$\mu$m, which are coming from rotational transition from the ground vibrational level, 0-0 S(1) to S(7). %with the exception of the 0-0 S(2) transition at $12.2785 \mu$m. 
    These lines are all among the strongest lines predicted in the MIRI/MRS range from PDR Models Rw3o and Qm3o of \citet{Draine1996}, which are characterized by a large UV flux and high density. Our results are consistent with the modeling of \cite{Larsson2023} using near-IR H$_2$ lines.

    \item We assessed the morphological and spectral evolution of the interaction regions detected by \cite{Larsson2023} and J23, over the 384 day period between the two epochs. We isolated the broad, highly blue- and red-shifted components and found  evolving spatial and spectral properties in both the northern and southern interaction regions for several lines, including [Fe~{\sc ii}]~5.34~\micron, [Ar~{\sc ii}]~6.99~\micron, and [Ne~{\sc ii}]~12.81~\micron. In general, the lines are brighter and more spatially extended at Day~13311. We expect this trend to continue into the future as more and more of the inner ejecta interacts with the ER. 

    \item From the line ratios of the [Fe~{\sc ii}] NIR and mid-IR lines we find an electron density of $1300-15,000$~cm$^{-3}$
and an electron temperature of $1500-2000$~K in the interaction region of the metal rich ejecta, consistent with X-ray heating from the ejecta - ER collision. We also see broad lines from Ne, S, Cl, Ar and Ni, broad lines from H and He are weak or absent from the interaction region spectra.  %

\end{itemize}

\begin{acknowledgments}

The authors thank the anonymous referee whose comments and suggestions helped improve the paper. 
This work is based on observations made with the NASA/ESA/CSA James Webb Space Telescope. The data were obtained from the Mikulski Archive for Space Telescopes at the Space Telescope Science Institute, which is operated by the Association of Universities for Research in Astronomy, Inc., under NASA contract NAS 5-03127 for JWST. These observations are associated with programs \#1232 and \#2763. 

PJK, RW, and JJ acknowledge support from the Research Ireland Pathway programme under Grant Number 21/PATH-S/9360. JL acknowledges support from the Knut \& Alice Wallenberg Foundation. OCJ acknowledges support from an STFC Webb fellowship. MM and NH acknowledge support through a NASA/JWST grant 80NSSC22K0025, and MM and LL acknowledge support from the NSF through grant 2054178. MM and NH acknowledge that a portion of their research was carried out at the Jet Propulsion Laboratory, California Institute of Technology, under a contract with the National Aeronautics and Space Administration (80NM0018D0004). JB thanks the Belgian Federal Science Policy Office (BELSPO) for the provision of financial support in the framework of the PRODEX Programme of the European Space Agency (ESA). RDG was supported, in part, by the United States Air Force. T.T. acknowledges support from the NSF grant AST-2205314 and the NASA ADAP award 80NSSC23K1130.

\end{acknowledgments}

%--------------------------------------------------------------------------------------------------------------
\facilities{All of the MIRI/MRS data presented in this paper were obtained from the
Mikulski Archive for Space Telescopes (MAST) at the Space Telescope Science 
Institute. The specific observations analyzed can be accessed via
\dataset[DOI:10.17909/pz4v-ej33]{https://doi.org/10.17909/pz4v-ej33
[doi.org]} and \dataset[DOI:10.17909/6z9k-9194]{https://doi.org/10.17909/6z9k-9194
[doi.org]}.
}

%--------------------------------------------------------------------------------------------------------------

\software{Astropy \citep{Astropy2018, Astropy2022},  
          specutils \citep{specutils},
          Photutils \citep{Bradley2022},
          Regions \citep{Bradley2022b}.
          }

%--------------------------------------------------------------------------------------------------------------

%\bibliographystyle{aasjournal}

%--------------- APPENDIX ----------------%
%--------------- APPENDIX ----------------%

\vspace{2cm}

%-----BACKGROUND-REGION-SPECTRA-------%
\appendix
\section
{The line spectra of the Cycle 1 and 2 background regions}
In Table~\ref{tab:background_spectra} we list the relative line fluxes measured in the Cycle~1 and 2 background region spectra, as discussed in Section~\ref{sect:line_fluxes}.

\begin{table*}
\caption{Detected emission lines, their measured radial velocities\tablenotemark{a}, full-width half maxima, and line fluxes relative to [Ne~{\sc iii}] 15.5551~$\mu$m = 100.0, for the background regions used for Cycles 1 and 2. } 
\label{tab:background_spectra}
\centering
\begin{tabular}{llllllll}
\hline
\hline
 & & \multicolumn{3}{c}{Cycle~1 background region} & \multicolumn{3}{c}{Cycle~2 background region} \\
\cmidrule(lr){3-5}\cmidrule(lr){6-8}
Species & $\lambda_{\rm lab}$\tablenotemark{b} 
& Velocity & FWHM & Relative Flux & Velocity & FWHM & Relative Flux 
\\ 
 & $\mu$m & km~s$^{-1}$ & km~s$^{-1}$ &  & km~s$^{-1}$ & km~s$^{-1}$ &   \\
\hline
{[}Ar~{\sc ii}{]} & 6.985274 & 2.0$\pm$2.9 & 91.6$\pm$7.8 & 4.64$\pm$0.33 & -- & -- & -- \\
H~{\sc i}+He~{\sc i} 6-5 & 7.459858 & -11.6$\pm$5.6 & 86.4$\pm$12.5 & 2.32$\pm$0.31 & -- & -- & -- \\
{[}Ne~{\sc vi}{]} & 7.6524 & -13.3$\pm$4.0 & 74.0$\pm$22.1 & 4.47$\pm$0.66 & -- & -- & -- \\
{[}Ar~{\sc iii}{]} & 8.99138 & -24.1$\pm$2.0 & 96.7$\pm$2.0 & 19.3$\pm$0.34 & -26.2$\pm$2.1 & 93.6$\pm$5.1 & 18.3$\pm$0.87 \\
{[}S~{\sc iv}{]} & 10.51049 & -21.8$\pm$0.5 & 92.4$\pm$1.2 & 22.3$\pm$0.26 & -23.5$\pm$1.1 & 90.6$\pm$3.5 & 25.3$\pm$0.63 \\
H~{\sc i}+He~{\sc i} 7-6 & 12.371898 & -21.9$\pm$8.5 & 149.0$\pm$31.6 & 2.74$\pm$0.42 & -- & -- & -- \\
{[}Ne~{\sc ii}{]} & 12.813548 & -22.9$\pm$0.4 & 100.1$\pm$1.4 & 30.9$\pm$0.31 & -20.8$\pm$0.8 & 92.1$\pm$3.0 & 23.4$\pm$0.47 \\
{[}Ne~{\sc v}{]} & 14.32168 & -15.5$\pm$3.0 & 92.4$\pm$7.9 & 2.34$\pm$0.16 & -- & -- & -- \\
{[}Ne~{\sc iii}{]} & 15.5551 & 5.8$\pm$0.2 & 116.5$\pm$0.61 & 100.0$\pm$0.5 & 3.0$\pm$0.2 & 116.6$\pm$0.6 & 100.0$\pm$0.4 \\
{[}S~{\sc iii}{]} & 18.71303 & -22.5$\pm$0.4 & 139.6$\pm$0.6 & 86.6$\pm$0.44 & -18.5$\pm$1.3 & 137.8$\pm$2.4 & 74.2$\pm$1.27 \\
{[}Ne~{\sc v}{]} & 24.3175 & -12.0$\pm$7.2 & 120.1$\pm$14.6 & 5.21$\pm$0.60 & -- & -- & -- \\
{[}O~{\sc iv}{]} & 25.8903 & -24.1$\pm$2.3 & 132.3$\pm$5.7 & 35.1$\pm$1.33 & -- & --  & -- \\
\hline
\end{tabular}
\flushleft
\tablenotetext{a}{Radial velocities are with respect to the SN~1987A frame, defined by \citet{Groningsson2008} as corresponding to +286.7 km~s$^{-1}$ heliocentric.}
\tablenotetext{b}{Vacuum wavelengths from the compilation by \cite{vanHoof2018}.}
\end{table*}

\end{document}